\documentclass[twocolumn]{aastex631}
\usepackage{amsmath,apjfonts,amsfonts,
amssymb,enumerate, graphicx, natbib, 
times,textcomp,verbatim,url,nicefrac}                                   
\def\apjl{ApJ}%
\def\apjs{ApJS}%
\def\ao{Appl.~Opt.}%
\def\aap{A\&A}%
\newcommand{\tess}{\textit{TESS}}

\newcommand{\dynesty}{\texttt{dynesty}}

\newcommand{\NsupernovaTotal}{850}
\newcommand{\NsupernovaLCs}{307}
\newcommand{\NsupernovaRemoved}{543}
\newcommand{\NsupernovaHighDR}{74}
\newcommand{\NsupernovaFL}{56}

\newcommand{\NsupernovaHighDRHQ}{10}
\newcommand{\NsupernovaPeak}{46}

\newcommand{\meanPLAll}{1.93 $\pm$ 0.57}
\newcommand{\meanPLHQ}{2.06 $\pm$ 0.51}
\newcommand{\meanPLnoFL}{1.87 $\pm$ 0.51}

\newcommand{\meanTriseAll}{15.7 $\pm$ 3.5}
\newcommand{\meanTriseHQ}{15.2 $\pm$ 0.8	}
\newcommand{\meanTrisenoFL}{15.2 $\pm$ 2.6}

\newcommand{\threeSigmaLimit}{31 R$_\odot$ (separations  $> 5.7\times 10^{12}$~cm}
\newcommand{\twoSigmaLimit}{10 R$_\odot$ (separations  $> 1.9\times 10^{12}$~cm}
\newcommand{\NsupernovaLessTenRsun}{31}

\tabletypesize{\footnotesize}                                                             
\shorttitle{\tess\ Ia Supernovae}    
 \shortauthors{Fausnaugh et al.}    

\begin{document}

\title{  Four years of Type Ia Supernovae Observed by TESS: Early Time Light Curve Shapes and Constraints on Companion Interaction Models}

\author[0000-0002-9113-7162]{M.~M.~Fausnaugh}
\affil{Department of Physics and Kavli Institute for Astrophysics and Space Research, Massachusetts Institute of Technology, Cambridge, MA 02139, USA}
\affil{Department of Physics \& Astronomy, Texas Tech University, Lubbock TX, 79410-1051, USA}

%Core SNe authors
\author{P.~J.~Vallely}
\affil{Department of Astronomy, The Ohio State University, 140 West 18th Avenue, Columbus, OH 43210, USA}

\author[0000-0002-2471-8442]{M.~A.~Tucker}
\affil{Institute for Astronomy, University of Hawai'i, 2680 Woodlawn Drive, Honolulu, HI 96822, USA}
\affil{DOE CSGF Fellow}

\author{C.~S.~Kochanek}
\affil{Department of Astronomy, The Ohio State University, 140 West 18th Avenue, Columbus, OH 43210, USA}
\affil{Center for Cosmology and AstroParticle Physics, The Ohio State University, 191 W. Woodruff Ave., Columbus, OH 43210, USA}

\author[0000-0003-4631-1149]{B.~J.~Shappee}
\affil{Institute for Astronomy, University of Hawai'i, 2680 Woodlawn Drive, Honolulu, HI 96822, USA}

\author{K.~Z.~Stanek}
\affil{Department of Astronomy, The Ohio State University, 140 West 18th Avenue, Columbus, OH 43210, USA}
\affil{Center for Cosmology and AstroParticle Physics, The Ohio State University, 191 W. Woodruff Ave., Columbus, OH 43210, USA}

%tess team
\author{George R.\ Ricker}
\affil{Department of Physics and Kavli Institute for Astrophysics and Space Research, Massachusetts Institute of Technology, Cambridge, MA 02139, USA}

\author{Roland\ Vanderspek}
\affil{Department of Physics and Kavli Institute for Astrophysics and Space Research, Massachusetts Institute of Technology, Cambridge, MA 02139, USA}

\author{Manan~Agarwal}
\affil{Department of Physics and Kavli Institute for Astrophysics and Space Research, Massachusetts Institute of Technology, Cambridge, MA 02139, USA}

\author[0000-0002-6939-9211]{Tansu~Daylan}
\affiliation{Department of Astrophysical Sciences, Princeton University, Princeton, NJ 08544, USA}
\affiliation{LSSTC Catalyst Fellow}

\author[0000-0002-7778-3117]{Rahul~Jayaraman}
\affil{Department of Physics and Kavli Institute for Astrophysics and Space Research, Massachusetts Institute of Technology, Cambridge, MA 02139, USA}

\author{Rebekah~Hounsell}
\affiliation{University of Maryland, Baltimore County, Baltimore, MD 21250, USA}
\affiliation{NASA Goddard Space Flight Center, Greenbelt, MD 20771, USA}

\author[ 0000-0002-5788-9280]{Daniel~Muthukrishna}
\affiliation{Department of Physics and Kavli Institute for Astrophysics and Space Research, Massachusetts Institute of Technology, Cambridge, MA 02139, USA}

\begin{abstract}

We present \NsupernovaLCs\ Type Ia supernova light curves from the first four years of the \tess\ mission, sectors~1--50.  We use this sample to characterize the shapes of the early time light curves, measure the rise times from first light to peak, and  search for companion star interactions.  Using simulations, we show that light curves must have noise $<$\,10\% of the peak flux to avoid biases in the early time light curve shape, restricting our quantitative analysis to \NsupernovaHighDR\  light curves.  We find that the mean power law index $t^{\beta_1}$ of the early time light curves is $\beta_1 = $\meanPLAll\ and the mean rise time to peak is \meanTriseAll\ days. We also estimate the underlying population distribution for $\beta_1$ and find a Gaussian component with mean 2.29, width 0.34, and a long tail extending to values less than 1.0.  We  use model comparison techniques to test for the presence of companion interactions. In contrast to recent results in the literature, we find that the data can rarely distinguish between models with and without companion interaction models, and caution is needed when claiming detections of flux excesses in early time supernova light curves.  Nevertheless, we find three high-quality SN light curves that tentatively prefer the addition of a companion interaction model, but the statistical evidence for the companion interactions is not robust. We also find two SNe  that disfavor the addition of a companion interaction model to a curved power law model.  Taking the \NsupernovaHighDR\ SNe together, we calculate 3$\sigma$ upper limits on the presence of  companion signatures to control for orientation effects that can hide companions in individual SN light curves.  Our results rule out  common progenitor systems with companions having Roche lobe radii $>$\,\threeSigmaLimit,  99.9\% confidence level) and disfavor companions having Roche lobe radii $>$\,\twoSigmaLimit, 95\% confidence level).  Lastly, we discuss the implications of our results for the intrinsic fraction of single degenerate progenitor systems.

\end{abstract}

\section{Introduction \label{sec:intro}}

Transient surveys are pushing to larger and larger sky coverage and faster and faster cadence.  As a result, the number of supernovae (SNe) discovered per year is growing at an exponential rate, and the densely sampled light curves enable new scientific investigations.  SNe light curves that capture the time of explosion can address questions related to the mechanism of explosion, the characteristics of shock breakout and shock cooling, the circumstellar environment, and the nature of the progenitor system.  For Type Ia SNe in particular, the nature of the progenitor systems is a long-standing question \citep{Whelan1973,Nomoto1984, Iben1984,Webbink1984, Iwamoto1999, Hillebrandt2000,Podsiadlowski2008,Woosley2011}, with the relative rates of single- and double-degenerate progenitor systems remaining uncertain \citep{Hillebrandt2013,Hoflich2013,Maoz2014,Strolger2020}.

Early time Type Ia SN light curves constrain the progenitor systems because companion stars will produce an observable signal when struck by the supernova blast wave \citep{Marietta2000, Maoz2014}.  Models for the interaction of a companion star and SN blast wave were developed by \citet{Kasen2010}.  When the blast wave encounters a companion star, the ejecta form a bow shock.  The shocked ejecta are compressed and heated, producing thermal emission at X-ray, UV, and optical wavelengths  that lasts for 1--14 days after the initial explosion and reaches luminosities of $10^{43}$~erg~s$^{-1}$.  Combined with the flux from the SN ejecta powered by $^{56}$Ni decay, the flux from the companion interaction modifies the shape of the early time light curve.  

A complicated shape in the early time SN light curve might therefore suggest the presence of a companion star.  However, other effects can  mimic a companion interaction signature, such as the distribution of $^{56}$Ni in the SN ejecta or the presence of circumstellar material  \citep{Piro2012, Piro2016, Contreras2018, Maeda2018, Polin2019, Magee2020a,Magee2020b}.   The amplitude, timescale, and color of the early time light curve can distinguish between these different scenarios.  In practice, there are a handful Type Ia SN light curves with candidate companion interactions, but their interpretation is debated---see, for example,  SN2012cg \citep{Marion2016,Shappee2018}, iPTF14agt   \citep{Cao2015, Kromer2016}, and SN2018oh \citep{Shappee2019, Dimitriadis2019}.  In some cases, the UV/optical light curves have complicated shapes, but the color evolution cannot be fit by the \citet{Kasen2010} models, as for SN2017cbv, SN2019yvq, SN2020hvf, and 2021aefx  \citep{Hosseinzadeh2017, Miller2020a, Tucker2021, Jiang2021,Ashall2022,Hosseinzadeh2022}.  Using physical models to search for excess flux in early time Type Ia light curves, \citet{Magee2022} and \citet{Deckers2022} estimate that 18--28\% of Type Ia light curves show an early time flux component, some of which are consistent with companion interactions.  \citet{Burke2022a} and \citet{Burke2022b} took a similar approach but defined the excess relative to an empirical template and found a rate of 33\%.

On the other hand, the majority of early time Type Ia SNe light curves show a monotonic rise that is likely associated with the SN ejecta alone.  The smooth rise implies an upper limit on any companion star's orbital separation, or requires an unfavorable viewing angle such that the bow shock is oriented away from the observer.  For SN2011fe, there are very tight limits on the presence of companion interaction signatures within 4 hours of explosion.  Even if the bow shock was viewed at an unfavorable angle, it is difficult to hide the interaction signature of a main sequence companion and  SN2011fe was likely a product of the double degenerate scenario \citep{Bloom2012}.  A similar result was found for SN2014J \citep{Siverd2015, Goobar2015}.  For more distant SNe, the limits on early time light curves are less constraining and a large sample of Type Ia  light curves is needed to control for the possibility of unfavorable viewing angles.

In this work, we use 50 sectors of Transiting Exoplanet Survey Satellite (\tess) observations to analyze early time light curves of \NsupernovaLCs\ Type Ia SNe.  Before \tess, only {\it Kepler} could provide continuous coverage of early time SN light curves \citep[see also \citealt{Rest2018,Ridden-Harper2019,Ridden-Harper2020} for {\it Kepler} light curves of other transients]{Olling2015, Garnavich2016, Shappee2019,Dimitriadis2019,Wang2021}.  With its large field-of-view, \tess\  observes a large number of bright SNe, averaging six Type Ia SN light curves for every sector.    \tess\ has also made important contributions to  extragalactic transient studies more broadly:  \citet{Vallely2021} used \tess\ data to measure rise times of 22 core-collapse supernova from 2018--2020, while high cadence \tess\ light curves have enabled new investigations of  Tidal Disruption Events \citep{Holoien2019}, Gamma Ray Bursts \citep{Smith2021}, and Active Galactic Nuclei  \citep{,Burke2020,Weaver2020, Payne2021,Payne2022,Hinkle2022b,Hinkle2023}. 

Our goal in this contribution is to search for signatures of companion interactions that are consistent with the  \citet{Kasen2010} models.   We compare models with and without companion interaction signatures to identify light curves that show evidence for a companion star.  Our methods address issues with some previous searches for companion interactions that depend on assumptions about the shape of the underlying SN light curve and the choice of time windows to fit.  In the majority of cases, the \tess\ data cannot distinguish between models with and without companion interactions.  However, we found three candidate SN light curves that tentatively prefer the addition of a companion interaction to a single power law model, with candidate companion stars having Roche lobe radii of 1.9--3.4 R$_\odot$ (separations between 3--6$\times 10^{11}$ cm).  On the other hand, the statistical evidence is weak for these models, and we discuss the effect of sampling and correlated noise in identifying convincing interaction signatures.  We also found two other SN light curves that robustly disfavor adding companion interaction signatures to power law models. 

In \S\ref{sec:observations}, we  describe the \tess\ observations, SN sample, and light curve properties.   In \S\ref{sec:analysis}, we describe our model and fitting procedure.  In \S\ref{sec:results} we present our results.  In \S\ref{sec:discussion} we discuss our results, including possible avenues for confirming or falsifying the presence or absence of companion stars and the implications of our results for the intrinsic fraction of single degenerate progenitor systems.  We summarize our findings in \S\ref{sec:conclusion}.   We provide technical details about the light curve extraction and \tess\ systematic errors in Appendix~\ref{app:lightcurves}, details about our fitting procedure and model comparison techniques in Appendix~\ref{sec:fits}, details about subsamples of the \tess\ Type Ia SNe in Appendix~\ref{sec:subsamples}, and figures showing light curves with and without companion signatures in Appendix~\ref{sec:candidates}.  We use a consensus cosmology throughout, with $H_0 = 70$\, km s$^{-1}$ Mpc$^{-1}$.  We correct for Galactic extinction using a \citet*{Cardelli1989} extinction law and $E(B-V)$ values from \citet*{Schlafly2011}.

\section{Observations\label{sec:observations}}

The \tess\ primary mission sky survey operated from 2018 July--2020 July, and the first extended mission operated from July 2020--September 2022.  \tess\ sweeps out one ecliptic hemisphere each year, using 13 pointings that divide the hemisphere into 24$\times$96 square degree "sectors."  \tess\ stares at each sector continuously for about 27 days, with approximately 1 day interruptions halfway through to download the accumulated data.  The main exceptions to this pattern are for \tess\ sectors 42--46, which observed the Ecliptic plane.  The \tess\ Full-Frame Images (FFIs) were collected at a 30 minute cadence in the primary mission (sectors 1--26)  and at a 10 minute cadence in the extended mission (sectors 27--50).   The \tess\ passband ranges from  600 nm to 1,000 nm.

\subsection{Light Curves and Supernova Sample \label{sec:sample}}

We extracted light curves for Type Ia SNe in the \tess\ fields of view discovered by ground-based surveys while \tess\ was observing.  We also included SNe discovered up to 30 days after the \tess\ observations, because \tess\ may still capture the early time rise of the light curve.  In total, there are \NsupernovaTotal\ such Type Ia SNe, drawn from reports to the Transient Name Server (TNS).\footnote{\url{https://www.wis-tns.org/}}  All of these SNe have spectroscopic classifications, which provide estimates of the redshifts, distances, and luminosities.  If the SN has a host galaxy reported to TNS, we use the host galaxy redshift.  Otherwise, we use the reported SN redshift,  and we assume an uncertainty of 1,000 km s$^{-1}$ that we propagate to the distance and luminosity estimates.  There are two SNe with erroneous redshifts on TNS, SN2018kfv and SN2020ut. For SN2018kfv, the observed spectrum is consistent with $z \leq 0.03$ \citep{Fausnaugh2021}, while for SN2020ut the correct redshift is 0.035 (\citealt{Perley2020}, their Table~3).  

We used the same light curve extraction method as \citet{Fausnaugh2021}, which employs difference imaging and forced photometry with a model of the TESS point spread function (PSF), using the \texttt{ISIS} software package \citep{Alard1998,Alard2000}.  Difference imaging and forced photometry work well for faint sources in TESS images; besides \texttt{ISIS}, a pipeline specific to TESS using a similar approach was developed by \citet[\texttt{TESSreduce}]{Ridden-Harper2021}.  The primary challenge for the supernova work presented here is the TESS backgrounds.  The TESS backgrounds are typically 16th magnitude per pixel, but are highly variable when the Earth is above TESS's sunshade and can become as bright as 14th mag per pixel.  These bright, time-variable backgrounds are a major hindrance to analysis of faint early time supernova light curves (often fainter than 19th mag).  In contrast, studies that focus on brighter sources in TESS images are limited by systematic errors from pointing jitter and differential velocity aberration. In these cases, corrections using cotrending or detrending work well. These methods are used by the TESS mission pipeline \citep{Jenkins2016}, \texttt{lightkurve} \citep{lightkurve2018}, the TESS Quick Look pipeline \citep{Huang2020},  and \texttt{eleanor} \citep{Feinstein2019}. For bright sources, the background correction is less important, and over-fitting nuisance variations is acceptable as long as cotrending/detrending preserves the astrophysical signal of interest.  For example, these methods can flatten a light curve in a way that will preserve exoplanet transits, without distinguishing between stellar variability signals and residual systematic errors.  However, we found that these approaches severely over-fit supernova light curves, and so we make use of difference imaging to remove systematic trends at the pixel level.  We explored the reliability of our difference imaging approach in some detail; a full discussion is  given in Appendix~\ref{app:lightcurves},  including a  comparison of our \texttt{ISIS} light curves to those from \texttt{TESSreduce}.

 \begin{figure}
    \centering
    \includegraphics[width=0.5\textwidth]{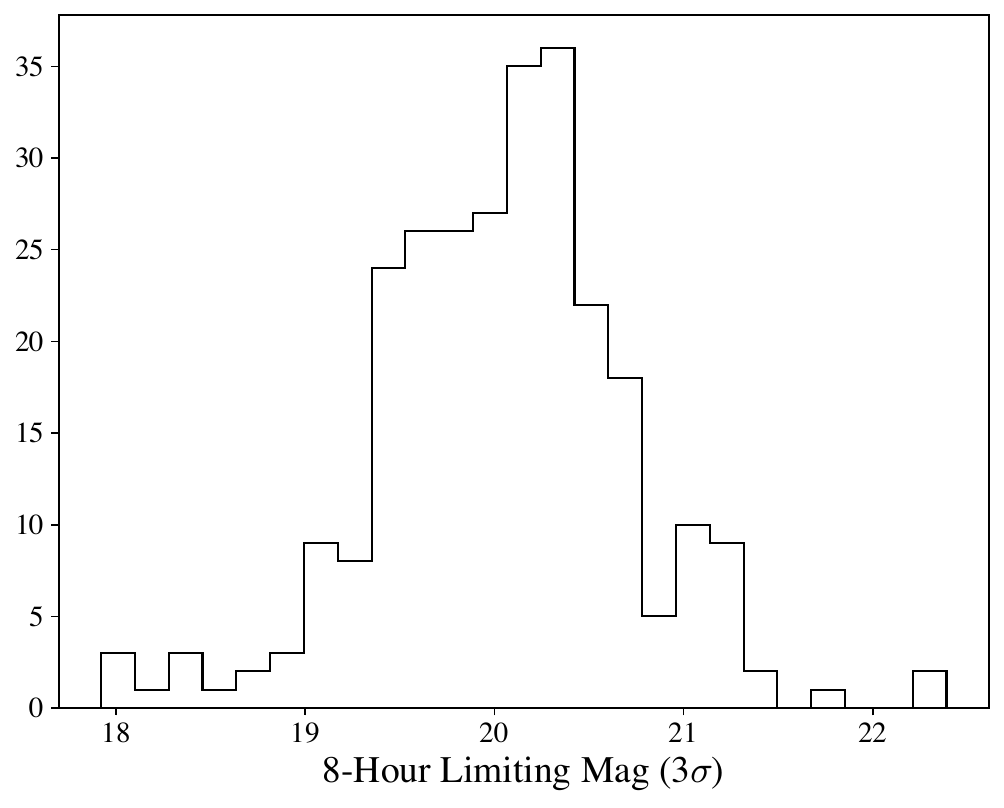}
    \caption{Distribution of 8-hour, 3$\sigma$ \tess\ limiting magnitudes for the sample of \NsupernovaLCs\ Type Ia SN light curves.  The limits are derived by binning the first 2 days of the light curves to 8 hours (to avoid the SN signals), calculating the RMS scatter, and converting to magnitudes with $T_{\rm lim} = -2.5\log\left( 3\times {\rm RMS}\right) + 20.44$ (see \S\ref{sec:sample}).  The tail of the distribution brighter than 19th mag is due to systematic errors in the light curves that increase the scatter.  The tail of the distribution fainter than 21st mag is statistical noise, due to the small number of data points used to calculate the 8-hour RMS scatter (typically six points or less).  The average limiting magnitude is 20.01, the median is 20.02, and the mode is  20.33.}
   \label{fig:maglimit}
\end{figure}

\begin{figure}
    \centering
    \includegraphics[width=0.5\textwidth]{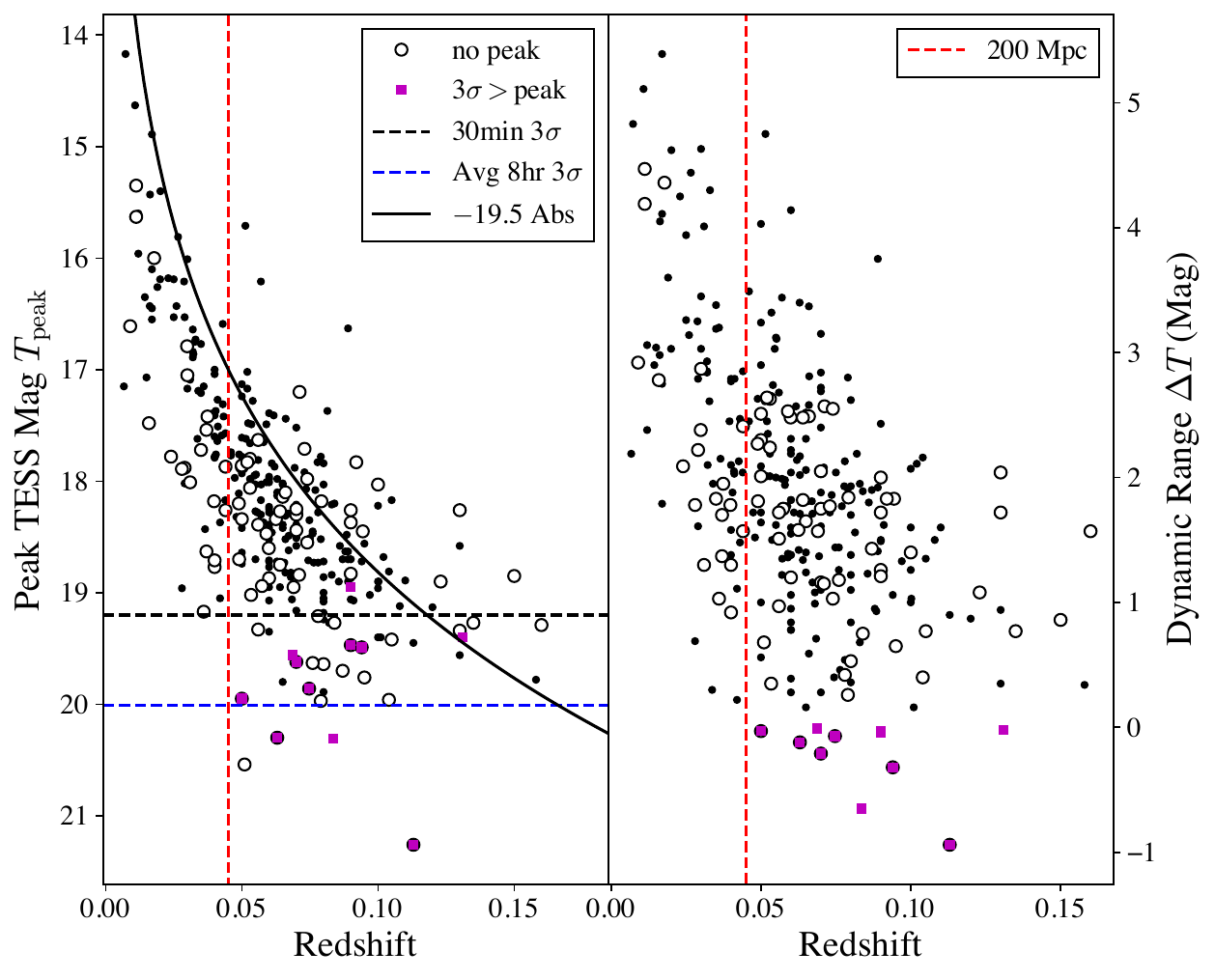}
    \caption{Peak magnitude $T_{\rm peak}$ and dynamic range $\Delta T$  of the SN sample as a function of redshift.  The value of $T_{\rm peak}$ is based on a parabolic fit to the peak of the light curve, and $\Delta T$ is defined as the difference between the limiting magnitude $T_{\rm lim}$ and $T_{\rm peak}$ (see \S\ref{sec:sample}).  For SN where \tess\ did not observe the peak, we use the maximum flux observed by \tess\ for $T_{\rm peak}$; these SNe are marked with open circles and represent lower limits on the peak brightness.  The 10 SNe  with limiting magnitudes brighter than $T_{\rm peak}$ have large systematic errors early in the light curve that produce a formal 3$\sigma$ limit brighter than $T_{\rm peak}$; however, a coherent SN signal is still evident at later times.  These SN are marked with magenta circles.  The apparent magnitude of a SN with peak absolute magnitude of $-$19.5 is shown by the solid black line in the left panel. }
   \label{fig:snproperties}
   \end{figure}

After extracting light curves for the \NsupernovaTotal\ SNe, we visually inspected the light curves.  We removed sources with no detection of the SN or where \tess\ did not observe the SN explosion.  No SN signal was detected if the SN is very faint, the explosion occurred after the TESS observations, or the light curve is particularly noisy (due to time-variable scattered light, crowding by nearby stars/galaxies, or nearby saturated pixels).   \tess\ may also miss the SN explosion if the SNe was discovered close to the start of the \tess\ observing sector so that the time of first light was before the \tess\ observations. Through visual inspection, we removed a total of \NsupernovaRemoved\ light curves, leaving \NsupernovaLCs\ objects in our sample.   For comparison, \citet{Fausnaugh2021} presented 24 Type Ia SN light curves from the first six months of \tess\ observations.  Of the \NsupernovaRemoved\ light curves that we removed, 333 were consistent with white noise.  The majority of these non-detections were for SNe discovered in the 30 day time window after \tess\ stopped observing, suggesting that \tess\ did not observe the time of first light.  Another 76 light curves showed a SN signal but the time of first light was before the \tess\ observations began.  The remaining 134 light curves of the \NsupernovaRemoved\ that were removed showed large residuals  that precluded the detection of any SN signal.   These cases result from especially noisy background corrections or strongly blended stellar variability, possibly combined with a weak supernova signal.  For the \NsupernovaLCs\ SNe that passed the visual inspection, the light curves show a clear SNe rise after a flat pre-explosion baseline.  

Table~\ref{tab:physical_data} summarizes the key properties of the SN sample.  We defined the limiting magnitude $T_{\rm lim}$ for each SNe by binning the light curve to 8 hours, calculating the root-mean-square (RMS) scatter of the light curve over the first two days of observations (to avoid the SN signal), and converting to a 3$\sigma$ limit with $T_{\rm lim} = -2.5\log\left( 3\times {\rm RMS}\right) + 20.44$.  Although $T_{\rm llim}$ is in general a function of time, light curves with strong changes in the noise properties were removed by our visual inspection step, and a single limit suffices to characterize the remaining light curves. Figure~\ref{fig:maglimit} shows the distribution of $T_{\rm lim}$ on 8 hour time scales.  

Table~\ref{tab:physical_data} gives an estimate of the time of peak $t_{\rm peak}$ for each SN light curve.  The time of peak $t_{\rm peak}$ and peak \tess\ magnitude $T_{\rm peak}$ were estimated by fitting a parabola to the SN light curve in a time window around the light curve maximum. In cases where the SN peak occurs during a \tess\ data gap, we used the maximum of the parabolic fit within the time window of the \tess\ observations.  For SNe that \tess\ did not observe at peak, the values of $T_{\rm peak}$ and $t_{\rm peak}$ correspond to the maximum observed by TESS; these objects are flagged in Table~\ref{tab:physical_data}.  A key property of \tess\ SNe light curves is also the "dynamic range" $\Delta T = T_{\rm lim} - T_{\rm peak} $, which indicates how close to the time of first light  each SNe was detected.  Figure~\ref{fig:snproperties} shows the redshifts, peak magnitudes $T_{\rm peak}$, and dynamic ranges $\Delta T$ of the SN sample.  Table~\ref{tab:lc_stub} gives machine-readable light curves of all \NsupernovaLCs\ SNe.  

\begin{rotatetable*}
\begin{deluxetable*}{llrrrrrrrrccc}
\tablewidth{0pt}\tablecaption{Properties of the Supernova Sample\label{tab:physical_data}}
\tablehead{\multicolumn{2}{c}{Name}& \colhead{$T_{\rm peak}$}&\colhead{$T_{\rm lim}$} & \colhead{$\Delta T$}&\colhead{$t_{\rm peak}$} & \colhead{$M_T$}&\colhead{$\nu L_{\nu}$}&\colhead{Redshift}&\colhead{$E(B-V)$} &\multicolumn{3}{c}{Bitmasks}\\
&&\colhead{(mag)\tablenotemark{a}}&\colhead{(mag)}&\colhead{(mag)}&\colhead{(days)}&\colhead{(mag)\tablenotemark{b}}&\colhead{(10$^{42}$\,erg s$^{-1}$)}&  &\colhead{(mag)} & \colhead{Reduction\tablenotemark{c}}& \colhead{\texttt{TESSred}\tablenotemark{d}} & \colhead{Quality\tablenotemark{e} }}
\startdata 
SN2018exc &     ATLAS18tne     &     16.21$\pm$0.05&     19.65     &     3.44      &    1352.91    &$-$          20.88$\pm$0.14           &     26.20     $\pm$           3.33&    0.0570     &     0.033     & 000  & 000  & 0001 \\ 
SN2018fhw &    ASASSN-18tb     &     16.10$\pm$0.05&     20.21     &     4.11      &    1360.41    &$-$          18.29$\pm$0.43           &     2.32      $\pm$           0.92&    0.0170     &     0.027     & 010  & 000  & 0001 \\ 
SN2018fpm &    ASASSN-18to     &     15.35$\pm$0.05&     19.54     &     4.19      &    1353.16    &$-$          18.09$\pm$0.05           &     1.92      $\pm$           0.10&    0.0112     &     0.011     & 000  & 000  & 1001 \\ 
SN2018fqn &    ASASSN-18tq     &     17.17$\pm$0.06&     19.62     &     2.45      &    1370.91    &$-$          19.73$\pm$0.15           &     9.01      $\pm$           1.25&    0.0520     &     0.038     & 001  & 000  & 0000 \\ 
SN2018fub &    ASASSN-18ty     &     16.21$\pm$0.05&     19.46     &     3.25      &    1378.24    &$-$          19.32$\pm$0.26           &     6.05      $\pm$           1.43&    0.0288     &     0.011     & 100  & 000  & 0001 \\ 
SN2018fvi &    ASASSN-18ug     &     17.37$\pm$0.07&     19.16     &     1.79      &    1364.95    &$-$          18.93$\pm$0.07           &     4.30      $\pm$           0.26&    0.0404     &     0.024     & 000  & 000  & 0100 \\ 
SN2018fzi &     ATLAS18uoo     &     19.18$\pm$0.07&     20.40     &     1.22      &    1369.87    &$-$          18.66$\pm$0.12           &     3.47      $\pm$           0.37&    0.0800     &     0.019     & 000  & 000  & 0010 \\ 
SN2018fzz &    ASASSN-18uo     &     17.87$\pm$0.05&     19.88     &     2.01      &    1352.47    &$-$          18.90$\pm$0.15           &     4.21      $\pm$           0.60&    0.0500     &     0.021     & 000  & 000  & 1000 \\ 
SN2018gjz &    ZTF18abtvgxv    &     18.21$\pm$0.05&     20.16     &     1.95      &    1352.91    &$-$          19.78$\pm$0.10           &     9.77      $\pm$           0.93&    0.0830     &     0.054     & 001  & 000  & 0000 \\ 
SN2018grv &    ZTF18abwerpm    &     18.51$\pm$0.06&     21.20     &     2.69      &    1380.56    &$-$          19.05$\pm$0.12           &     4.92      $\pm$           0.53&    0.0700     &     0.031     & 001  & 000  & 0100 \\ 
SN2018gyr &    ASASSN-18wf     &     17.72$\pm$0.05&     19.55     &     1.83      &    1381.22    &$-$          18.24$\pm$0.05           &     2.26      $\pm$           0.11&    0.0350     &     0.012     & 000  & 000  & 1000 \\ 
SN2018hgc &    ASASSN-18xr     &     17.02$\pm$0.05&     19.03     &     2.01      &    1406.04    &$-$          19.86$\pm$0.05           &     10.20     $\pm$           0.52&    0.0520     &     0.026     & 001  & 000  & 0100 \\ 
SN2018hib &     Gaia18czg      &     15.43$\pm$0.06&     19.48     &     4.05      &    1415.02    &$-$          18.85$\pm$0.45           &     3.89      $\pm$           1.60&    0.0163     &     0.018     & 010  & 000  & 0001 \\ 
SN2018hka &     ATLAS18wwt     &     18.63$\pm$0.06&     20.00     &     1.37      &    1406.04    &$-$          17.49$\pm$0.20           &     1.13      $\pm$           0.21&    0.0370     &     0.028     & 001  & 000  & 1000 \\ 
SN2018hkb &     ATLAS18wwv     &     18.26$\pm$0.06&     19.83     &     1.57      &    1405.33    &$-$          18.39$\pm$0.17           &     2.60      $\pm$           0.42&    0.0440     &     0.103     & 001  & 000  & 1100 \\ 
\ldots&\dots&\dots&\dots&\dots&\dots&\dots&\dots&\dots&\dots&\dots&\dots&\ldots\\ 
\enddata 
\tablecomments{ A machine readable version of this table with the full SN sample is available in the online version of this article.     \tablenotetext{a}{Uncertainty includes the \tess\ instrument flux calibration (0.05 mag, Appendix~\ref{app:lightcurves}), uncertainty in the fit to the baseline flux (\S\ref{sec:model}), and any uncertainty in the calibration offset of the second sector, if applicable (\S\ref{sec:model}).}\tablenotetext{b}{Absolute \tess\ magnitude.  Includes uncertainty in $T_{\rm peak}$ (column 2) and an assumed uncertainty of 1000 km s$^{-1}$ in the redshift if the host galaxy is not reported to TNS.}    \tablenotetext{c}{The data reduction bitmask marks SN where we applied the "background model" correction (first bit, "100"), where we detrended variable star signals (second bit, "010"), or where the light curve is affected by asteroids (third bit, "001").  See \S\ref{sec:systematics} for details.}    \tablenotetext{d}{The \texttt{TESSreduce} bitmask marks SN where we extracted a \texttt{TESSreduce} light curve (first bit, "100"), where we elected to use the \texttt{TESSreduce} light curve rather than the \texttt{ISIS} light curve (second bit, "010"), or where the two light curves show different shapes/amplitudes (third bit, "001"). See \S\ref{sec:tessred} for details.}    \tablenotetext{e}{The quality bitmask marks SN where TESS did not observe the peak flux (first bit, "1000", see Appendix~\ref{sec:subsamples}), the SN is part of the "No First Light" subsample (second bit, "0100", see Appendix~\ref{sec:subsamples}), the SN is part of the "High Quality" subsample (third bit, "0010",  see Appendix~\ref{sec:noise}), or the SN is part of the "High Dynamic Range" subsample (fourth bit, "0001", see Appendix~\ref{sec:obs_bias}).} }
\end{deluxetable*}
\end{rotatetable*}

 \begin{deluxetable*}{lrrrrrrrr}
\tablewidth{0pt}\tablecaption{SN Light Curves\label{tab:lc_stub}}
\tablehead{\colhead{Name}& \colhead{BJD $-$245700.0}&\colhead{counts s$^{-1}$ }& \colhead{Uncertainty}&  \colhead{Fraction of Peak} & \colhead{Uncertainty}& \colhead{$T_{\rm mag}$}& \colhead{Unceratinty}& \colhead{Calib. Offset}}
\startdata 
SN2018exc&  1325.32891   &$-$     5.63      &     6.03      &$-$    0.0382     &    0.0409     &     19.65     &--&--\\ 
SN2018exc&  1325.34974   &     9.35      &     6.07      &    0.0634     &    0.0412     &     19.21     &     0.70      &--\\ 
SN2018exc&  1325.37057   &$-$     1.34      &     6.14      &$-$    0.0091     &    0.0416     &     19.65     &--&--\\ 
\ldots&\dots&\dots&\dots&\dots&\dots&\dots&\dots&\dots \\ 
\enddata 
\tablecomments{ A machine readable version of this table is available in the online version of this article.  For fluxes below the 8 hour detection limits (including zero and and negative flux due to noise), the magnitudes are represented by $T_{\rm lim}$  from Table~\ref{tab:physical_data} and have uncertainties marked as NaN. The Calibration Offset gives the shift applied to flux calibrate the second sector of \tess\ observations for supernova observed near the ecliptic poles (see Appendix~\ref{sec:fits}).  }
\end{deluxetable*}

\section{Analysis \label{sec:analysis}}

There are  two major issues with previous work on high cadence, early time light curves of SN---one related to the choice of time windows to fit, and the other related to the choice of model to fit.  We designed our analysis to deal with both issues.

The first issue is the choice of time windows to fit around the time of first light.  \citet{Firth2015} and \citet{Olling2015} fit the early rise of their SN light curves up to 40--50\% of peak brightness in order to avoid changes in the light curve curvature near peak.  Many studies fit a similar fraction of peak brightness for consistency  (e.g., \citealt{Dimitriadis2019, Miller2020b, Fausnaugh2021, Dimitriadis2023}).  However,  this choice affects fits to the light curve shape, with smaller fractions of peak flux leading to smaller power law indices and rise times. Similarly, \citet{Magee2020a} showed that data near both first light and SN peak are important for understanding the shape of the rising light curve.   

To deal with this issue,  we adopt the parametric model of SN light curves introduced by \citet{Vallely2021}.  The \citet{Vallely2021} model includes a time-dependent power law index that allows the light curve to turn over as the SN approaches peak flux.   With this model, we can fit the light curves from the earliest TESS observations through the SN peak, and so our results do not depend on the choice of time windows or  flux fraction of peak to fit.

Another difficulty is that previous work approached the analysis of early time light curves by searching for a flux excess relative to a fitted power law model (e.g., \citealt{Olling2015,Dimitriadis2019}).  However, the properties of the flux excesses recovered in this way depend sensitively on the shape of the subtracted model. The issue is that  there are degeneracies between the light curve shape and the presence of a companion interaction.  For example,  \citet{Fausnaugh2021} showed that the combination of a large companion interaction model (which rises as $t^{1/2}$)  and a fireball model (which rises as $t^2$), will lead to a fitted power law index of  unity, and the residuals from such a power law fit will not clearly show a flux excess.  Similarly, for SN2018oh it is possible to fit the \textit{K2} light curve with a two-component power law \citep{Shappee2019} or a fireball model with a flux excess \citep{Dimitriadis2019}, but little work has gone into a direct comparison of these models.

To solve this problem, we fit rising power law models both with and without companion interaction components, and we use  model comparison techniques to determine if a given light curve prefers the presence of a companion star signature.  We use two metrics to perform the model comparison.  First, we use Dynamic Nested Sampling \citep{Skilling2006} to estimate the Bayesian evidence of each model.  We use the  \dynesty\ package  \citep{Higson2019,Speagle2020} to implement the Dynamic Nested Sampling algorithm, and we used the posterior distributions sampled by \dynesty\  to estimate the model parameters.  Second, we calculated the Bayesian Information Criterion (BIC) for each model, which imposes a stronger penalty on models with extra parameters and serves as a check on the robustness of our results.  Further details about the model comparison tests are given in Appendix~\ref{sec:fits}.

A different approach is to use physical models for the rising light curve, such as those calculated by \citet{Magee2018} and \citet{Magee2020a}.  Large residuals relative to a physical model can identify flux excesses, as shown by \citet{Deckers2022} and \citet{Magee2022}. These models can also  reproduce complex behavior in the light curves, such as bumps and wiggles caused by blobs of $^{56}$Ni in the outer ejecta.  However, a particular grid of models may not capture the full range of early time SN light curve behaviors, and it may be unclear if small residuals indicate a shortcoming in the model physics or an additional flux component.  We therefore chose to use the curved power law model and perform comparisons with and without \citet{Kasen2010} models to identify potential companion interactions.  We restrict our analysis to companion interaction models here, and plan to incorporate other flux components in future work ($^{56}$Ni distribution in the ejecta, circumstellar interactions, etc.).
\begin{figure*}
    \centering
	\begin{tabular}{cc}	
	    \includegraphics[width=0.5\textwidth]{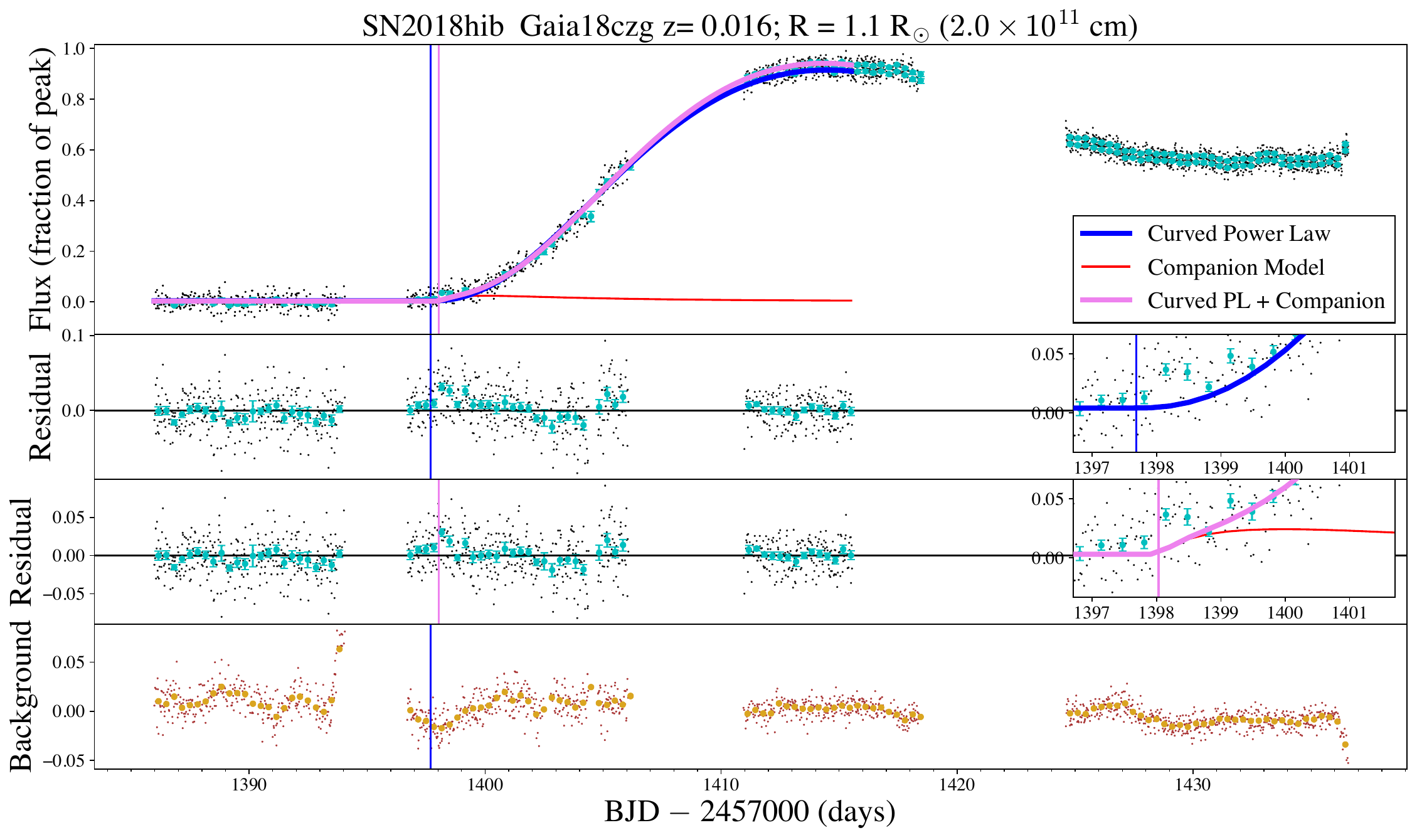}    & \includegraphics[width=0.5\textwidth]{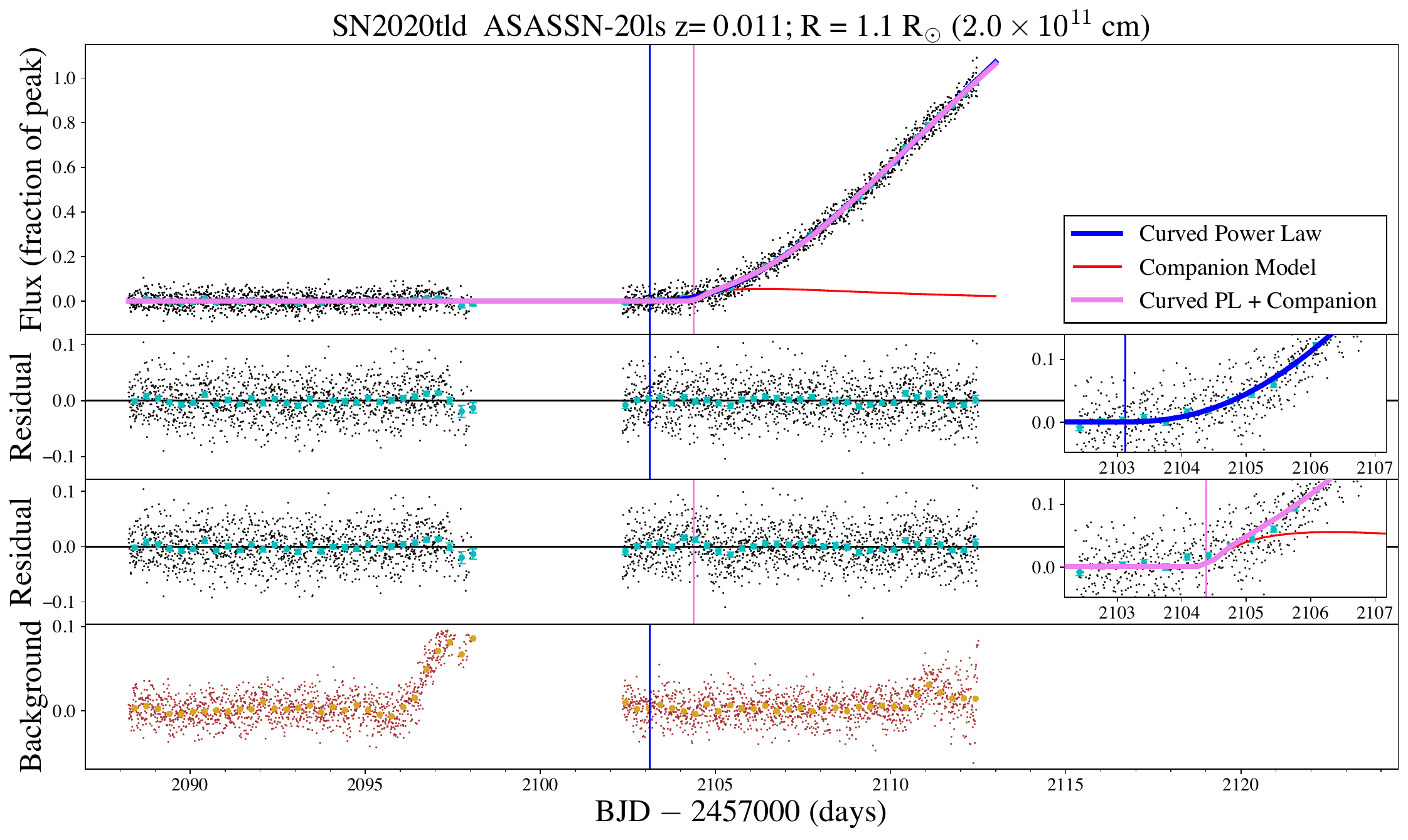}  \\
	    \includegraphics[width=0.5\textwidth]{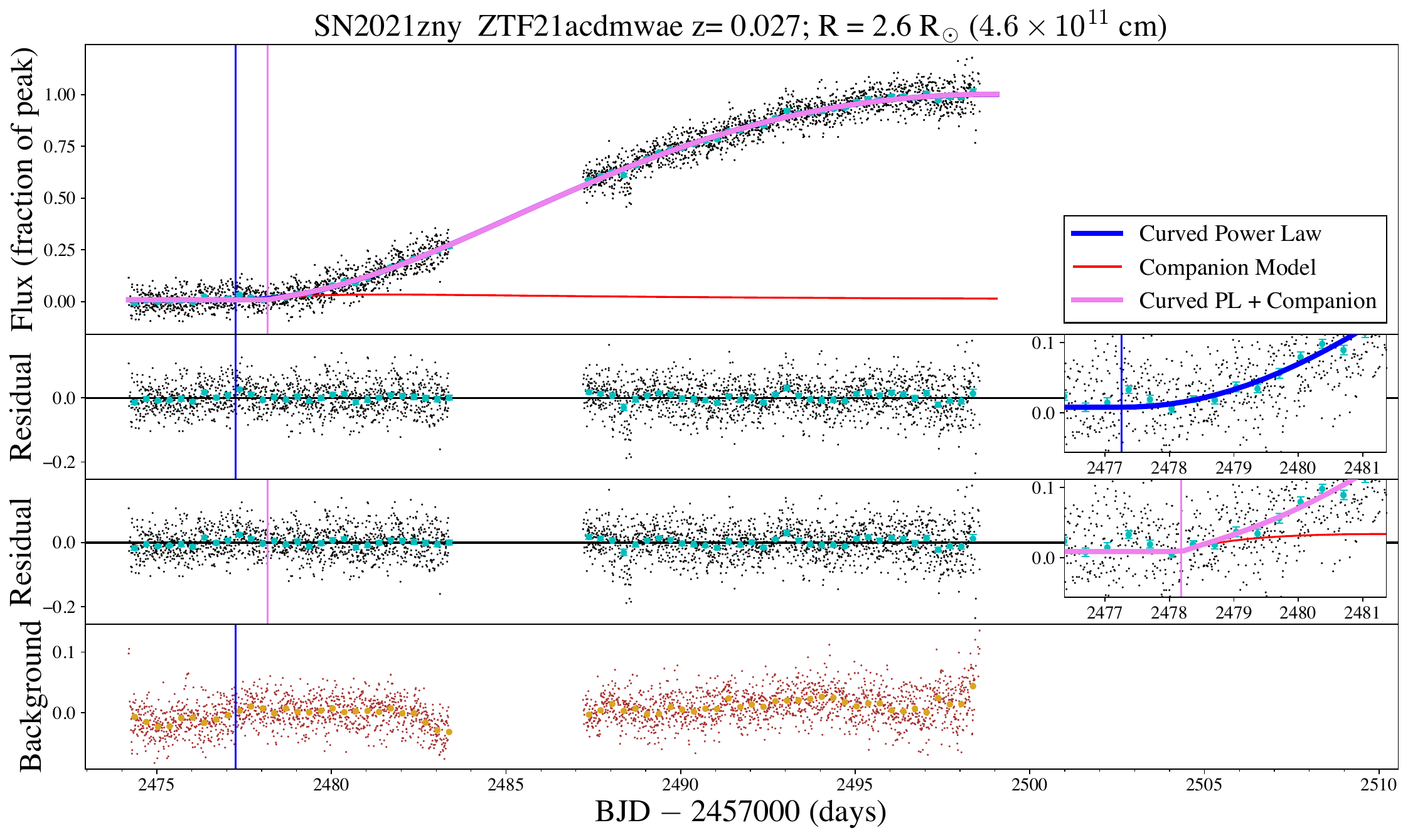}   & \includegraphics[width=0.5\textwidth]{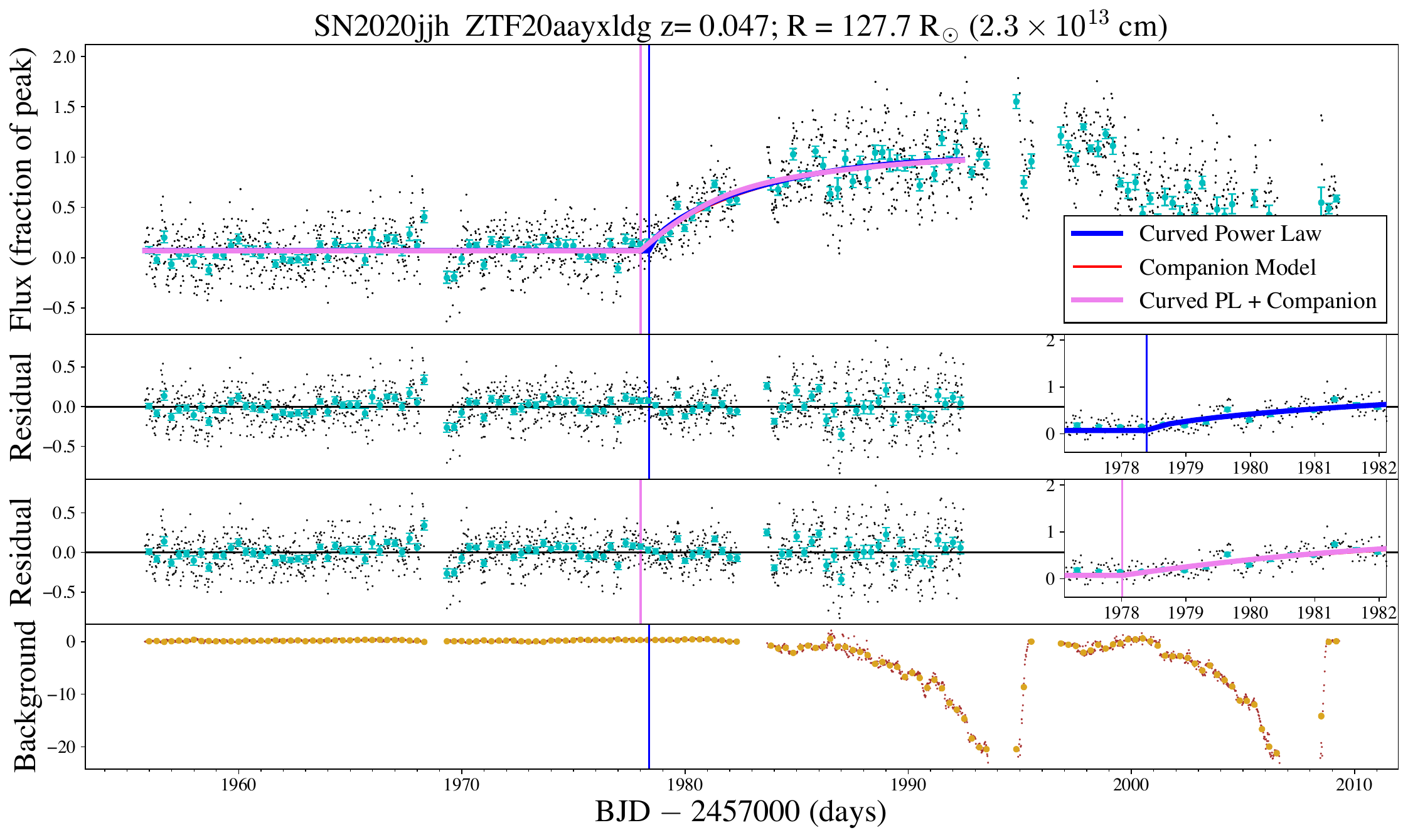}  \\
	    \end{tabular}
    \caption{Examples of  SN light curves and maximum likelihood models from the \texttt{dynesty} posterior distributions. The top panels show the original light curve in black and the average over 8 hour bins in cyan (we fit the unbinned light curves and only show binned data to guide the eye).  The maximum likelihood curved power law models without a companion interaction are shown in blue, and the curved power law models with a companion interaction are shown in violet.  The companion interaction component alone is shown in red.  For light curves with two sectors, the second sector is calibrated to the first by fitting an offset that best aligns with the curved power law---this offset can be different for the two models, and so we show two versions of the second sector with each calibration offset.     The second-down panels show the residuals from the curved power law model, and the third-down panels show the residuals for the curved power law model plus a companion interaction.  The insets show the 5 days near the time of first light in the original data (identical to the top panel).  The bottom panels show the local backgrounds in brown and binned to 8 hours in yellow---errors in the background estimation are the main systematic error in these light curves, and so correlated variations in the residuals and backgrounds suggest times with unreliable data.  The vertical lines mark the times of first light for each model.  The complete figure set is available in the online journal article.
    }
   \label{fig:SN_model_fits}
\end{figure*}

\subsection{Model\label{sec:model}}

We fit the light curves using the curved power law model proposed by \citet{Vallely2021},
\begin{align}
\label{equ:model}
f(\Delta u) &= H(\Delta u) \cdot C \cdot  (\Delta u) ^{\beta_1(1 + \beta_2(\Delta u))} + f_0 
\end{align}
where
\begin{align}
\Delta u &= \frac{t - t_0}{1 + z},
\end{align}
 $t_0$ is the time of first light, $z$ is the SN redshift, $\beta_1$ and $\beta_2$ govern the shape of the light curve, $H(\Delta u)$ is the Heaviside function, $C$ is a constant of normalization, and $f_0$ is the residual background at early times. The parameters $t_0$, $\beta_1$, $\beta_2$, $C$, and $f_0$ are left free to vary in the fit while $z$ is fixed to the value in Table~\ref{tab:physical_data}.  Before fitting the light curves, we normalized the peak flux to unity (using the parabolic fits described in \S\ref{sec:sample}) and so we treat $C$ and $f_0$ as nuisance parameters.  The parameter $\beta_1$ represents the early time rise of the light curve when $\Delta u$ is small.  The parameter $\beta_2$ governs the turnover of the light curve near the time of peak luminosity.  For SNe light curves that span two sectors, we added another nuisance parameter for an additive calibration offset that aligns the second sector with the flux scale of the first sector.  Aligning the light curves in this way allows us to use information from the second sector to constrain the light curve shape parameters while marginalizing over the uncertainty in the relative flux calibration.  These calibration offsets have been applied to the light curves in Table~\ref{tab:lc_stub}, and the offsets are given in the light curves of each supernovae.

Besides the curved  power law model, we also fit the light curves with a composite model \begin{align}
\label{equ:model2}
F(\Delta u) &= f(\Delta u) + K(\Delta u)
\end{align}
where $f(\Delta u)$ is the same as in Equation~\ref{equ:model} and $K(\Delta u)$ is the response in the \tess\ filter of the emission from a bow shock caused by the interaction of the SN blast wave with a companion star.  We used Equations~22 and 25 from  \citet{Kasen2010} to calculate the luminosity and temperature of the bow shock as a function of time: 
\begin{align}
L_{c,iso} &= 10^{43} 
\left( \frac{a}{10^{13}\,{\rm cm} } \right) 
\left( \frac{M_{\rm ej}}{1.4\,{\rm M}_\odot } \right) ^{1/4} 
\left( \frac{v_{\rm ej} }{10^9\,{\rm cm\ s^{-1}} } \right)^{7/4} \\
&\left( \frac{\kappa}{0.2\,{\rm cm^2\ g^{-1}} } \right) ^{3/4}
\left( \frac{\Delta u}{1\, {\rm day} } \right)^{-1/2}
 {\rm erg\ s}^{-1}\\
T_{\rm eff} &= 2.5\times10^4 
\left( \frac{a}{10^{13}\,{\rm cm} } \right)^{1/4} 
\left( \frac{\kappa}{0.2\,{\rm cm^2\ g^{-1}} } \right)^{-35/36}\\
&\left( \frac{\Delta u}{1\,{\rm day} } \right)^{-37/72}
{\rm K,}
\end{align}
where $a$ is the separation between the exploding star and its companion, $M_{\rm ej}$ is the mass of the ejecta, $v_{\rm ej}$ is the velocity of the ejecta, and $\kappa$ is the opacity of the ejecta.  For a blackbody of the appropriate temperature and luminosity, we used synthetic photometry (\texttt{pysynphot}, \citealt{pysynphot}) to calculate the flux $K(\Delta u)$ in the \tess\ bandpass, including the effects of cosmological redshift and Galactic extinction.  We evaluated a grid of 20 models logarithmically spaced in $a$, from (0.02--4.0)$\times10^{13}$\ cm.  The \citet{Kasen2010} models assume that the companion fills its Roche lobe.   For a binary system with stellar masses of 1.4 and 1.0 M$_\odot$ (for the exploding white dwarf and companion star, respectively), our grid of separations translates to a range of Roche lobe radii from 1.1 to 225~R$_\odot$.  We fixed $v_{\rm ej}$ to 10$^9$ cm s$^{-1}$ and $\kappa$ to 0.2 cm$^2$ g$^{-1}$, and we assumed an observer viewing angle of 45 degrees.  We discuss the impact of these assumptions on our results in \S\ref{sec:caveats}.

 \movetabledown=2.55in
\begin{deluxetable*}{lrrrrrrcc}
\tablewidth{0pt}\tablecaption{Curved Power Law Posterior Medians\label{tab:fit_results_no_comp}}
\tablehead{ 
\colhead{Name}& \colhead{First Light $t_0$\tablenotemark{a}}&\colhead{$t_{\rm rise}$\tablenotemark{b}}& \colhead{PL Index $\beta_1$ }& \colhead{PL Index $\beta_2$} &  \colhead{$\ln Z$\tablenotemark{c}}&   \colhead{BIC\tablenotemark{d}} & \colhead{rms$_{\rm Gauss}/$rms$_{\rm observed}$\tablenotemark{e}}& \colhead{Bitmask}\tablenotemark{f} }
\startdata 
SN2018exc &  1336.02   $\pm$0.10&   15.98&  2.27  $\pm$0.04&$-$ 0.0152 $\pm$0.0003&$-$ 1511.34  $\pm$0.14& 3005.22  &  0.2708&  0001  \\ 
SN2018fhw &  1343.08   $\pm$0.04&   17.04&  2.32  $\pm$0.01&$-$ 0.0160 $\pm$0.0001& 3129.42  $\pm$0.17&$-$ 6286.56  &  0.3378&  0001  \\ 
SN2018fpm &  1338.69   $\pm$0.04&--&  2.31  $\pm$0.01&$-$ 0.0146 $\pm$0.0002&$-$ 3203.14  $\pm$0.15& 6381.70  &  0.1972&  1001  \\ 
SN2018fqn &  1354.65   $\pm$0.31&   15.46&  1.87  $\pm$0.15&$-$ 0.0179 $\pm$0.0005&$-$ 1020.87  $\pm$0.12& 2033.69  &  0.5038&  0000  \\ 
SN2018fub &  1360.42   $\pm$0.20&   17.32&  2.13  $\pm$0.10&$-$ 0.0157 $\pm$0.0002&$-$ 1454.45  $\pm$0.13& 2893.40  &  0.2808&  0001  \\ 
SN2018fvi &  1348.21   $\pm$0.25&   16.09&  2.36  $\pm$0.12&$-$ 0.0181 $\pm$0.0006& 1528.62  $\pm$0.14&$-$ 3064.03  &  0.3531&  0100  \\ 
SN2018fzi &  1359.60   $\pm$0.95&    9.51&  1.04  $\pm$0.52&$-$ 0.0262 $\pm$0.0089&$-$  891.51  $\pm$0.09& 1791.25  &  0.8210&  0010  \\ 
SN2018fzz &  1341.91   $\pm$0.41&--&  2.28  $\pm$0.28&$-$ 0.0211 $\pm$0.0024&$-$ 1804.73  $\pm$0.11& 3608.12  &  0.2960&  1000  \\ 
SN2018gjz &  1330.74   $\pm$0.46&   20.47&  2.02  $\pm$0.09&$-$ 0.0099 $\pm$0.0010&$-$ 1834.79  $\pm$0.13& 3662.10  &  0.3010&  0000  \\ 
SN2018grv &  1366.66   $\pm$0.54&   12.99&  1.99  $\pm$0.34&$-$ 0.0240 $\pm$0.0018&$-$ 2105.76  $\pm$0.10& 4213.27  &  0.3864&  0100  \\ 
SN2018gyr &  1377.35   $\pm$0.17&--&  3.75  $\pm$0.28&$-$ 0.0947 $\pm$0.0066&$-$ 1935.84  $\pm$0.11& 3870.41  &  0.2811&  1000  \\ 
SN2018hgc &  1394.08   $\pm$0.57&   11.37&  1.94  $\pm$0.30&$-$ 0.0139 $\pm$0.0028&$-$ 1046.42  $\pm$0.11& 2089.11  &  0.3022&  0100  \\ 
SN2018hib &  1397.71   $\pm$0.06&   17.03&  2.28  $\pm$0.02&$-$ 0.0159 $\pm$0.0002& 2036.22  $\pm$0.17&$-$ 4097.83  &  0.1690&  0001  \\ 
SN2018hka &  1400.34   $\pm$0.67&--&  2.00  $\pm$0.64&$-$ 0.0289 $\pm$0.0173&$-$  996.04  $\pm$0.08& 2003.11  &  0.3957&  1000  \\ 
SN2018hkb &  1395.60   $\pm$0.58&--&  1.94  $\pm$0.16&$-$ 0.0021 $\pm$0.0024&$-$ 1136.11  $\pm$0.10& 2273.69  &  0.1029&  1100  \\ 
\ldots&\dots&\dots&\dots&\dots&\dots&\dots&\dots&\dots\\ 
\enddata 
\tablecomments{ A machine readable version of this table with the full SN sample is available in the online version of this article.  Uncertainties correspond to the widths of the  68\% credible regions of the posterior distributions. \tablenotetext{a}{Time of first light in units of BJD $-$ 247000.0 (days).}\tablenotetext{b}{Rise time in units of days. The uncertainty on the rise time is determined by the uncertainty on the time of first light (column 2).}\tablenotetext{c}{Natural log of the Bayesian Evidence $Z$ from \texttt{dynesty}.} \tablenotetext{d}{Bayesian Information Criterion.}\tablenotetext{e}{Noise metric defined in \S\ref{sec:results}, which quantifies departures of the light curves from random Gaussian noise. Values near zero indicate very little improvement from binning (systematic errors dominate over random noise), while 1 indicates perfect Gaussian noise scaling (random noise only). }\tablenotetext{f}{The "Bitmask" column is the same Quality Bitmask from Table~\ref{tab:physical_data}, which marks the subsamples of SNe: \tess\ did not observe the peak flux of this SN (first bit, "1000", see Appendix~\ref{sec:subsamples}), the SN is part of the "No First Light" subsample (second bit, "0100", see Appendix~\ref{sec:subsamples}), the SN is part of the "High Quality" subsample (third bit, "0010",  see Appendix~\ref{sec:noise}), or the SN is part of the "High Dynamic Range" subsample (fourth bit, "0001", see Appendix~\ref{sec:obs_bias}).}  }
\end{deluxetable*}

 \begin{rotatetable*}
\begin{deluxetable*}{lrrrrrrrrrrc}
\tablewidth{0pt}\tablecaption{Curved Power Law with Companion Interaction Posterior Medians\label{tab:fit_results}}
\tablehead{ 
\colhead{Name}& \colhead{First light $t_0$\tablenotemark{a}} & \colhead{$t_{\rm rise}$\tablenotemark{b}} & \colhead{PL Index $\beta_1$}& \colhead{PL Index $\beta_2$} &  \colhead{Med Rad\tablenotemark{c}}&\colhead{Max Rad\tablenotemark{c}}&\colhead{Med Sep\tablenotemark{d}}&\colhead{Max Sep\tablenotemark{d}}  & \colhead{$\ln Z$\tablenotemark{e}}& \colhead{BIC\tablenotemark{f}}&\colhead{Bitmask\tablenotemark{g}}  }
\startdata 
SN2018exc &     1336.36$\pm$0.17&   15.66&  2.29  $\pm$0.03&$-$ 0.0157 $\pm$0.0004& 13.69  $\pm$17.78& 127.68 &  2.46  $\pm$3.20& 22.99  &$-$ 1506.11  $\pm$0.14& 3000.11  &  0001  \\ 
SN2018fhw &     1343.67$\pm$0.04&   16.46&  2.37  $\pm$0.02&$-$ 0.0171 $\pm$0.0001&  1.11  $\pm$0.00&  1.11  &  0.20  $\pm$0.00&  0.20  & 3149.79  $\pm$0.17&$-$ 6326.53  &  0001  \\ 
SN2018fpm &     1339.56$\pm$0.04&--&  2.44  $\pm$0.01&$-$ 0.0176 $\pm$0.0002&  1.47  $\pm$0.00&  1.94  &  0.26  $\pm$0.00&  0.35  &$-$ 3070.82  $\pm$0.16& 6117.82  &  1001  \\ 
SN2018fqn &     1355.55$\pm$0.24&   14.60&  2.17  $\pm$0.16&$-$ 0.0200 $\pm$0.0010& 96.58  $\pm$54.66& 223.12 & 17.39  $\pm$9.85& 40.18  &$-$ 1018.27  $\pm$0.13& 2031.52  &  0000  \\ 
SN2018fub &     1360.90$\pm$0.18&   16.86&  2.04  $\pm$0.09&$-$ 0.0162 $\pm$0.0002&  1.47  $\pm$0.57&  5.93  &  0.26  $\pm$0.10&  1.07  &$-$ 1454.76  $\pm$0.14& 2896.26  &  0001  \\ 
SN2018fvi &     1348.83$\pm$0.29&   15.49&  2.40  $\pm$0.10&$-$ 0.0191 $\pm$0.0006&  3.39  $\pm$2.31& 18.10  &  0.61  $\pm$0.42&  3.26  & 1530.89  $\pm$0.14&$-$ 3062.98  &  0100  \\ 
SN2018fzi &     1359.99$\pm$0.39&    9.15&  0.87  $\pm$0.60&$-$ 0.0491 $\pm$0.0237& 168.78 $\pm$41.08& 223.12 & 30.39  $\pm$7.39& 40.18  &$-$  888.99  $\pm$0.08& 1797.45  &  0010  \\ 
SN2018fzz &     1342.85$\pm$0.39&--&  2.39  $\pm$0.27&$-$ 0.0243 $\pm$0.0027& 10.36  $\pm$10.18& 73.06  &  1.86  $\pm$1.83& 13.16  &$-$ 1804.84  $\pm$0.11& 3612.86  &  1000  \\ 
SN2018gjz &     1331.15$\pm$0.42&   20.09&  2.06  $\pm$0.07&$-$ 0.0104 $\pm$0.0009&  3.39  $\pm$3.15& 31.63  &  0.61  $\pm$0.57&  5.70  &$-$ 1833.01  $\pm$0.12& 3662.20  &  0000  \\ 
SN2018grv &     1367.31$\pm$0.40&   12.38&  2.23  $\pm$0.30&$-$ 0.0260 $\pm$0.0027& 23.93  $\pm$31.64& 223.12 &  4.31  $\pm$5.70& 40.18  &$-$ 2105.13  $\pm$0.11& 4217.65  &  0100  \\ 
SN2018gyr &     1377.37$\pm$0.20&--&  3.89  $\pm$0.14&$-$ 0.0914 $\pm$0.0099&  2.57  $\pm$1.68& 10.36  &  0.46  $\pm$0.30&  1.86  &$-$ 1932.27  $\pm$0.12& 3866.65  &  1000  \\ 
SN2018hgc &     1394.83$\pm$0.49&   10.65&  2.05  $\pm$0.28&$-$ 0.0154 $\pm$0.0032& 23.93  $\pm$55.50& 223.12 &  4.31  $\pm$9.99& 40.18  &$-$ 1045.96  $\pm$0.11& 2094.46  &  0100  \\ 
SN2018hib &     1398.08$\pm$0.09&   16.66&  2.32  $\pm$0.01&$-$ 0.0165 $\pm$0.0002&  1.11  $\pm$0.21&  2.57  &  0.20  $\pm$0.04&  0.46  & 2060.87  $\pm$0.17&$-$ 4145.30  &  0001  \\ 
SN2018hka &     1401.26$\pm$0.43&--&  2.44  $\pm$0.70&$-$ 0.0382 $\pm$0.0220&  5.93  $\pm$7.63& 41.81  &  1.07  $\pm$1.37&  7.53  &$-$  996.12  $\pm$0.08& 2009.25  &  1000  \\ 
SN2018hkb &     1396.79$\pm$0.59&--&  2.09  $\pm$0.15&$-$ 0.0024 $\pm$0.0029&  3.39  $\pm$2.31& 13.69  &  0.61  $\pm$0.42&  2.46  &$-$ 1135.53  $\pm$0.11& 2275.57  &  1100  \\ 
\ldots&\dots&\dots&\dots&\dots&\dots&\dots&\dots&\dots&\dots&\dots \\ 
\enddata 
\tablecomments{ A machine readable version of this table with the full SN sample is available in the online version of this article. Uncertainties correspond to the widths of the 68\% credible regions of the posterior distributions.   \tablenotetext{a}{Time of first light in units of BJD $-$ 247000.0 (days).}\tablenotetext{b}{Rise time in units of days. The uncertainty on the rise time is determined by the uncertainty on the time of first light  (column 2).}  \tablenotetext{c}{ Companion Roche lobe radii in units of solar radii.  "Med Rad" gives the median (50th percentile) of the posterior distribution and "Max Rad" gives the 3$\sigma$ upper limit (99.9th percentile).  } \tablenotetext{d}{Companion separations in units of 10$^{12}$~cm.  "Med Sep" gives the median (50th percentile) of the posterior distribution and "Max Sep" gives the 3$\sigma$ upper limit (99.9th percentile).} \tablenotetext{e}{Natural log of the Bayesian Evidence $Z$ from \texttt{dynesty}.}\tablenotetext{f}{Bayesian Information Criterion.}\tablenotetext{g}{The "Bitmask" column is the same Quality Bitmask from Table~\ref{tab:physical_data}, which marks the subsamples of SNe: \tess\ did not observe the peak flux of this SN (first bit, "1000", see Appendix~\ref{sec:subsamples}), the SN is part of the "No First Light" subsample (second bit, "0100", see Appendix~\ref{sec:subsamples}), the SN is part of the "High Quality" subsample (third bit, "0010",  see Appendix~\ref{sec:noise}), or the SN is part of the "High Dynamic Range" subsample (fourth bit, "0001", see Appendix~\ref{sec:obs_bias}).}  }
\end{deluxetable*}
\end{rotatetable*}

\section{Results\label{sec:results}}

Table~\ref{tab:fit_results_no_comp} gives the results of the \texttt{dynesty} fits for each SN with the curved power law model alone, and Table~\ref{tab:fit_results} gives the results for the curved power law model with a companion interaction component.  We adopt the median of the marginalized posterior distributions as the best estimate of each parameter and the width of the 68\% percent credible region as an estimate of the uncertainty.  Figure~\ref{fig:SN_model_fits} shows the data, maximum likelihood models, and residuals from the maximum likelihood models for each SNe.  We show the posterior parameter distributions for each model in Appendix~\ref{sec:fits}.  We found that there are strong correlations between the parameters $t_0$, $C$, and $\beta_1$.  These correlations exist because the same data can be fit with a later start time if the power law index is smaller, while the normalization is tied to the observed flux one day after after $t_0$ and therefore must be larger if $t_0$ is later.

The rise times $t_{\rm rise}$ in Tables~\ref{tab:fit_results_no_comp} and \ref{tab:fit_results} were calculated by subtracting the fitted time of first light $t_0$ from the time of peak $t_{\rm peak}$ and correcting for cosmological time dilation.  For SNe where \tess\  did not observe the peak flux, we do not report a value of $t_{\rm rise}$; these SNe are flagged in in Tables~\ref{tab:physical_data}, \ref{tab:fit_results_no_comp} and \ref{tab:fit_results}.

For some fits, the time of first light $t_0$ may fall in data gaps, either due to breaks during data downlink or data removed by our screening procedures in \S\ref{sec:systematics}.   These gaps introduce ambiguity in the derived values of $t_0$ and $\beta_1$ because the model must interpolate the data to the time of first light.  However, the uncertainties from the posterior distributions should capture the range of possible values for $t_0$ and $\beta_1$ in these cases, and the curved power law may do a reasonable job of extrapolating back to $t_0$ based on the changing shape of the light curve near the data gap.  We therefore include these objects in our analysis, but we flag them in Tables~\ref{tab:physical_data}, \ref{tab:fit_results_no_comp} and \ref{tab:fit_results}. We refer to this subset as the "No First Light" subsample.

After inspecting the residuals of each SN fit, we found that the residuals are often correlated in time, typically on time scales of 8 hours or shorter.  Correlated residuals suggest the presence of residual systematic errors in some of our light curves.   To identify problematic light curves, we defined a "noise metric" for each light curve.  The noise metric is based on the ratio of the expected to observed decrease in the root-mean-square (RMS) scatter of the residuals after binning on timescales of 2, 4, 8, 16, 24, 32, 64, and 128 hours (we include 24 hours because we are specifically worried about scattered light signals from the Earth).  First, we calculate the RMS scatter of the light curve residuals at each timescale, and then divide the observed RMS scatter by the RMS scatter of the native 30 or 10 minute samples.  This ratio RMS$_t$/RMS$_{\rm native} = {\rm rms}_t$ gives the rate at which random noise is averaging out of the light curves on longer and longer timescales.  We compare this observed scatter rms$_{t,{\rm observed}}$ to the Gaussian expectation rms$_{t,{\rm Gauss}} = N^{-1/2}$, where $N$ is the number of FFIs in a given time bin. We then divide the Gaussian scaled scatter rms$_{t,{\rm Gauss}}$ by the observed scaled scatter rms$_{t,{\rm observed}}$, and the noise metric is the minimum across timescales min(rms$_{t,{\rm Gauss}}/$rms$_{t,{\rm observed}}$) = rms$_{\rm Gauss}/$rms$_{\rm observed}$. The metric rms$_{\rm Gauss}/$rms$_{\rm observed}$ captures the largest departure of the SN from Gaussian noise, and should be sensitive to a variety of systematic errors that appear on different timescales.  This metric also has an intuitive scaling from 0 to 1, where values near zero indicate very little improvement from binning (systematic errors dominate over random noise), while 1 indicates perfect scaling according to the Gaussian expectation (random noise only). 

We then define a "High Quality" subsample as the SNe with noise metrics in the top quartile.  This threshold is at  rms$_{\rm Gauss}/$rms$_{\rm observed} \ge 0.49$, i.e., within a factor of about 2 of the Gaussian expectation.  There are 77 SNe in this subsample, which are the least affected by correlated noise. The "High Quality" SN sample is marked by a flag in Table~\ref{tab:physical_data} and Table~\ref{tab:fit_results}, and rms$_{\rm Gauss}/$rms$_{\rm observed}$ for each light curve is given in Table~\ref{tab:fit_results_no_comp}. We give more details and show the distributions of the noise properties in Appendix~\ref{sec:noise}.

Lastly, \citet{Miller2020b} found an observational bias such that the rising power law index and rise time are correlated with redshift.  This bias is easy to understand as a result of signal-to-noise ratio:  the lower the signal-to-noise ratio, the later the SN will be detected during its initial rise.   A later detection results in a later value of $t_0$ and a bias in $t_{\rm rise}$  to smaller values. We find the same bias in our sample of of \tess\ Type Ia SN light curves.  In Appendix~\ref{sec:obs_bias}, we use simulations to show that the \tess\ light curves require a peak flux with signal-to-noise ratio $>$\,10 on 30 minute time scales to obtain unbiased light curve parameters.  We therefore define a subsample of SNe with RMS scatter $\leq$\,10\% of peak flux, which we refer to as the "High Dynamic Range" subsample. There are \NsupernovaHighDR\ SN light curves in this subsample, which are marked in Tables~\ref{tab:physical_data}, \ref{tab:fit_results_no_comp} and \ref{tab:fit_results}.  We restrict our quantitative analysis to the "High Dynamic Range" subsample in the following analysis.  For comparison with ground-based light curves, this threshold corresponds to a dynamic range $\Delta T> 2.8$\, mag from detection to peak (on 8 hour timescales). More details are given in Appendix~\ref{sec:obs_bias}.

\begin{figure}
    \centering
    \includegraphics[width=0.5\textwidth]{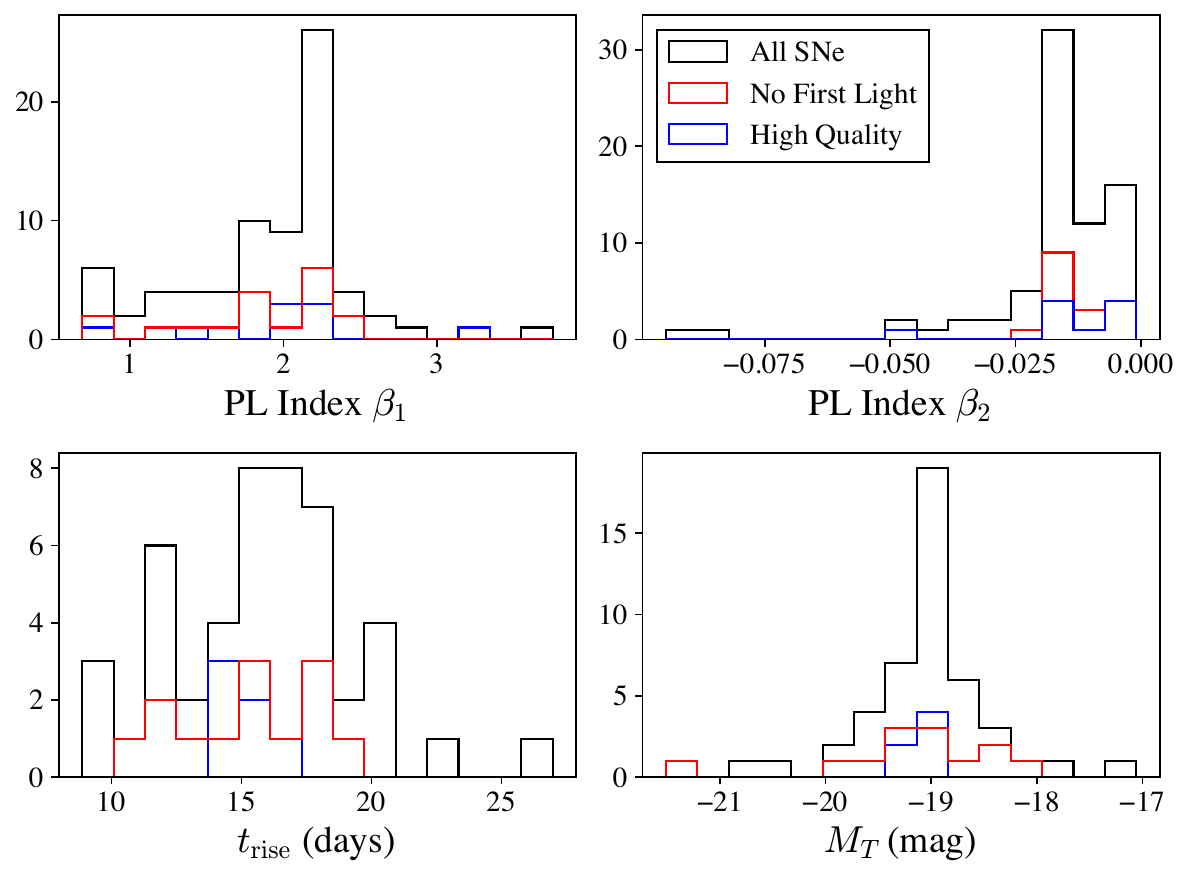}
    \caption{Distributions of power law indices $\beta_1$ and $\beta_2$ (which determine the shape of the early rise and peak of the light curves, respectively), the rise time $t_{\rm rise} = (t_{\rm peak} - t_0)/(1 + z)$, and absolute magnitude $M_T$.  We only show the \NsupernovaHighDR\ SNe in the "High Dynamic Range" sample, and we only show values of $t_{\rm rise}$ for the \NsupernovaPeak\ SNe that  \tess\ observed the peak (see \S\ref{sec:sample} and Appendix~\ref{sec:subsamples}).  The distributions of various subsets of SN are consistent with each other, and the mean values of $\beta_1$, $t_{\rm rise}$, and $M_T$ are consistent with the results from previous studies.}
   \label{fig:paramdists}
\end{figure}

\begin{figure}
    \centering
    \includegraphics[width=0.5\textwidth]{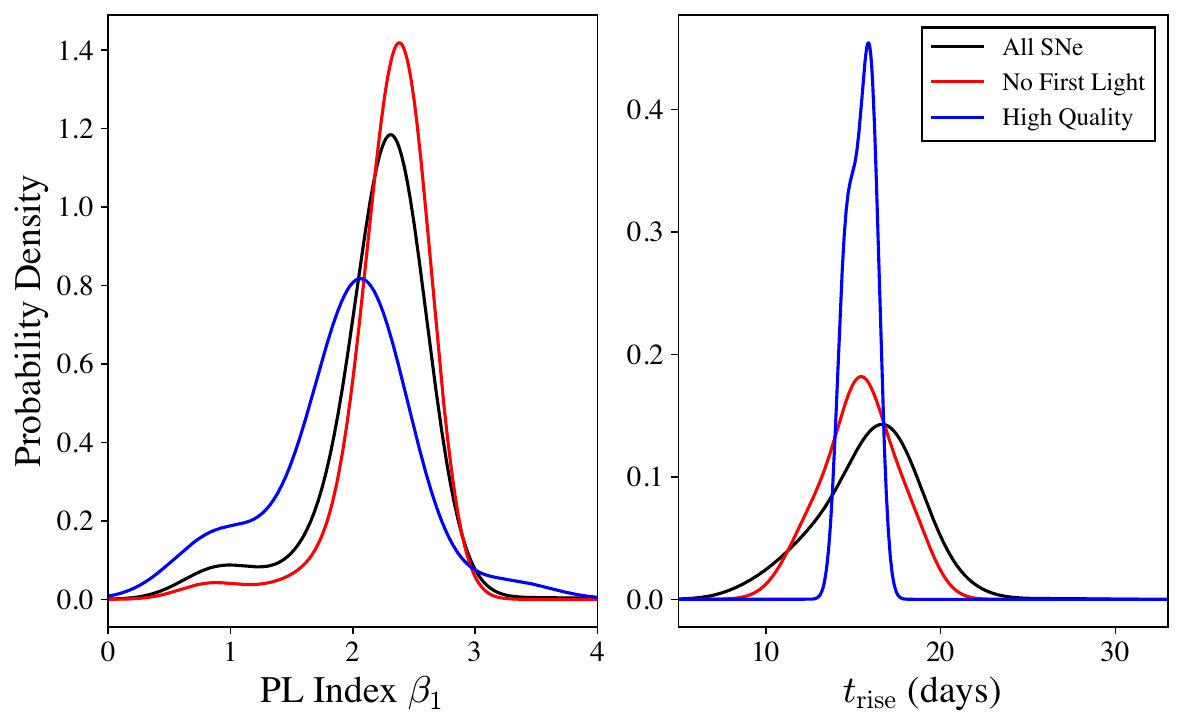}
    \caption{  Kernel density estimates (KDE) of the the underlying distribution of power law index $\beta_1$ and rise time $t_{\rm rise} = (t_{\rm peak} - t_0)/(1 + z)$ for the subsamples shown in Figure~\ref{fig:paramdists}.  For the KDE, we used a Gaussian kernel and bandwidth of 0.63, and the data were weighted by the inverse square of the uncertainties in  Table~\ref{tab:fit_results_no_comp}.      The estimates of the underlying distributions from various subsets are consistent with each other.  For $\beta_1$, the distribution is well characterized by a Gaussian  with mean 2.29 and width 0.34, and a long tail that extends below 1.0.  For $t_{\rm rise}$, the distribution is well characterized by a Gaussian with mean 15.89 days and width 3.82 days. }
   \label{fig:kde}
\end{figure}

\subsection{Distribution of Model Parameters\label{sec:paramdists}}

Figure~\ref{fig:paramdists} shows the distributions of power law indices $\beta_1$ and $\beta_2$, rise times $t_{\rm rise}$, and absolute magnitudes $M_T$.  Only the \NsupernovaHighDR\ SNe in the "High Dynamic Range" subsample are shown because the other SNe likely do not have reliable parameter estimates.  As a reminder, we do not calculate $t_{\rm rise}$ for SNe where \tess\ did not observe the peak flux, so there are only \NsupernovaPeak\ SN with measurements of $t_{\rm rise}$.  The "High Quality" and "No First Light"  subsamples that overlap with the "High Dynamic Range" subsample are also shown in Figure~\ref{fig:paramdists}.

The early time power law index spans a wide range of values, with most of the light curves ranging from power law index $\beta_1 = 0.7$ to 2.8.  The highest value is $\beta_1 = 3.75 \pm$ 0.57 for SN2018gyr.  The average value and population dispersion is \meanPLAll.  The median uncertainty on $\beta_1$ is 0.14 (from Table~\ref{tab:fit_results_no_comp}) , so the observed widths of the histograms in Figure~\ref{fig:paramdists} are dominated by the population dispersion rather than measurement noise.   The "High Quality"  and "No First Light" subsamples are consistent with the mean value of the full sample, with average values of $\beta_1 =$\,\meanPLHQ\ and \meanPLnoFL, respectively, and the widths of the distributions are again dominated by the population dispersion rather than measurement uncertainty.  These averages are in close agreement with previous studies, for example,  $2.01\pm0.5$ for the average $r$-band rising power law index found by \citet{Miller2020b}.  

We also estimated the intrinsic population dispersion using a kernel density estimate (KDE, \citealt{Silverman1986, Scott2015}).  We used a Gaussian kernel and weighted the measurements by the inverse squares of their uncertainties from Table~\ref{tab:fit_results_no_comp}.  We used a kernel bandwidth of 0.63, derived from the "rule of thumb" estimate for optimal bandwidth parameters \citep{Scott2015}.  KDE results are very sensitive to the bandwidth parameter, and so we also tested a distribution with a KDE bandwidth parameter of 0.20. The smaller bandwidth produces multiple modes associated with individual measurements of $\beta_1$, suggesting a lower limit to the KDE bandwidth.  The results for the larger bandwidth of 0.63 are shown in Figure~\ref{fig:kde}.  The distributions are well characterized by a Gaussian  with mean 2.29 and width 0.34, and a long tail that extends below 1.0. The long tail towards low values in Figure~\ref{fig:kde} may suggest heterogeneous behavior in early time Type Ia SN light curves, which is consistent with previous studies that find  diverse behavior in early time Type Ia SN light curves.   We discuss possible interpretations of the KDE analysis in \S\ref{sec:discuss_sample}.

The values of $t_{\rm rise}$ typically range from 8.6 to 22.9 days, with a mean value and population dispersion of \meanTriseAll\ days.  The longest rise time is 27.0 days for SN2019tfa; however, there is a large data gap between the first and second sector of this light curve, and the location of the light curve peak is ambiguous. Our value for  $t_{\rm rise}$ may therefore be overestimated. The second largest value of $t_{\rm rise}$ is 22.9 days for SN2020bqr, which is consistent with previous samples (e.g., \citealt{Miller2020b}).  The SNe with rise times less than 10 days tend to have gaps at the time of first light or near peak, so $t_{\rm rise}$ is likely underestimated for these sources.  However, at least one light curve, SN2019gqv, has a nearly complete light curve from $t_0$ to peak, and the curved power law fit yields $t_{\rm rise} = 8.4 $ days and $\beta_1 = 0.81 $ for this object.  Although classified as a normal Type Ia SN, the unusual light curve parameters for this source may indicate that it is a strange target in some other respect.  The "High Quality" and "No First Light" subsamples of SN are in agreement with the full sample, with mean values of $t_{\rm rise} =$ \meanTriseHQ\ and \meanTrisenoFL\ days, respectively.  Figure~\ref{fig:kde} shows distributions from weighted KDEs for the underlying distribution of $t_{\rm rise}$.  The population distribution is well characterized by a Gaussian with mean 15.89 days and width 3.82 days, consistent with the histograms in Figure~\ref{fig:paramdists}. 

The power law index $\beta_2$ characterizes the shape of the peak of the light curve, and is generally greater than $-0.04$.  The two SN with extreme values of $\beta_2 < -0.08$ did not have their peaks observed by \tess, and the results are a consequence of limited sampling only 2--3 days after first light.   There is no substantive difference in absolute magnitude between the "High Dynamic Range" sample, "High Quality" subsample,  "No First Light" subsample, and luminosity distribution of normal Type Ia SNe.  The most luminous SN is SN2019ugr at  $-21.51\pm 0.27$ absolute magnitude, while the least luminous SNe are SN2020cdj and SN2020fcw with absolute magnitudes of $-17.06\pm 0.49$ and  $-17.66\pm 0.51$, respectively.    For SN2019ugr, the uncertainty on the peak luminosity is relatively large, due to the uncertainty in the calibration offset of the second TESS sector. The large uncertainty on the calibration offset is due to data gaps between the first and second sector of \tess\ observations near the time of first light.  SN2019ugr is classified as a SNIa-91T-like  on TNS, so it likely has a higher luminosity than normal Type Ia SNe and we do not have good evidence that this source is significantly super-luminous.  For SN2020cdj and SN2020fcw, both sources have redshifts of 0.014--0.015, but no host redshifts were reported to TNS.  The uncertainties on the absolute magnitudes are therefore dominated by uncertainties in the redshifts (which we assume have uncertainties of 1,000 km s$^{-1}$) and, it is unclear if these objects are actually sub-luminous Type Ia SNe.

\begin{figure}
    \centering
    \includegraphics[width=0.5\textwidth]{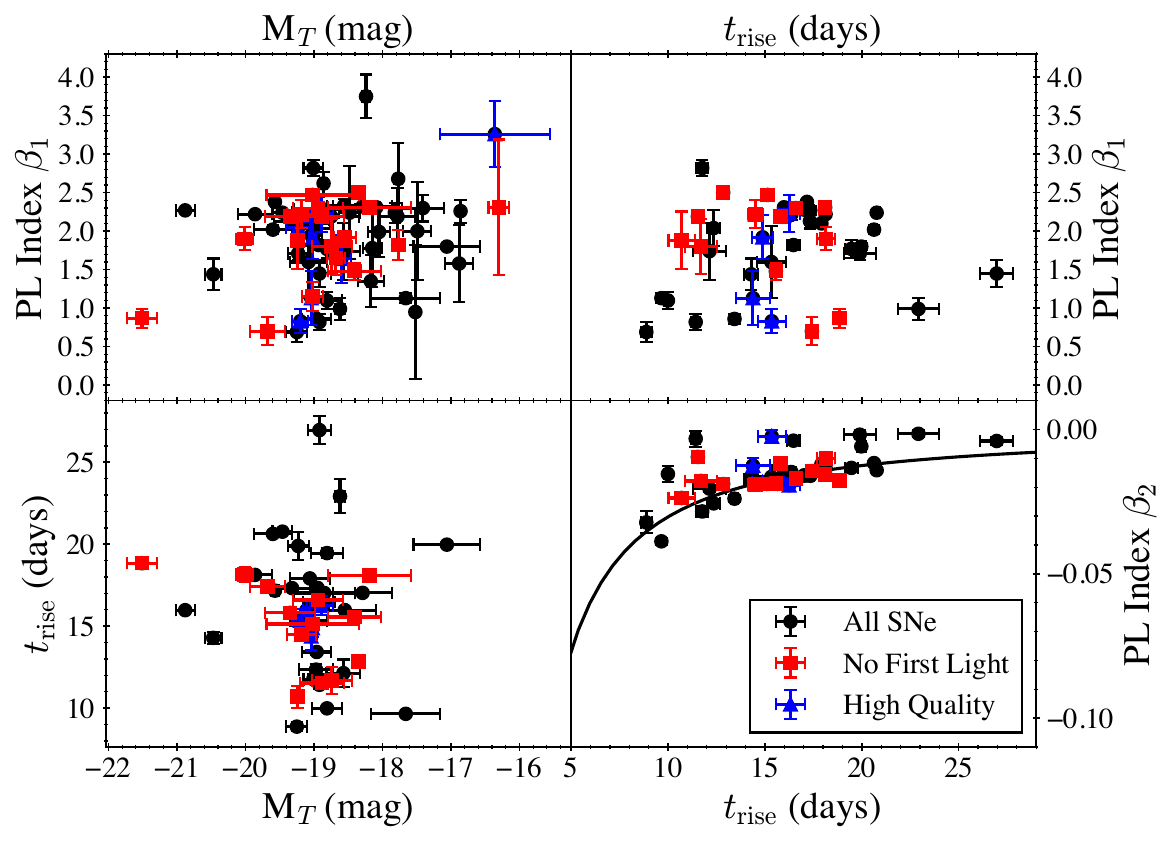}
    \caption{Correlations of power law indices $\beta_1$ and $\beta_2$,  rise time $t_{\rm rise}$, and absolute magnitude $M_T$.   We only show the \NsupernovaHighDR\ SNe in the "High Dynamic Range" sample, and we only show values of $t_{\rm rise}$ for the \NsupernovaPeak\ SNe that  \tess\ observed the peak  (see \S\ref{sec:sample} and Appendix~\ref{sec:subsamples}).  We do not find any significant correlations.  The solid line shows the analytic relationship between $\beta_2$ and $t_{\rm rise}$ (see \S\ref{sec:paramdists}).}
   \label{fig:physcorr}
\end{figure}

Finally, we searched for correlations between the power law indices $\beta_1$ and $\beta_2$, $t_{\rm rise}$, and absolute magnitude.  We did not find any strong correlations, as shown in Figure~\ref{fig:physcorr}.    The values of $\beta_2$ and $t_{\rm rise}$ match the mathematical relationship $-\beta_{2}^{-1} = t_{\rm rise} \left[ 1 + \ln (t_{\rm rise}) \right]  $, obtained by differentiating Equation~\ref{equ:model} and finding the maximum \citep{Vallely2021}.

\begin{figure*}
    \centering
    \includegraphics[width=\textwidth]{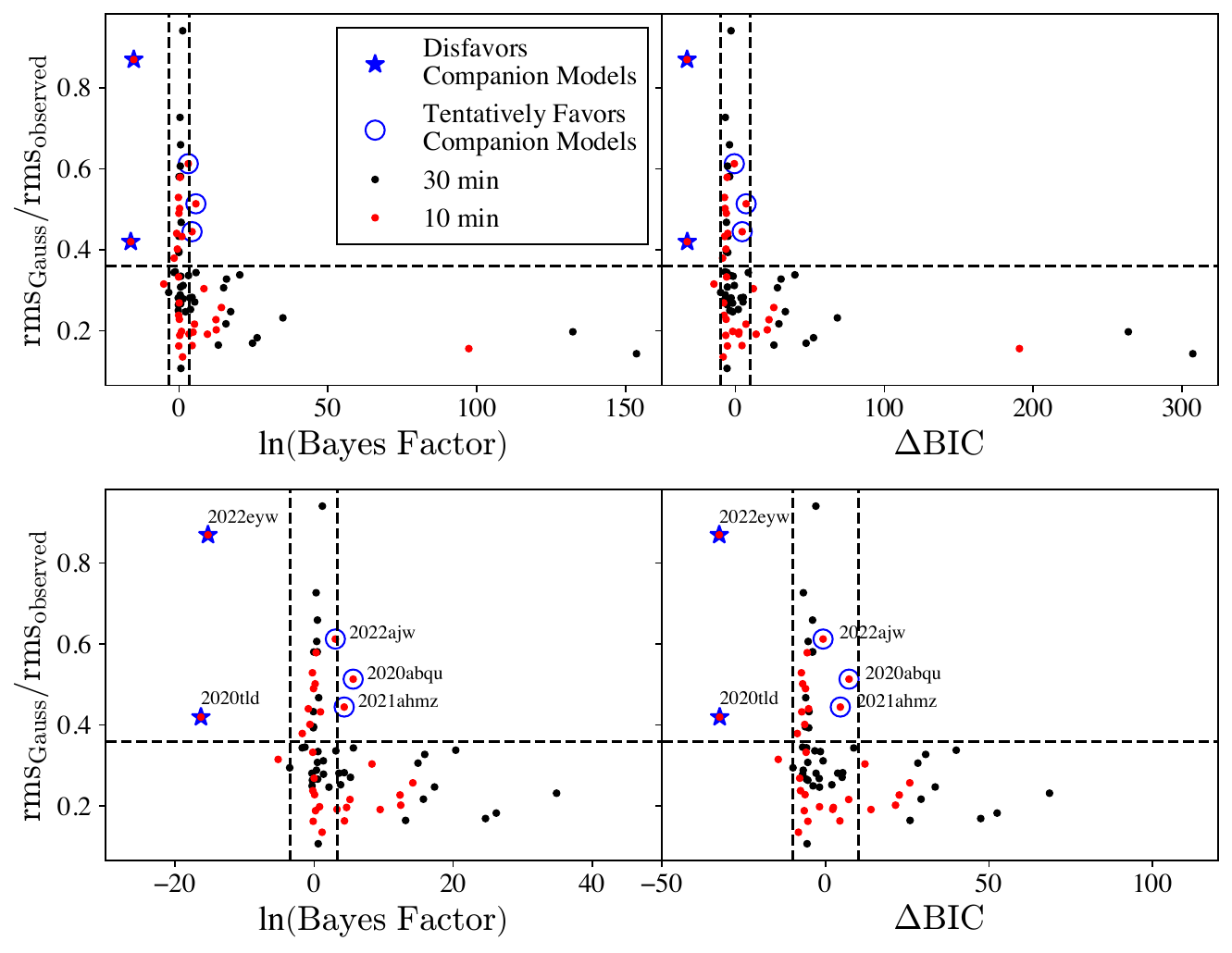}
    \caption{Light curve noise metric rms$_{\rm Gauss}/$rms$_{\rm observed}$ as a function of the Bayes Factors $K$ and the differences in Bayesian Information Criterion $\Delta {\rm BIC}$.  Larger values of the $K$ or the $\Delta {\rm BIC}$ indicate a preference for adding a companion interaction model  to the curved power law model, while smaller values disfavor the addition of a companion interaction.  The vertical dashed lines show $| \ln K | > 3.4$ and $|\Delta {\rm BIC}| > 10$.  Between these lines, the data cannot clearly distinguish between the two models.  The noise metric quantifies the departure of the light curve from random Gaussian noise, with values near 1 approaching Gaussian statistics (see \S\ref{sec:results}).  The horizontal dashed lines show a break in the distribution at rms$_{\rm Gauss}/$rms$_{\rm observed} = 0.36$, which we use as a threshold for reporting candidates that favor or disfavor companion interactions.  The bottom two panels are the same as the top panels, but with a restricted range on the horizontal axis to better visualize the data near zero. SN2020tld and SN2022eyw robustly disfavor the addition of companion interaction models.  On the other hand, SN2020abqu, SN2021ahmz, and SN2022ajw show a weak preference for companion interactions based on the Bayes Factor (SN2022ajw is just below the threshold with $\ln K = 3.01$).  However, the preference for a companion interaction model is not robust when using $\Delta$BIC.  See \S\ref{sec:model_comp} for details. } 
   \label{fig:bayes_hist}
\end{figure*}

\subsection{Model Comparison\label{sec:model_comp}}

We define the Bayes factor $K$ as the ratio of the Bayesian evidence $Z_1$ for the model with companion signatures to $Z_2$ for the model with no companion signatures. Large values of  $K$ favor a companion signature, while small values disfavor a companion.  We also compare the Bayesian Information Criterion ${\rm BIC_1}$ for the model with companion signatures to ${\rm BIC_2}$ for the model without companion signatures.  A smaller value of the BIC indicates a preference for a given model, and so we define $\Delta {\rm BIC} = {\rm BIC_2 } - {\rm BIC_1}$ such that larger values of $\Delta {\rm BIC}$ favor a companion signature.

The majority of $\ln K$ values cluster around 0, showing that the data usually cannot distinguish between models with and without a companion interaction component.  For $\Delta {\rm BIC}$, the values cluster around a median value of $-5.8$.  There are of order 1000 \tess\ data points fitted in each light curve, so we would expect $\Delta {\rm BIC} \approx -\ln 1000 = - 6.9$  for equal values of the likelihood from both models.  The cluster of $\Delta {\rm BIC} \approx -5.8$ therefore suggest that both models fit the data equally well and that the addition of a companion interaction signature is usually not warranted by the data.

There are a smaller number of SN at very large or small values of $K$ and $\Delta {\rm BIC}$.  We adopt a threshold Bayes factor $| \ln K| > 3.40$ and $|\Delta {\rm BIC}| > 10.0$ to distinguish between light curves that prefer or disfavor a companion signature and light curves that are too close to call.

There are 27 sources with both $\ln K$ and $\Delta {\rm BIC}$ above our thresholds.  Of these, SN2018fpm, SN2020ftl, and SN2022exc have exceptionally high values of $\ln K$ and $\Delta {\rm BIC}$, indicating a very strong preference for adding a companion interaction to the curved power law model.  The light curves for these three SNe are shown in Appendix~\ref{sec:candidates}; the model residuals are strongly correlated, suggesting that many of the 27 candidate companion interactions are false positives triggered by residual systematic errors.  

Using the noise metric rms$_{\rm Gauss}/$rms$_{\rm observed}$ defined in \S\ref{sec:results} and Appendix~\ref{sec:noise}, we can identify sources strongly affected by systematic noise based on their departure from expectations for scaling of random Gaussian noise.  Figure~\ref{fig:bayes_hist} shows rms$_{\rm Gauss}/$rms$_{\rm observed}$ as a function of $\ln K$ and $\Delta {\rm BIC}$. The majority of the SNe lie well below the expectation for Gaussian scaling, suggesting that the \citet{Kasen2010} models are fitting correlated noise rather than a physical flux component.

There is a natural break in the distribution of rms$_{\rm Gauss}/$rms$_{\rm observed}$ at 0.36 (noise properties within a factor of $\sim$3 of Gaussian scaling), with two SNe in this range above our Bayes factor threshold: SN2020abqu and SN2021ahmz.  If the \citet{Kasen2010} signals are real,  then these SNe could be candidate single degenerate progenitor systems.  However, the evidence for the companion interaction based on $\Delta$BIC is weaker and and so our results do not robustly show evidence for the companion interactions. Along similar lines, SN2022ajw is a high quality light curve (rms$_{\rm Gauss}/$rms$_{\rm observed}$  = 0.6) that stands out in the distribution of Bayes factors, but is just below our threshold ($\ln K = 3.06$ instead of 3.4). Again, the evidence for the companion model based on $\Delta$BIC is weaker than using $\ln K$. We  identify these three SNe as tentative companion interaction candidates, with the caveat that the evidence for the extra flux component is intriguing but not overwhelming. The median Roche lobe radii of the posterior distributions would be 3.39,  1.94, and 2.57 R$_\odot$ for SN2020abqu, SN2021ahmz, and SN2022ajw, respectively, which would correspond to companion stars at separations between $(3$--$6)\times10^{11}$~cm.  However, the preference for \citet{Kasen2010} models could also be caused by other physical processes that are not included in our analysis, such as $^{56}$Ni mixing or circumstellar material.   All three of these SNe are classified as normal Type Ia and have early time power law indices $\beta_1$ between  2.0 and 2.3,  so it is not obvious that they are subject to physical processes atypical for Type Ia SNe. Light curves for these SNe are shown in Appendix~\ref{sec:candidates}. 

There are three sources with  both $\ln K$ and $\Delta {\rm BIC}$ that indicate a preference against the addition of a companion signature: SN2020tld, SN2022eyw, and SN2021abko. SN2020tld is the only one of the three with rms$_{\rm Gauss}/$rms$_{\rm observed} \ge 0.36$.  In addition, SN2022eyw is affected by scattered light signals that drive the low value of  rms$_{\rm Gauss}/$rms$_{\rm observed}$, but the SN signal is isolated from these times and rms$_{\rm Gauss}/$rms$_{\rm observed}$  = 0.9 during the time most relevant to analysis of the SN (SN2022eyw is the second-best quality light curve if we use this restricted range).  The last source is SN2021abko, which has rms$_{\rm Gauss}/$rms$_{\rm observed} =0.32$.  In this case, the light curve is affected by correlated noise, and we cannot clearly rule out the presence of companion interaction signatures.  Light curves for all three sources are shown in Appendix~\ref{sec:candidates}.  

For SN2020tld and SN2022eyw, the observer's viewing angle may be much larger than 45 degrees so that the shock from the companion interaction is obscured from the line of sight.  Alternatively, the smallest companion Roche lobe radius that we tested was 1.1\,R$_\odot$ ($2 \times 10^{11}$~cm), and tighter binary systems with weaker companion interaction signals may still be consistent with these light curves.  The last possibility is that these SNe do not have surviving companion stars,  as would be expected for many double degenerate progenitor scenarios.  SN2020tld is spectroscopically classified as a normal Type Ia on TNS, while SN2022eyw is classified as a SNIax-2002cx-like.  

Table~\ref{tab:candidates} summaries these results, with properties of the five supernovae that prefer or disfavor the addition of a companion model to the curved power law.

Lastly, SN2021zny bears special mention because it was the subject of a recent multiwavelength analysis by \citet{Dimitriadis2023}.  These authors claim to detect a flux excess within the first 1.5 days after explosion that they attribute to circumstellar interaction. Our \texttt{ISIS} light curve shows a similar variation as the \citet{Dimitriadis2023} \texttt{TESSreduce} light durve at BJD$-$2,457,000 = 2476 to 2478 days, which they interpret as a flux excess. However, our model comparison formalism does not show any evidence for an extra flux component---a curved power law can fit the data just as well as a model with a companion interaction component.  This difference shows the advantage of using our model comparison approach, which evaluates whether or not the data prefer additional flux components. In a similar way, \citet{Dimitriadis2023} claim that they detect the excess in ZTF {\it g}-, {\it r}-, and {\it i}-bands, but it is possible that the ZTF data would be fit just as well by a flat baseline (see their Figures~14 and 15).

SN2021zny lies on a TESS CCD "strap," which amplifies the time variable background and systematic errors from spacecraft jitter.  Although \texttt{TESSreduce} and \texttt{ISIS} do a reasonable job of correcting these issues, the corrections have finite accuracy.  We found good agreement between our \texttt{ISIS} light curve and the \citet{Dimitriadis2023} \texttt{TESSreduce} light curve, but there are residuals near the peak of the light curve in our fits that show similar amplitudes to the putative excess near the time of first light.  Given the complicated issues and the small amplitudes, it is uncertain whether or not these features are astrophysical in origin.

\movetabledown=2.55in
\begin{deluxetable*}{lrrccc}
\tablewidth{0pt}\tablecaption{SN with Evidence for or Against Companion Interactions \label{tab:candidates}}
\tablehead{ 
\colhead{Name}& \colhead{$\ln K$}\tablenotemark{a}& \colhead{$\Delta {\rm BIC}$}\tablenotemark{b}& \colhead{rms$_{\rm Gauss}/$rms$_{\rm observed}$}\tablenotemark{c} & \colhead{Median Radius\tablenotemark{d}} &\colhead{Median Separation\tablenotemark{e}}}
\startdata 
SN2020abqu&   5.62   &   7.22   & 0.5133 &  3.39  $\pm$1.68&  0.61  $\pm$0.30\\ 
SN2021ahmz&   4.33   &   4.56   & 0.4444 &  1.94  $\pm$0.84&  0.35  $\pm$0.15\\ 
SN2022ajw &   3.06   &  -0.71   & 0.6124 &  2.57  $\pm$1.21&  0.46  $\pm$0.22\\ 
\hline
SN2020tld &  -16.27  &  -32.31  & 0.4196 &      --       &      --       \\ 
SN2022eyw &  -15.26  &  -32.44  & 0.8696 &      --       &      --       \\ 
\enddata 
\tablecomments{ SN2020abqu, SN2021ahmz, and SN2022ajw show a slight preference for adding a companion interaction model to a curved power law, though the statistical evidence is not robust when using $\Delta$BIC. SN2020tld and SN2022eyw show robust evidence against the addition of a companion interaction model.   \tablenotetext{a}{Natural log of Bayes Factor $K$.  Larger values of K indicate a preference for companion interactions.}\tablenotetext{b}{Difference between Bayesian Information Criteria for models with and without companion interactions.  Larger values of $\Delta {\rm BIC}$ indicate a preference for companion interactions.}  \tablenotetext{c}{ Noise metric defined in \S\ref{sec:results}, which quantifies departures of the light curves from random Gaussian noise. Values near zero indicate very little improvement from binning (systematic errors dominate over random noise), while 1 indicates perfect Gaussian noise scaling  (random noise only). For SN2022eyw, the light curve is affected by scattered light but the SN signal occurs at a time isolated from the excess noise.} \tablenotetext{d}{ Companion Roche lobe radii in units of solar radii.  "Median Radius" gives the median (50th percentile) and 68\% credible region of the posterior distribution.} \tablenotetext{e}{Companion separations in units of 10$^{12}$~cm.  "Median Separation" gives the median (50th percentile) and 68\% credible region  of the posterior distribution.}}
\end{deluxetable*}

\subsection{Companion Upper Limits\label{sec:limits}}

Even if the data for a given light curve does not prefer one model or the other, the \texttt{dynesty} posterior distributions provide  upper limits on the presence of any companion interaction signatures.  In particular, the large sample of upper limits from \tess\ allows us to control for geometric effects that can obscure companion interactions from the observer's line of sight.  Following \citet{Fausnaugh2021}, the  number of SNe needed to put limits on the companion Roche lobe radii/separations at the 64\%, 95\%, and 99.9\% confidence levels (1$\sigma$, 2$\sigma$, and 3$\sigma$ limits) are 14, 29, and 44 SNe, respectively.  This calculation is based on the assumption that the observer's viewing angles are randomly distributed, and that non-detections correspond to systems with companions at large separations but viewed at angles greater than 45 degrees. \footnote{The amplitude of the companion interaction signature decreases with increasing viewing angle, and a more realistic approach would take into account the fact that companion interactions are still detectable at viewing angles greater than 45 degrees.  However, the amplitude falls off relatively quickly with viewing angle, while all companion interactions at smaller angles are brighter and so easier to detect.  Fixing the viewing angle also simplifies the analysis because the viewing angle is degenerate with companion separation. }

\begin{figure}
    \centering
    \includegraphics[width=0.5\textwidth]{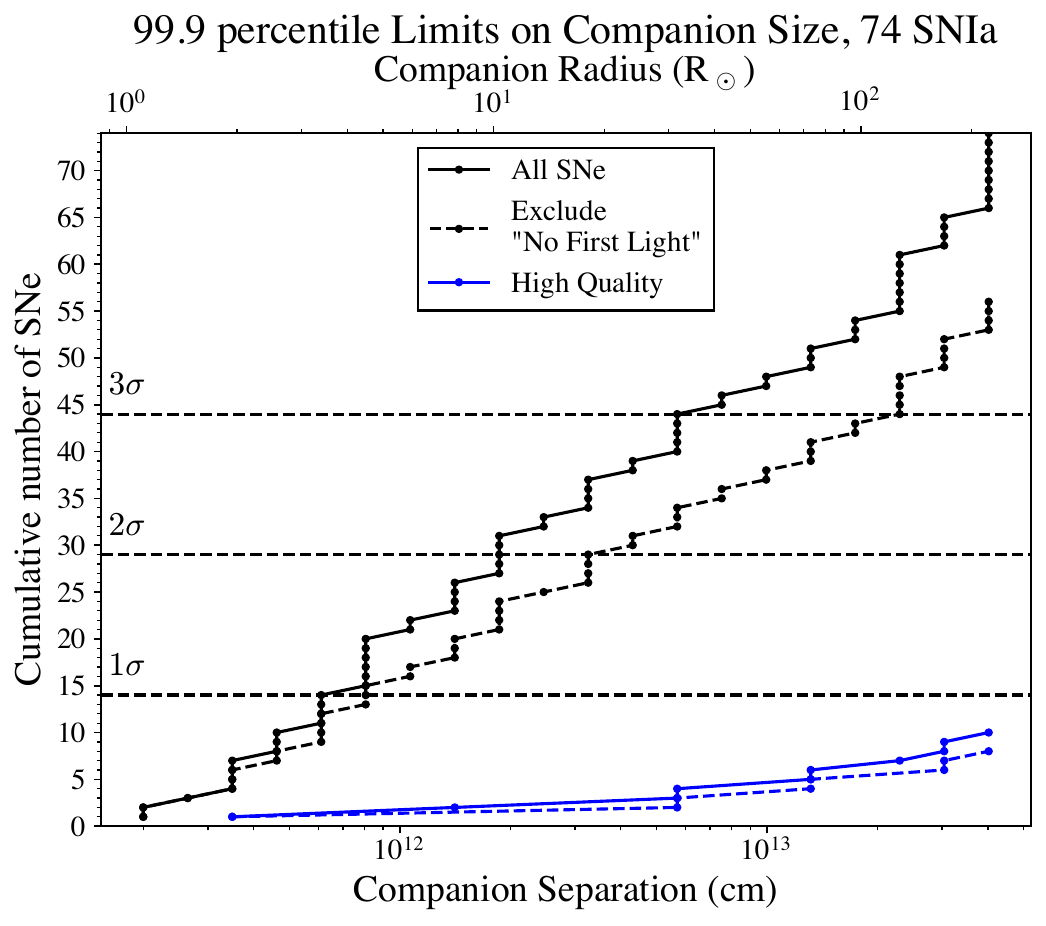}
    \caption{Cumulative number of SNe as a function of 3$\sigma$ companion limit.  The horizontal dashed lines show the probability that a given number of SN were all observed at viewing angles greater than 45 degrees and would not show a significant companion interaction signature.  The intersection of these lines and the cumulative distribution therefore represents upper limits on typical Type Ia SN companion stars.  The "No First Light" sample requires extrapolation to the time of first light and so introduces ambiguity into the fits for the companion interaction models.  The "High Quality" sample 	has a noise metric rms$_{\rm Gauss}/$rms$_{\rm observed}$ within a factor of 2 of Gaussian statistics and constitutes a more homogeneous data set, although there are only 10 objects with high enough signal-to-noise to yield unbiased fits (see~\S\ref{sec:results}).}    
   \label{fig:companion_limit}
\end{figure}

Figure~\ref{fig:companion_limit} shows the cumulative distribution of the 3$\sigma$ upper limits on  companion separations and radii.  Assuming that companions are common, the horizontal lines show the probabilities of finding a given number of SN with all companions viewed at angles greater than 45 degrees.  The 2$\sigma$ line crosses the cumulative distributions at \twoSigmaLimit), and the 3$\sigma$ line crosses the distribution at \threeSigmaLimit).  There is more uncertainty in the companion upper limits for light curves with gaps near the time of first light, so we also show the geometric limits after excluding the "No First Light" subsample of light curves.  In this case, the limits shift to about 18 R$_\odot$ and 125 R$_\odot$ for the 2$\sigma$ and 3$\sigma$ limits, respectively ($> 3\times 10^{12}$ cm and $> 2\times 10^{13}$ cm).  

There only 10 "High Quality" SNe that overlap with the "High Dynamic Range" subsample, so the geometric limits are not statistically significant in this case. However, this subsample is very homogeneous---all light curves are from our \texttt{ISIS} reduction pipeline, and only three SN (SN2020dxr, SN2021hup, and SN2022eyw) used the background model correction.

These results disfavor stars at large separations that  fill their Roche lobes, in agreement with previous results \citep{Hayden2010,Bianco2011}.  The companion star limits presented here are about the same as previous limits based on stacking, but are presented for a relatively large number of individual SN light curves.  In total, \NsupernovaLessTenRsun\ light curves have very tight limits $\leq10\,$R$_\odot$ (separations $<1.4\times 10^{12}$~cm).

The 3$\sigma$ limit at about \threeSigmaLimit) does not preclude the candidate companion interactions from \S\ref{sec:model_comp}, which have estimated companion Roche lobe radii of 1.9--3.4 R$_\odot$ (separations of  3 to $6\times10^{11}$~cm).  A confirmed companion would do the most to inform these statistical limits, which we discuss in \S\ref{sec:pop}. 

\section{Discussion\label{sec:discussion}}

Despite the large number of Type Ia SNe observed by \tess, three-quarters of the light curves do not have a high enough signal-to-noise ratio to reliably measure the time of first light $t_0$ and rising power law index $\beta_1$.  For \tess\ light curves, the noise requirement is  $\lesssim$\,10\% of peak on 30 minute time scales, equivalent to a dynamic range $\Delta T$ between peak and detection of $ > 2.8$ mag on 8 hour timescales.  A prior on the light curve shape (i.e., assuming a value for $\beta_1$) would allow us to extrapolate noisy \tess\ SN light curves detected near peak back to the time of first light.  The population distribution from KDE in \S\ref{sec:paramdists} shows a dominant Gaussian with mean $\beta_1 = 2.29$.  Assuming this value for $\beta_1$ would allow  \tess\ to put constraints on the time of first light even for SNe  detected near peak at 19th or 20th magnitude.  Although the population dispersion of $\beta_1$ is quite broad with a width of 0.34 and a longer tail towards small values less than 1.0, errors in the assumed value of $\beta_1$ add only a modest amount of noise to the estimates of the time of first light.  For example, if we assumed $\beta_1 = 2.29$ when it in fact was 2.0, the error on the time of first light is $<$\,1.6 days. The error grows to about 2.0 days if the true value of $\beta_1 = 1.95$ (1$\sigma$ deviation) and 6.2 days if the true value of  $\beta_1 = 1.0$.  \tess\ estimates of the time of first light may therefore provide a useful supplement to SNe discovered by surveys such as ASAS-SN, ATLAS, ZTF, and eventually bright SN discovered by the Rubin Observatory and the \textit{Roman Space Telescope}. 

\begin{figure*}
    \centering	
        \includegraphics[width=\textwidth]{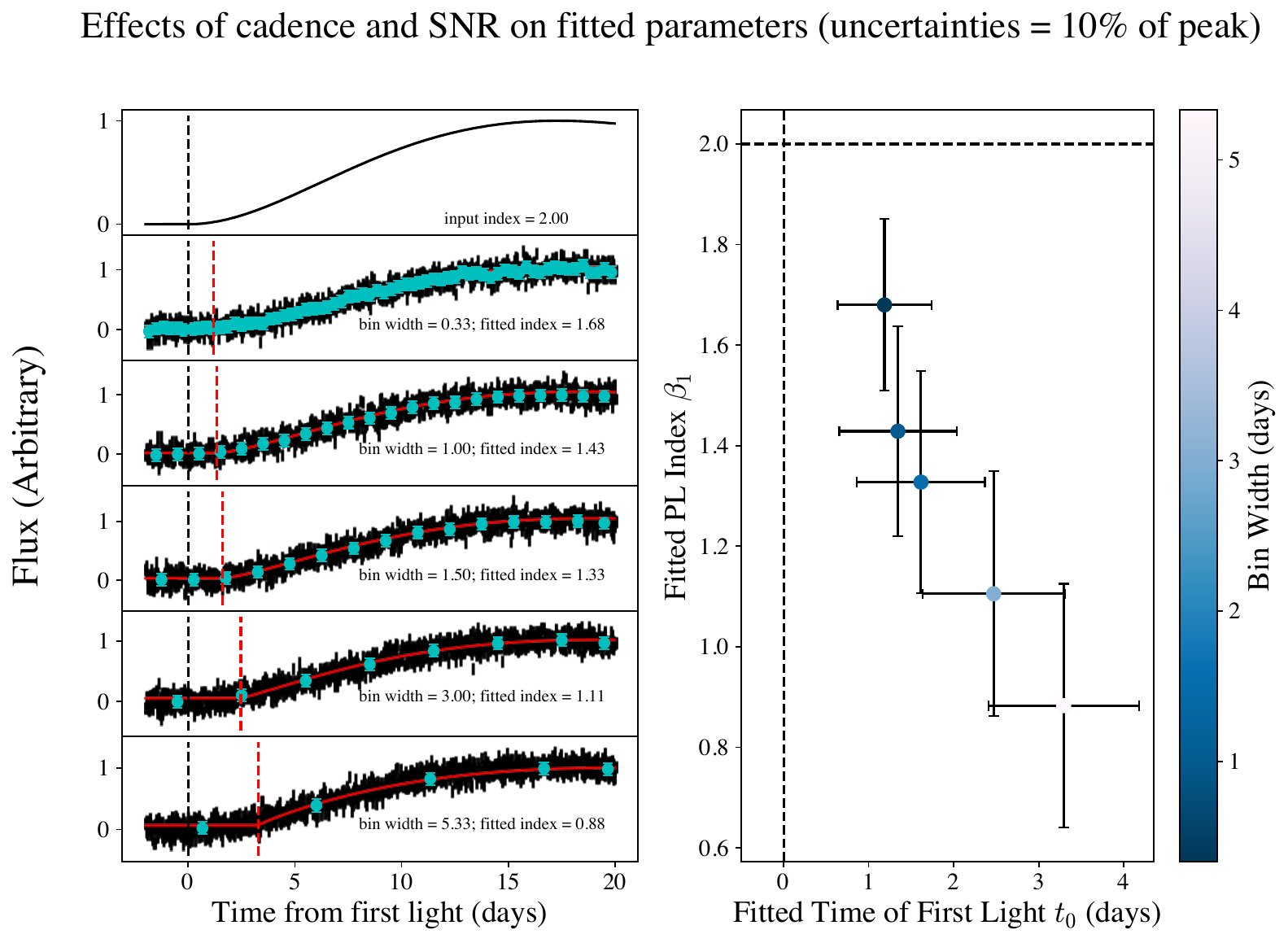}
    \caption{Simulations that quantify the effect of cadence and signal-to-noise ratio on fitted parameters for typical ground-based light curve  properties.  The left panels show simulated light curves with 30 minute sampling and Gaussian noise with $\sigma$ equal to 10\% of peak flux.  The input model is shown in the top panel, and has a time of first light $t_0 = 0.0$, power law index $\beta_1 = 2.0$, power law index $\beta_2 = -0.015$, corresponding to a rise time $t_{\rm rise} = 17.25$ days.  In the lower panels, the light curve is binned to 0.33, 1.0, 1.5, 3.0, and 5.3 days (cyan points), and assigned a 10\% uncertainty relative to peak.  The maximum likelihood models from \texttt{dynesty} are shown with red lines, and the recovered value of $t_0$ is shown with the vertical dashed red lines.  The right panel shows the recovered values of power law index $\beta_1$ as a function of the recovered time of first light $t_0$, with the true values of the input model shown by dashed black lines.   There is a  bias in the recovered parameters as the cadence decreases and the uncertainty on the measurements at early times increases. }
   \label{fig:simbinning}
\end{figure*}

 We caution that the requirements for light curves from other facilities for measuring $t_0$ and $\beta_1$ may be different, depending on their point-to-point precision and cadence. We have experimented with simulations that rebin the TESS light curve to the typical cadence of ground-based light curves (1 to 5 days), and assigned varying levels of uncertainties. We found that for cadence $>$\,5 days or uncertainties $>$10\% of peak light curve flux, there are significant biases in the fitted values of $t_0$ and $\beta_1$---an example is shown in Figure~\ref{fig:simbinning}. This result suggests that ground-based studies should be careful about the data they use to investigate early-time SN light curve behaviors; generally, high signal-to-noise and fast cadence are required to recover unbiased parameters.  In addition, the biases from Appendix~\ref{sec:obs_bias} show that the method used to fit the light curves can introduce substantial biases, which we discuss in the next section. 

\subsection{Comparison to Previous Results \label{sec:discuss_sample}}
It is difficult to compare studies of early time Type Ia light curve shapes and rise times across the literature (see the detailed discussion in \citealt{Miller2020b}, their \S9.2).   The primary difficulty is whether to  shape-correct the light curves and apply $K$-corrections (e.g., \citealt{Riess1999,Hayden2010,Ganeshalingam2011}), or to measure the rise times directly from the data (\citealt{Firth2015,Zheng2017a, Zheng2017b, Papadogiannaki2019, Miller2020b, Fausnaugh2021}).    Measuring the rise times directly from the data is independent of the choice of a Type Ia shape template or the early time SED, and leaves open the possibility of correlating the rise times against shape parameters, observed colors, and luminosities.  We therefore took the latter approach in \S\ref{sec:paramdists}.

Our measurement of the average power law index $\beta_1$ is \meanPLAll, while our estimate of the underlying population distribution using KDE finds a Gaussian component with $\beta_1 = 2.29 \pm 0.34$ and a long tail towards lower values of $\beta_1$.  In either case, the typical uncertainty on individual estimates of $\beta_1$ is small, so the relatively large widths of the distributions indicate a range of Type Ia SN properties.  Our estimates of the mean value of $\beta_1$ are consistent with previous results.  \citet{Papadogiannaki2019} found an early time  power law index of 1.97$\pm$0.06 and \citet{Miller2020b} found a value of $2.0 \pm 0.5$  (in {\it r}-band).\footnote{\citet{Firth2015} found a power law index of 2.44$\pm$0.13, which is formally inconsistent with these estimates.  However, \citet{Miller2020b} suggest that this may be an artifact of their fitting procedure.}   The differences in $\beta_1$ may result from the \tess\ bandpass (600-1,000 nm), which extends to significantly redder wavelengths than the PTF/ZTF {\it r}-band (500 to 700 nm). SNe colors also evolve in the early phase of the light curve, so we do not expect the power law rise in different filters to be identical.

The wide range of observed values of $\beta_1$ is consistent with previous work that finds a diversity of behavior in early time Type Ia SN light curves \citep{Papadogiannaki2019, Miller2020b}.  However, the broad tail towards small values of $\beta_1$ and the dominant Gaussian component at $\beta_1 = 2.29 \pm 0.34$  suggest structure in the underlying distribution. Previous work  has found a mix of shapes in rising SN light curves, some that rise linearly and some that rise parabolically.  However, the tail towards small values of $\beta_1$ suggests additional structure in the population distribution and a continuum of rising light curve shapes.  A larger sample of early time light curves is needed to verify the existence of these components and to better characterize the intrinsic distribution's properties.  The KDE estimates of the distribution should be interpreted qualitatively because the width of the Gaussian component strongly depends on the KDE bandwidth parameter.  However, we can speculate about possible interpretations of these components.  The dominant Gaussian component might correspond to normal Type Ia SNe, with the width representing some range of physical properties such as explosion energy, $^{56}$Ni mixing,  ejecta density profile, viewing angle, or some combination of the these parameters.  The long tail towards small values of $\beta_1$ may represent a different population of Type Ia SNe, perhaps that have distinct explosion mechanisms or ejecta characteristics.  Alternatively, the tail may represent a range of extrinsic properties, such as the presence and distribution of circumstellar material around the exploding star.  However, all of the SNe in the tail are classified as normal Type Ia on TNS except for SN2019ugr (SNIa-91T-like), so structure in the underlying distribution does not trivially map onto simple classification schemes.   \tess\ will provide a larger sample as it continues to observe, and with additional data such as color information at early times (by using, for example, ground-based data from existing surveys), it should be possible to determine if the tail towards smaller values of $\beta_1$ represents a distinct population of Type Ia SNe or not.

Our measurement of an average rise time $t_{\rm rise}=$\,\meanTriseAll\ days is very close to the values from  \citet{Zheng2017b}, \citet{Papadogiannaki2019}, and \citet{Burke2022a}, who found $t_{\rm rise} = $16.0$\pm$1.8 days (56 Type Ia SNe), 16.8$\pm$0.6 days (265 Type Ia SNe), and 16.8$\pm$1.3 days (9 Type Ia SNe), respectively.  Our estimate of the underlying population distribution using KDE results in a consistent value with slightly smaller dispersion, $15.89\pm 3.82$\, days.  These estimates of $t_{\rm rise}$ were derived by fitting the full extent of the light curve from before first light to peak with the broken power law model from \citet{Zheng2017a}. \citet{Firth2015} and \citet{Miller2020b} found larger values of $t_{\rm rise}$,  18.98$\pm$0.54 days (18 Type Ia SNe) and 18.5$\pm$1.6 days (127 Type Ia SNe), respectively.   These larger estimates were derived by instead fitting a single power law up to fixed flux fractions of  peak (50\% and 40\%, respectively).  As discussed by \citet{Miller2020b} in their Appendix~D, the fitted power law index and rise time depends on the choice of flux fraction of peak, with smaller fractions of peak (down to 25\%) resulting in smaller power law indices and rise times.  The dependence on flux fraction of peak exists because the power law index $\beta_1$ and time of first light $t_0$ are correlated, and the light curve curvature is already starting to change at 20--50\% of the peak flux.  Thus, fitting different flux fractions leads to different average power law indices, which changes the derived values of $t_0$ and  $t_{\rm rise}$.   The curved power law model that we employ and the broken power law from \citet{Zheng2017a} take the changing curvature into account, but this makes comparison of the rise times with \citet{Firth2015} and \citet{Miller2020b} difficult.  Color evolution may also contribute to differences in $t_{\rm rise}$ because SNe peak at different times in different filters and so must have a different value of $t_{\rm rise}$ in the \tess\ passband than the {\it r}-band.

\subsection{Caveats on Companion Interaction Models\label{sec:caveats}}

There are two key parameters in the \citet{Kasen2010} models that we did not explore here: the velocity and opacity of the SN ejecta.  There are large uncertainties on both of these physical parameters.  Higher ejecta velocities convert more energy into thermal radiation when the SNe ejecta encounter the companion and produce a bow shock. The companion interaction signal is proportional to the square of the ejecta velocity, so even a factor of 3 change in ejecta velocity can change the companion interaction signal by an order of magnitude.   However, we have no way of directly measuring the ejecta velocity at these early times (within a few hours or minutes of explosion).  Instead, we fix the ejecta velocity  in our \citet{Kasen2010} models by assuming a typical  explosion energy of 10$^{51}$ erg s$^{-1}$ and ejecta mass of 1.4 M$_\odot$.  As a consequence, our results do not take into account variations in the explosion energy, although the assumed value is in reasonable agreement with the few early spectroscopic observations that exist (e.g., \citealt{Ashall2022}).  Opacity calculations, on the other hand, are subject to  theoretical uncertainties from ejecta composition and absorption cross-section, especially in the near-infrared.  Furthermore, the population dispersion of the early time ejecta opacity is almost entirely unconstrained because there are only a handful of SN spectral templates and none of these templates are valid at the earliest times of the light curve.  The \citet{Kasen2010} models assume Thomson opacity, and large deviations from this value would impact our results.  However, the \citet{Kasen2010} models also depend less sensitively on the opacity than the ejecta velocity, so variation in this parameter likely contributes a more subtle effect. The assumed primary-secondary mass ratio of 1.4:1.0 also affects the relationship between the companion separation and Roche lobe radius, but enters as the 1/3 power and so has a very small effect relative to uncertainties in $v_{\rm ej}$~and~$\kappa$.

Another limitation of our analysis is that complicated early time light curve shapes can also be produced by interactions with circumstellar material or the distribution of $^{56}$Ni in the SN ejecta.  Color information at the early times of the light curves would probably do the most to constrain the nature of the candidate companion interactions in \tess\ light curves.  Independent data from other telescopes can also help constrain the shape of the early time light curves (see \S\ref{sec:model_comp} and Appendix~\ref{sec:noise}).  We will pursue a larger project that draws on multi-band light curves from ground-based surveys (ASAS-SN, ATLAS, and ZTF) to investigate the nature of the tentative candidate companion interactions presented in \S\ref{sec:model_comp}.  Additional data, particularly spectra, are also important for distinguishing between competing interpretations of early time Type Ia SN light curves \citep{Ashall2022, Dimitriadis2023}. 

\subsection{Confirming or Falsifying Companion Star Candidates\label{sec:confirmdeny}}
We now turn to the question of how to confirm or falsify the five SNe reported in \S\ref{sec:model_comp} that tentatively show evidence for or against companion signatures.   As a reminder, the light curves all show some level of correlated noise (rms$_{\rm Gauss}/$rms$_{\rm observed} \ge 0.36$), which can be difficult to interpret.  However, if the companion signals are real, we would expect to find hydrogen emission in the nebular spectra of the SNe.  Hydrogen emission is expected for single degenerate systems because material is stripped from the envelope of the companion star by the collision of the supernova ejecta \citep{Chugai1986, Marietta2000}.  Detection of hydrogen in the SNe that favor companion interactions would go a long way towards building confidence in the interpretation of these objects as single degenerate progenitor system.  The hydrogen signature is also expected to exist regardless of viewing angle, and so the presence of hydrogen in a SN for which the \tess\ light curve disfavors a companion would rule out a double degenerate interpretation and the \tess\ light curves would instead  constrain the viewing angle.  

On the other hand, non-detections  of hydrogen put upper limits on the amount of stripped material from the companion's envelope.  There are theoretical uncertainties on the predicted strength of the hydrogen emission lines, but detectable H$\alpha$ emission is expected for stripped masses of 0.01~M$_\odot$ or greater \citep{Liu2017, Botyanszki2018, Dessart2020}.   Observations of nebular-phase Type Ia SNe  have set limits on stripped hydrogen of less than 0.001 M$_\odot$ or smaller for about 120 SNe within $\sim$80 Mpc \citep{Tucker2020, Graham2022}. 

 The five SNe that show evidence for or against companion interactions are at redshifts from $z=0.009$ to 0.049, (39--217 Mpc).  At these distances,  nebular-phase spectroscopy might be within the reach of 6--8 meter class telescopes.  However, at the time of this writing, all the supernovae are 450 days or longer past the time of discovery; combined with the large distances, they are not suitable for nebular phase spectroscopic observations.  
 
Another means of following up these results might be to search for surviving luminous companion stars with high resolution imaging from {\it HST} or {\it JWST}. Ten hours of integration with WFC3 could potentially push limits on surviving companion stars to absolute magnitudes of $-$5.5 ({\it F814W}) at 40 Mpc, similar to the magnitude of the candidate He star donor for SN2012Z \citep[although this SN is a Type Iax]{McCully2014}.  However, our closest SN with a companion interaction candidate is at 140 Mpc (SN2021ahmz) and the closest SN that disfavors a companion interaction candidates is at 38~Mpc (SN2022eyw),  so follow-up with {\it HST} is not feasible.  {\it JWST} NIRCam can provide the same limit at 70--80 Mpc  in 10 ks ({\it F150W} or {\it F200W}).  Such a limit would test for the presence of high-mass companion stars \citep{Li2011}, and both SN2020tld and SN2022eyw are viable candidates for high-resolution imaging within 70 Mpc. While both of these light curves disfavor companion star signatures, surviving giants could exist due to orientation effects.    This rough estimate for NIRCam does not take into account potential differences in the companion SEDs, extinction by the host galaxy, or reprocessing of optical light into the IR by dust formed in the SN ejecta.  Crowding will also be a difficulty for such observations, because the angular resolution of NIRCam corresponds to a physical scale of  about 20 pc at 50~Mpc.

 \subsection{Type Ia SN Progenitor Population\label{sec:pop}}
Although correlated noise significantly affect the \tess\ light curves and results in at least 27 false positives out of 74 light curves, the \citet{Kasen2010} models usually do not improve the light curve fits.  We therefore should have detected companions above the upper limits of our posterior distributions if they exist in our sample.  This is most clearly the case for very high signal-to-noise ratio light curves, such as those with dynamic range $\Delta T > 3.5$\, mag.  Examples in Figure~\ref{fig:SN_model_fits} include SN2018hib, SN2018koy, SN2019ltt, SN2020aoi, SN2021iff, SN2021zny, and SN2022exc (see the online article figureset).  In these cases, the light curve flux limits rule out large companion interaction models, and the limits would be even smaller without the presence of correlated noise.  The geometric $3\sigma$ limit on typical companion star Roche lobe radii $<$\,31 R$_\odot$ (separations $<$\,$5.7\times 10^{12}$ cm) should therefore be robust.  

 Galactic red giant-white dwarf binary systems such as Omicron Ceti, RS Ophiuchi, and T Coronae Borealis have long orbital periods between about 200 and 500 days, corresponding orbital separations $\gtrsim$\,$2.7\times 10^{13}$ cm and Roche lobe radii $\gtrsim$\,150 R$_\odot$.  Our sample has 61 SN with upper limits below this value, which means that geometric effects cannot be invoked to explain non-detections of analogous progenitor systems in the \tess\ sample. The alternative explanations are that the average SN explosion energies are significantly smaller than what we assume here, the ejecta are more opaque, or similar red giant-white dwarf binary systems are not common progenitors of Type Ia supernovae.  The Galactic binary systems may still result in Type Ia SN as long as these kinds of systems do not represent a sizable fraction of the Type Ia population.
 
A confirmed companion would do the most to constrain the fraction of single degenerate Type Ia progenitor systems.  Intuitively,  viewing angles $<$\,45 degrees represent a relatively small fraction of the available parameter space, so even a single detection implies a large fraction of systems with companion stars.  To illustrate how detections constrain this fraction, we use the following statistical model.   Assuming that the \citet{Kasen2010} model holds for the fraction $f=0.146$ of viewing angles within $45^\circ$, at a given orbital separation we have a binomial probability of
\begin{equation}
  { d P \over d F} \propto \left( f F \right)^{N_d} \left ( (1 - f )F \right)^{N_n}
\end{equation}
for the intrinsic fraction of systems with companions $0 \leq F \leq 1$, where $N_d$ are the number of companion signature detections and $N_n$ are the number of non-detections.  We consider three scenarios.  First, for each companion radius/orbital separation, we consider SN2020abqu, SN2021ahmz, and SN2022ajw as detections at companion radii smaller than the 99.9th percentile of the posterior distribution.   Second, we assume that all of these detections are false positives created by other physical processes (e.g. $^{56}$Ni mixing) or \tess\ systematics. We count the other 71 SNe from the "High Dynamic Range" subsample as non-detections for these two calculations. In the third scenario, we calculate $F$ allowing for 5 additional companion signatures that we may not have detected because our tests did not favor one model or the other.  We arbitrarily choose $N_d + 5$  only to demonstrate the sensitivity of the inferred value of $F$ to the number of detected companions.  We assumed that these hypothetical companions have Roche lobe radii less than 10 R$_\odot$ (separations less than $1.9 \times 10^{12}$ cm), i.e., below the geometric 2$\sigma$ limit from~\S\ref{sec:limits}.

\begin{figure}
    \centering
    \includegraphics[width=0.5\textwidth]{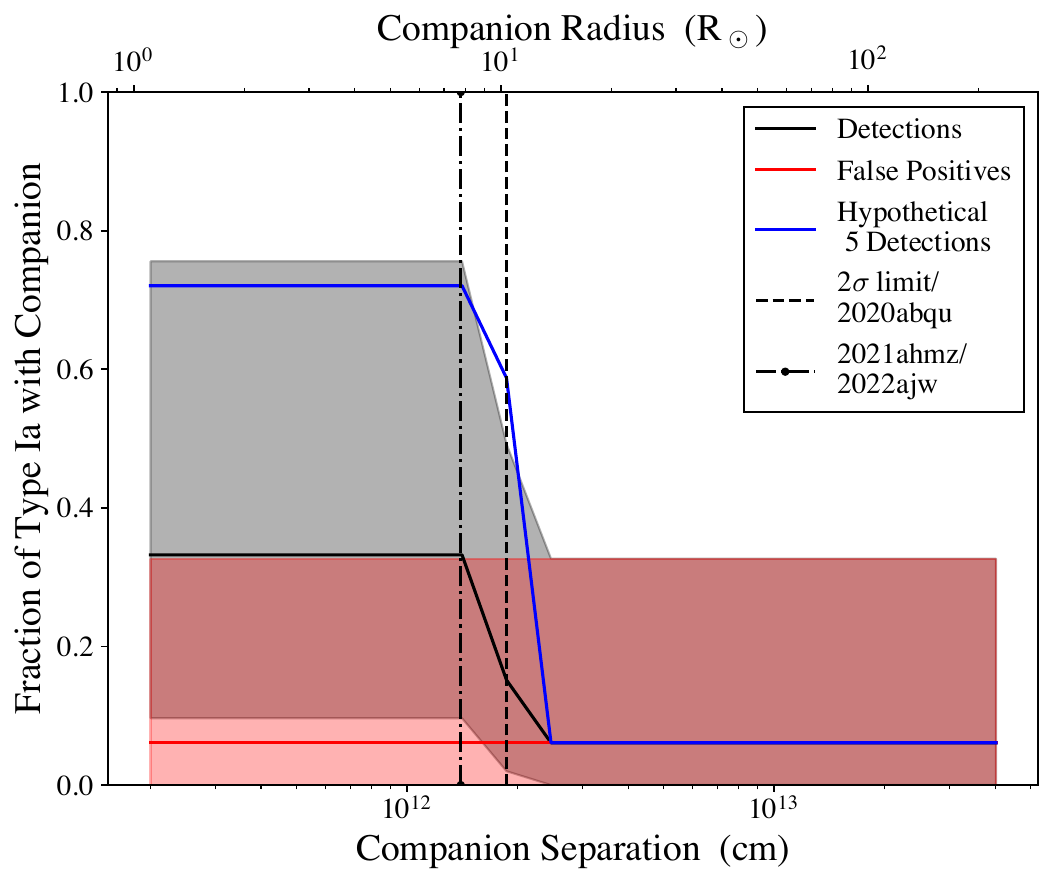}
    \caption{Constraints on the fraction $F$ of Type Ia SNe with companion stars, as a function of companion star Roche lobe radius/orbital separation.  Three scenarios are shown.  The black line shows the case where  the candidate companion interactions from \S\ref{sec:model_comp} are considered detections.  The red line shows the case where the candidates are false positives, due to  \tess\ systematic errors or physical processes such as $^{56}$Ni mixing.  The blue line shows a hypothetical case in which we add five additional companion signature detections below 10 R$_\odot$ (the geometric 2$\sigma$ limit from Figure~\ref{fig:companion_limit}), to demonstrate the sensitivity of the inferred value of $F$ on the number of detections.  The solid lines show the median values of $F$ calculated from a binomial distribution and the shaded regions show the 95\% confidence intervals.    }
   \label{fig:companion_population}
\end{figure}

Figure~\ref{fig:companion_population} shows the constraints on $F$ as a function or orbital separation for the three fiducial cases.  Where $N_d > 0$ we show the median and 95\% confidence range on $F$, and where $N_d = 0$ we show the 97.5\% confidence ($\sim$2$\sigma$) upper limit on $F$.   If the companion signatures are false positives,  the $\sim$2$\sigma$ upper limit is $F\leq 0.33$.  However, only two detections bring the 2$\sigma$ upper limit to $F\lesssim 0.76$ (at small companion Roche lobe radii/orbital separations).  If 5 of the other 72 Type Ia light curves contain companion interactions below our upper limits but were not identified by the model comparison tests, the median value of $F$ would be 0.72 and the 2$\sigma$ upper limit would be 0.98.  There may be additional systematic uncertainties on $F$ from the selection function of our sample, which depends on both  the efficiency of the ground-based surveys and the detectability of SN signals in \tess\ FFIs.

Given the sensitivity of $F$ to detections, it is important to determine if the three tentative companion interactions from \S\ref{sec:model_comp} are real.  Unfortunately, we cannot be confident about this based on the \tess\ data alone.  The difficulty is that the amplitudes of the companion interactions are very small and so detections are very sensitive to the noise properties of \tess\ light curves.  Larger amplitude \citet{Kasen2010} models are  easier to detect, and SN2020abqu, SN2021ahmz, and SN2022ajw are at relatively large distances for the local sample (about 140--217 Mpc. $z = 0.032$ to 0.049).  If the candidate companion signatures are real,  we would therefore expect to find stronger companion signals in future \tess\ observations, especially for brighter SNe at lower redshift.
 
The low detection rate of \citet{Kasen2010} models in \tess\ light curves at first glance appears to be at odds with recent work that find 10--33\% of Type Ia light curves have early time flux excesses \citep{Deckers2022,Magee2022,Burke2022a,Burke2022b}. However, if the companion stars are at relatively small separations, the companion interaction signals are relatively small and \tess\ is not sensitive enough to detect the flux excess unless the system is at a very small cosmological distance. If the rate of flux excesses is in fact 30\%, we should expect to detect a companion interaction signature from \tess\ in the near future.  If \tess\ instead fails to find such an interaction, there are several possible explanations.

The first possibility is that the flux excesses found by \citet{Deckers2022}, \citet{Magee2022},  \citet{Burke2022a}, and \citet{Burke2022b} have very blue colors, such that there is no appreciable signal in the \tess\ bandpass.  If the colors are bluer than the \citet{Kasen2010} model used here, it might suggest that the flux excesses are caused by some effect other than companion interaction signatures.  For example, most models with $^{56}$Ni mixed into the outer  SN ejecta have a bluer color than the predictions for the \citet{Kasen2010} models. 

The second possibility is that systematic errors in the \citet{Kasen2010} models over-predict the companion interaction signature in the \tess\ filter.  For example, the \citet{Kasen2010} models assume electron opacity, and \citet{Burke2022a} suggest that line blanketing could explain the poor matches that are observed between the \citet{Kasen2010} models and real Type Ia SN light curves at UV wavelengths.  In the same way, a color-dependent term could suppress the \citet{Kasen2010} predictions in the \tess\ bandpass.  Assumptions about the opacity are an obvious possibility; however, Fe-peak elements cannot  provide enough opacity at near-IR wavelengths to suppress the signal in the \tess\ bandpass, so it is unclear what effect would be in play that suppresses the signal in the \tess\ bandpass.

The last possibility is that the discrepancy is caused by different methods for fitting the models. Our model comparison approach is very skeptical of bumps and wiggles in early time light curves unless they are well matched in shape to the \citet{Kasen2010} models.  On the other hand, the claims of flux excess in some objects are sometimes based on only 1 or 2 data points, especially in the ZTF-2018 sample.  It is therefore unclear if there is good evidence for including the companion component in these cases, and so the true rate may not be as high as 30\%.  Our method should be an improvement, in that we explicitly check whether or not adding a companion is merited by the data.

 \section{Conclusion\label{sec:conclusion}}
 
 We extracted and analyzed  \NsupernovaLCs\ light curves of Type Ia SN from \tess\ Sectors 1--50.  Measurement bias driven by the signal-to-noise ratio limits our sample to \NsupernovaHighDR\ SNe with reliable measurements of the time of first light and the rising power law index.  We also searched the \tess\ light curves for evidence of companion interaction signatures.  Our main results are:
 \begin{itemize}
 \item The average power law index $\beta_1$ is \meanPLAll\ and the average rise time $t_{\rm rise}$ is \meanTriseAll\ days.  Both values are consistent with previous results.  The smallest reliable rise time is 8.4 days for SN2019gqv, and the largest is 22.9 days for SN2020bqr.

\item We find an underlying population distribution for the power law index $\beta_1$ that consists of a Gaussian component with mean 2.29 and width 0.34, and a long tail extending to small values of $\beta_1 < 1.0$.  These components may represent two distinct populations of Type Ia SNe, or may represent the ranges of SNe explosion parameters and extrinsic factors such as the presence of circumstellar material.  For $t_{\rm rise}$, we find an underlying Gaussian distribution with mean 15.89 days and width 3.82 days.

 \item Correlated noise in the \tess\ light curves makes the rate of companion star false positives high, but we identify problematic light curves with residual systematic errors by examining the noise on different timescales (from 2 hours to 5.3 days).  The presence of correlated noise also requires caution when analyzing individual objects. An example is SN2021zny, where our analysis does not find evidence for an additional flux component and we attributes features in the early time light curve to systematic errors from the data reduction. 
 
 \item We found three SNe with noise properties within a factor of 3.0 of the expectation for Gaussian statistics that show a slight preference for the addition of a companion interaction to a curved power law model: SN2020abqu, SN2021ahmz, and SN2022ajw.   The candidate companion stars have Roche lobe radii of 3.4, 1.9,  and 2.6 R$_\odot$ for SN2020abqu, SN202ahmz, and SN2022ajw respectively (orbital separations between 3--6$\times 10^{11}$ cm).  If these candidates are not due to residual systematic errors, they may have had single degenerate progenitor systems, or the shapes of the light curve may be influenced by additional physical effects (e.g., $^{56}$Ni mixing or circumstellar interactions). However, the evidence in favor of the companion interactions is based on the Bayes factor, and the detections are not robust for the Bayesian Information Criterion. We therefore label these as tentative interaction candidates.

 \item We found two SNe that statistically disfavor adding a companion interaction to a curved power law model: SN2020tld and SN2022eyw.  If the companion signatures are ruled out, these SNe must have had companion stars oriented at $>$\,45 degrees from the observer's line of sight, have had companion stars smaller than 1.1 R$_\odot$ (separations $< 2\times 10^{11}$ cm), or have no surviving companion stars.
 
 \item Controlling for orientation effects, we found 3$\sigma$ limits ruling out typical companion star Roche lobe radii greater than \threeSigmaLimit)  and 2$\sigma$ limits disfavoring Roche lobe radii greater than \twoSigmaLimit).  These results are in line with previous limits from stacking, and rule out red giants as common Type Ia SN companion stars.  We also found \NsupernovaLessTenRsun\ individual light curves where the  upper limits on companion star Roche lobe radii are $<$\,10 R$_\odot$ (separations $<2\times 10^{12}$~cm).
 
 \end{itemize}
  
\facility{TESS}

\software{ 
  Matplotlib \citep{matplotlib},
  Numpy \citep{numpy},
  Scipy \citep{scipy}, Astropy \citep{astropy},
  lmfit v1.0.3 \citep{lmfit},
  dynesty \citep{Higson2019,Speagle2020},
  pysynphot \citep{pysynphot}, SNooPy \citep{Burns2011},
  ISIS \citep{Alard1998}
 }
\bigskip

\begin{acknowledgments}

  This paper includes data collected by the \tess\ mission, which are publicly available from the Mikulski Archive for Space Telescopes (MAST). Funding for the \tess\ mission is provided by NASA's Science Mission directorate. This research has made use of NASA's Astrophysics Data System, as well as the NASA/IPAC Extragalactic Database (NED) which is operated by the Jet Propulsion Laboratory, California Institute of Technology, under contract with the National Aeronautics and Space Administration.

  CSK and KZS TESS research is supported by NASA grant 80NSSC22K0128.  The material is based upon work supported by NASA under award number 80GSFC21M0002. PJV is supported by the National Science Foundation Graduate Research Fellowship Program Under Grant No. DGE-1343012. MAT acknowledges support from the DOE CSGF through grant DE-SC0019323.  TD acknowledges support from MIT's Kavli Institute as a Kavli postdoctoral fellow.  This work was supported by an LSSTC Catalyst Fellowship awarded by LSST Corporation to TD with funding from the John Templeton Foundation grant ID \#62192.

\end{acknowledgments}

\bibliography{refs}

\appendix

\section{Light curves and data reduction.\label{app:lightcurves}}

We used the same light curve extraction method as \citet{Fausnaugh2021}, which employs difference imaging and point spread function (PSF) photometry with the \texttt{ISIS} software package \citep{Alard1998,Alard2000}.  For difference imaging, \texttt{ISIS} solves for a spatially variable kernel that transforms a reference image to match a target FFI as closely as possible.  The spatially variable kernel corrects systematic errors in \tess\ FFIs at the pixel level, such as those caused by differential velocity aberration and pointing jitter.  We constructed reference images for each \tess\ observing sector from the sigma-clipped average of 20 FFIs taken at times with the lowest backgrounds.  We configured \texttt{ISIS} to process each FFI in sixteen parallel subimages, which improves the speed of the difference imaging and controls for strong variations in the PSF across the focal plane.   Each subimage was 512$\times$512 pixels, which is a spatial scale over which the PSF does not substantially change.   \texttt{ISIS} also removed a global estimate of the background for each subimage by fitting and subtracting a two-dimensional polynomial before the image subtraction step.  We found that the quality of the image subtractions improved by first smoothing the \tess\ FFIs with a 2D Gaussian of full width at half maximum equal to 2.1 pixels.  Finally, we removed the TESS CCD "straps" (\citealt{TIH}, \S6.6.1) from the difference image by subtracting a median-filtered image of the affected columns from the difference images and interpolating the local background along rows.

For PSF photometry, we provided \texttt{ISIS} the location of each SN in the \tess\ FFIs based on the coordinates reported to TNS.  We then gave \texttt{ISIS}  the mission-supplied model of the \tess\ pixel response function (PRF)\footnote{https://archive.stsci.edu/missions/\tess\/models/prf\_fitsfiles/. } corresponding to the appropriate 512$\times$512 subimage to perform PSF photometry on each SN. The PRF is a convolution of the PSF with the spacecraft pointing jitter profile and  detector intrapixel responses.  \texttt{ISIS}  convolves the PRF model with the spatially variable kernel for each FFI at the location of the SN, fits the PRF model to the differenced FFI at that location, and  integrates the fitted PRF to measure the differential flux.   At this step, \texttt{ISIS} also estimates a local residual background for each SN from the median pixel value in an annulus with inner/outer radius of 4/8 pixels.  Errors in TNS coordinates can affect the forced photometry, though owing to the large \tess\ pixels shifts up to 8 arcseconds (0.4 pixels) typically do not significantly affect the resulting light curve.  The main exceptions are when the SN is near a bright star or the host galaxy is relatively bright.

This procedure resulted  in differential light curves relative to the SN flux in the reference image.  We have not been able to  measure the SNe flux in the \tess\ reference images to better than 10\% because of crowding by nearby stars and contamination by host galaxies in the detectors' large pixels.  Instead, we flux-calibrated the light curves by aligning the pre-explosion baseline of each SN light curve to zero flux.  We converted to magnitudes using the instrumental zeropoint of 20.44\,$\pm$\,0.05 \citep{TIH}.  For SNe observed across multiple sectors, the second sector is aligned to the same flux scale as the first sector by introducing a calibration parameter to the rising light curve model, which is described in Appendix~\ref{sec:fits}.  By treating the calibration offset as a nuisance parameter, we leverage the full dataset to infer the light curve shape while marginalizing over uncertainty in the flux calibration.

\begin{figure*}
    \centering
    \includegraphics[width=\textwidth]{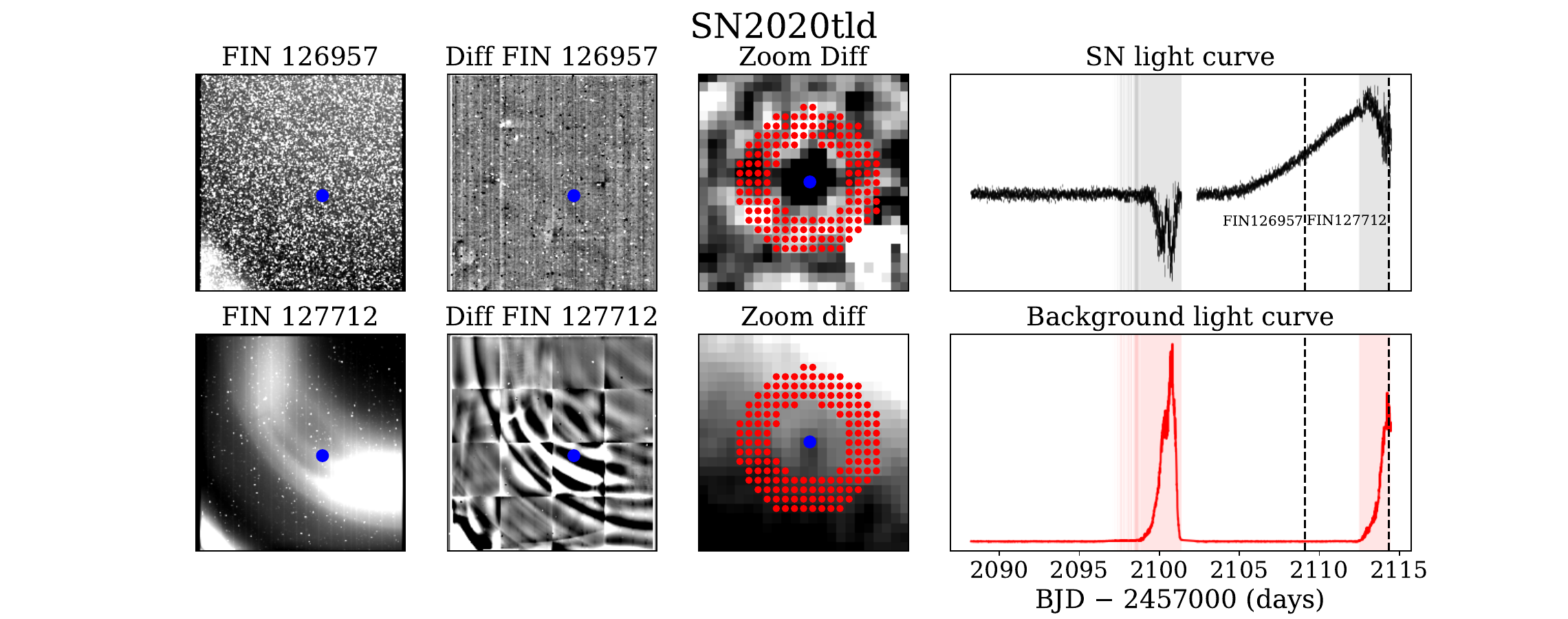}
    \caption{An example of our background trimming procedure from \S\ref{sec:systematics}.  A low background FFI is shown in the top panels and an FFI with bright scattered light is shown in the bottom panels.  The left-most column shows the original FFIs.  The second column shows the difference images.  \texttt{ISIS} fits and subtracts a 2D polynomial to remove the global background during difference imaging, which leaves behind high-frequency background variations in the bottom row.  The 4x4 "grid effect" in the bottom row is caused by the 16 subimages used in the difference imaging (see Appendix~\ref{app:lightcurves}).  The background correction is not forced to be continuous across subimages.  The position of SN2020tld is shown with the blue dot, and the third column shows a 23x23 pixel zoom-in of the difference image around SN2020tld.  The red dots show the annulus in which we calculate a local background correction  in the difference image.  The rightmost column shows the SN light curve (top) and local background estimates from the annulus (bottom), with vertical dashed lines marking the epochs of the two FFIs (FIN stands for "FFI Index Number").  The gradient across the SN background causes large systematic errors in FIN~127712, and we removed epochs such as these from the light curves with three rounds of 5$\sigma$ clipping on the local background (the red light curve in the bottom right panel).  The shaded regions show the time windows removed by this procedure.}
   \label{fig:bkg_clip}
\end{figure*}

\begin{figure*}
    \centering
    \includegraphics[width=\textwidth]{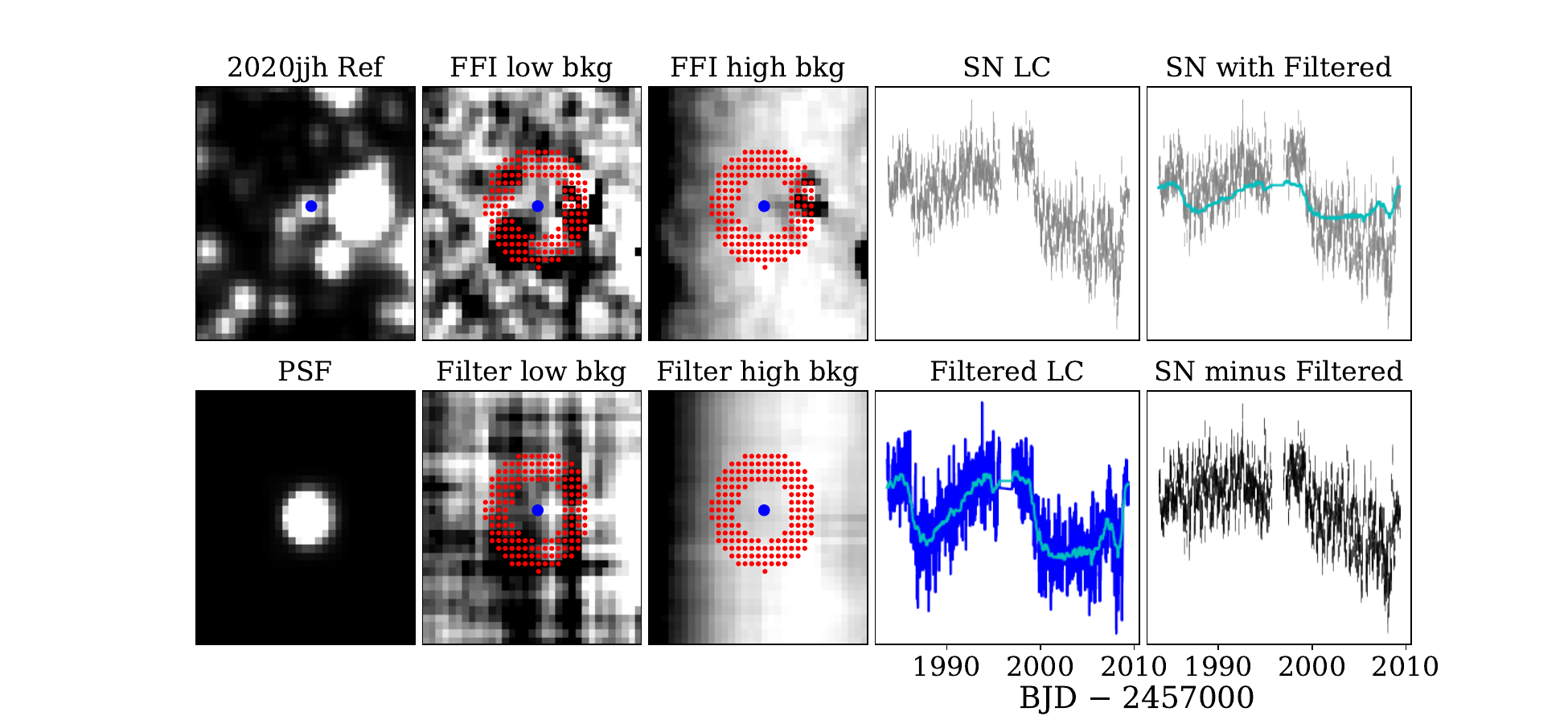}
    \caption{An example of our "background correction" procedure described in \S\ref{sec:systematics}.  The left column shows a 33x33 pixel zoom-in around SN2020jjh in the reference image of Sector~25 (top) and a model of the PSF used for photometry (bottom).  The position of SN2020jjh is marked by the blue dot.  The second column shows an FFI difference image with a low background (top) and the same image after filtering with a 100 pixel median along rows and columns (bottom).   The red dots show the annulus in which we calculated a local background correction. The images in the third column are the same as for the second column, but for an FFI difference image with a high background level.  The fourth column shows the light curve for PSF photometry at the location of the SN in the FFI difference images (top panel) and filtered images (bottom panel).  After removing the SN and star residuals in the difference images with the median filter, the blue light curve measures residual errors in the photometric aperture.    The cyan line shows a smoothed version of the blue light curve (to reduce statistical noise).  We refer to this smoothed light curve as the "background model," which in some cases identifies residual systematic errors in our light curves.  In the right column, the top panel shows the original SN light curve  with the background model light curve (the grey and cyan data are identical to the data in the fourth column).    The bottom-right panel shows the original SN light curve minus the  "background model" light curve (gray points minus cyan line), which better captures the shape of SN2020jjh around peak. }
   \label{fig:bkg_mod}
\end{figure*}

 \begin{figure*}
    \centering
	\begin{tabular}{c}	
	     \includegraphics[width=\textwidth]{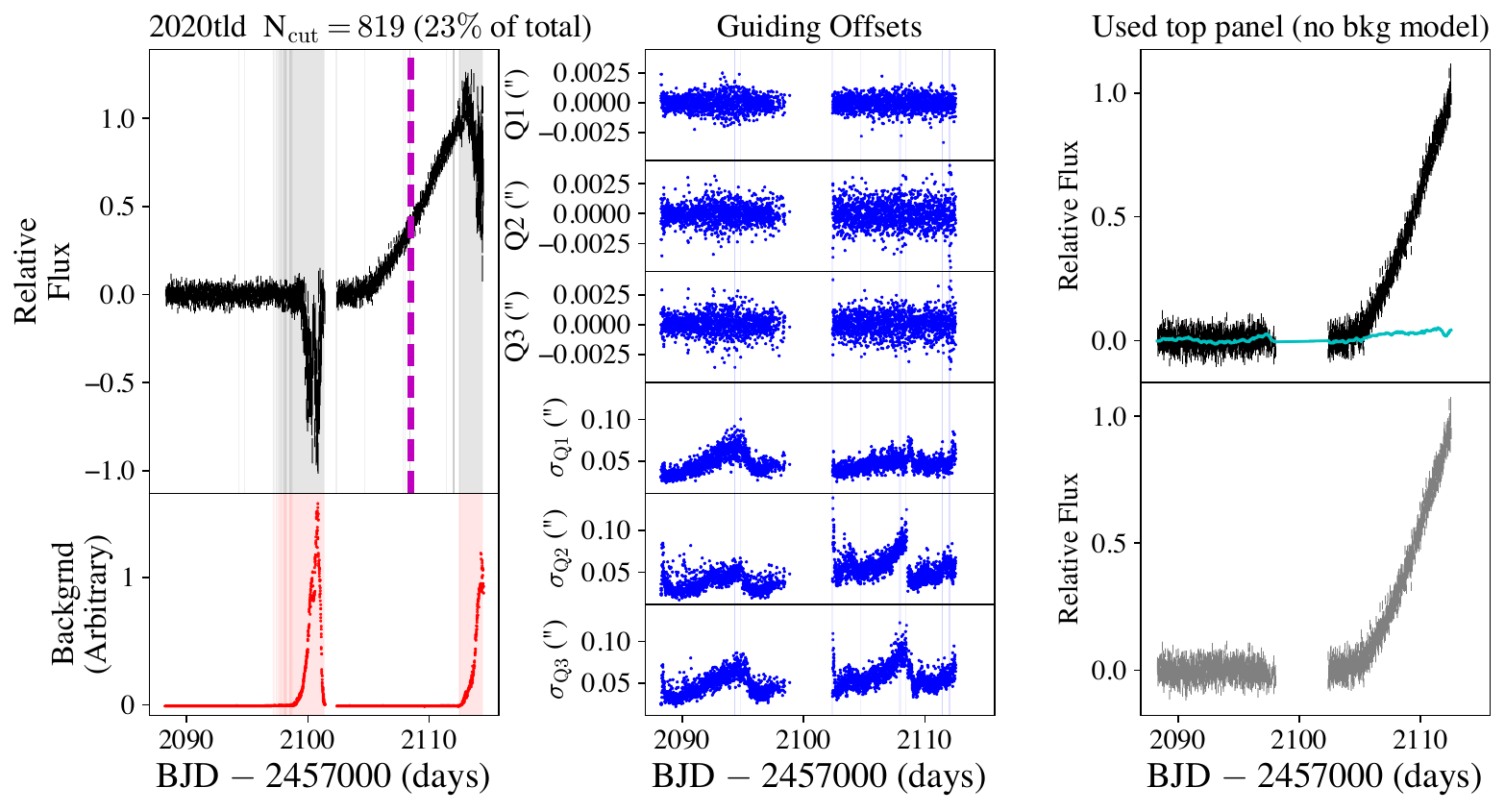}  \\
	     \includegraphics[width=\textwidth]{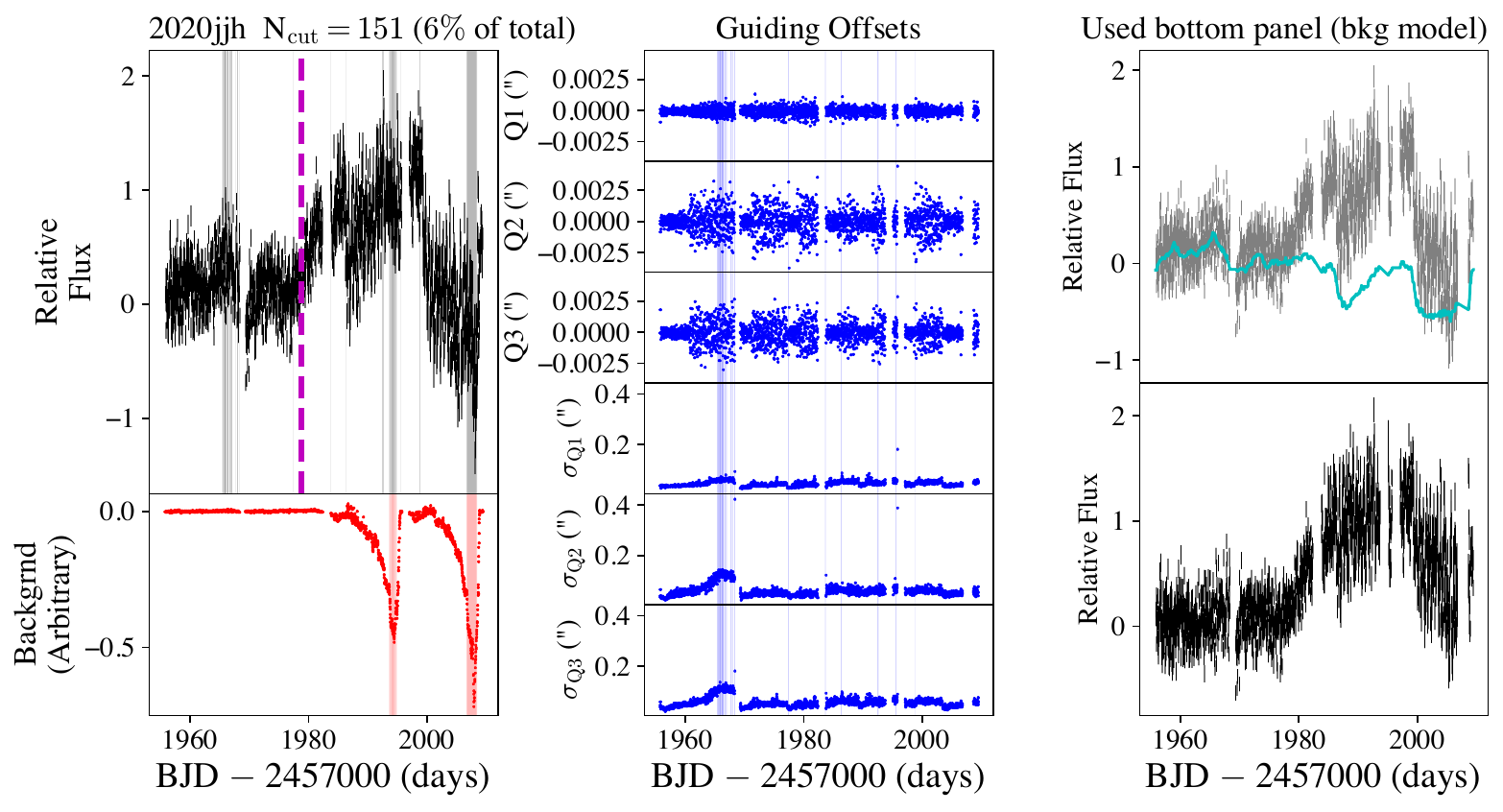}  \\
	    \end{tabular}

    \caption{Two examples of raw light curves, data removed based on background estimates and guiding offsets, and corrections from the "background model" procedure described in \S\ref{sec:systematics}.  The top left panels shows the SN flux in units relative to peak.  The vertical magenta line shows the time of discovery from TNS, and the shaded regions shows time ranges removed from $\sigma$-clipping against the local background estimates and guiding offsets (see \S\ref{sec:systematics}).  The bottom left panel shows the local background estimates in red, and the shaded regions show the times removed by $\sigma$-clipping.  The middle columns shows the average 3-dimensional \tess\ guiding system offsets (top three panels) and their standard deviations over FFI exposures (bottom three panels).  The shaded  regions again show the times removed by $\sigma$-clipping.  The right panel shows the "background model" light curve in cyan, in the same units as the SN flux light curve (black or grey points).  The top panel shows the SN light curve without the "background model" correction, and the bottom panel shows the SN light curve with the background model correction.  The final light curve that we used for analysis is highlighted in black in the rightmost column; for SN2020tld, we did not apply the background correction, but for SN2020jjh we did apply the background correction.  The complete figure set is available in the online journal article.}
   \label{fig:SN_lc}
\end{figure*}
\begin{figure*}
    \centering
    \includegraphics[width=0.33\textwidth]{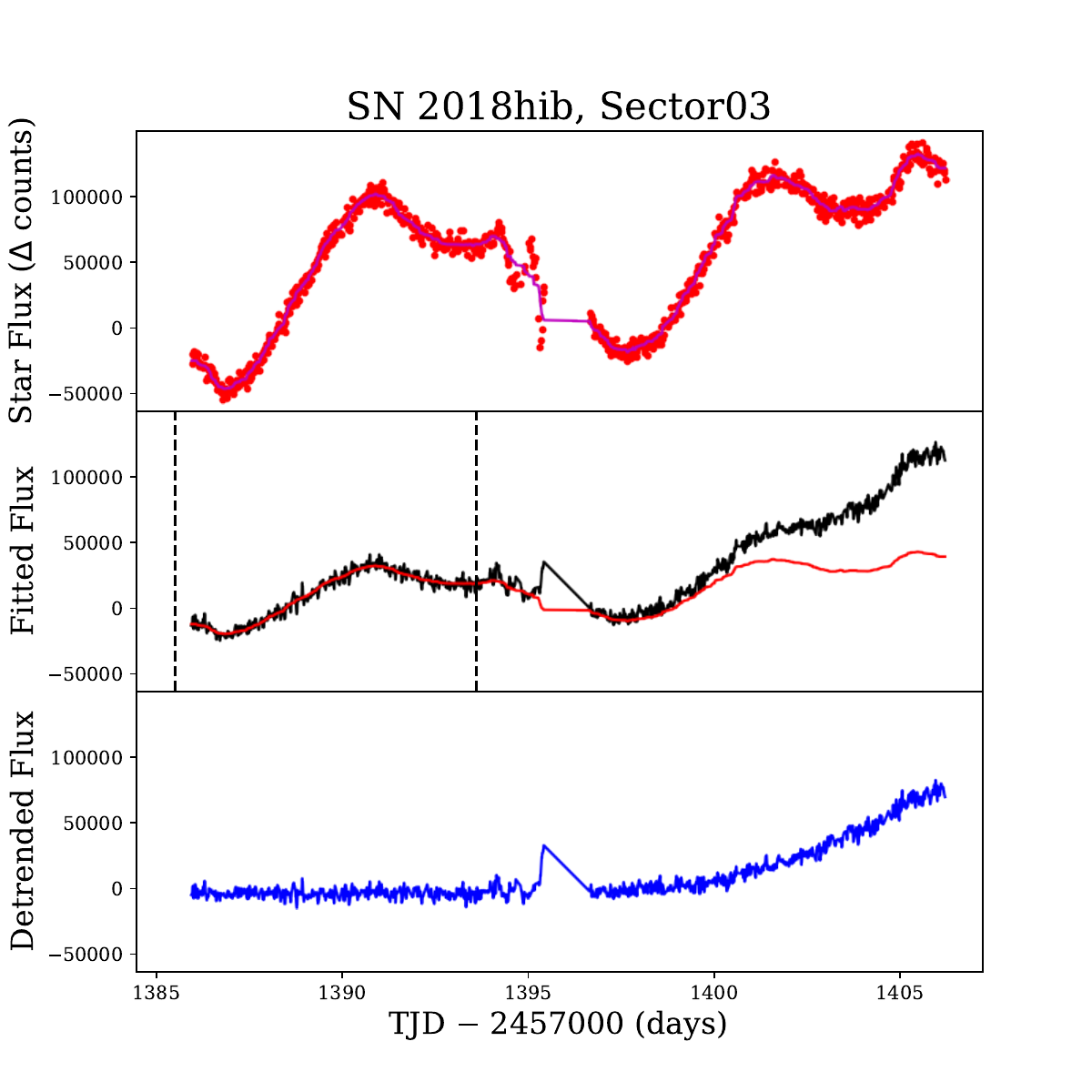}
    \includegraphics[width=0.33\textwidth]{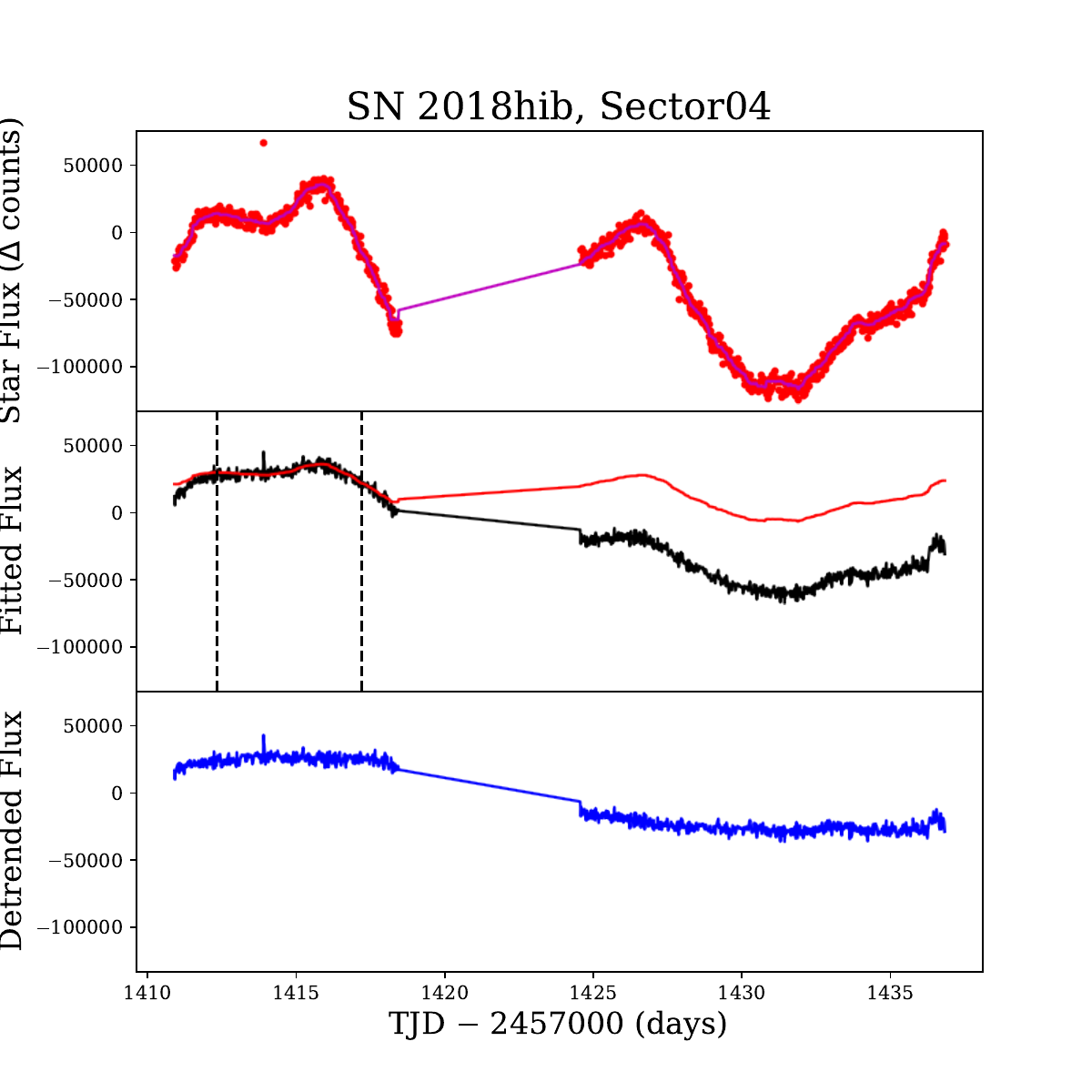}
    \includegraphics[width=0.33\textwidth]{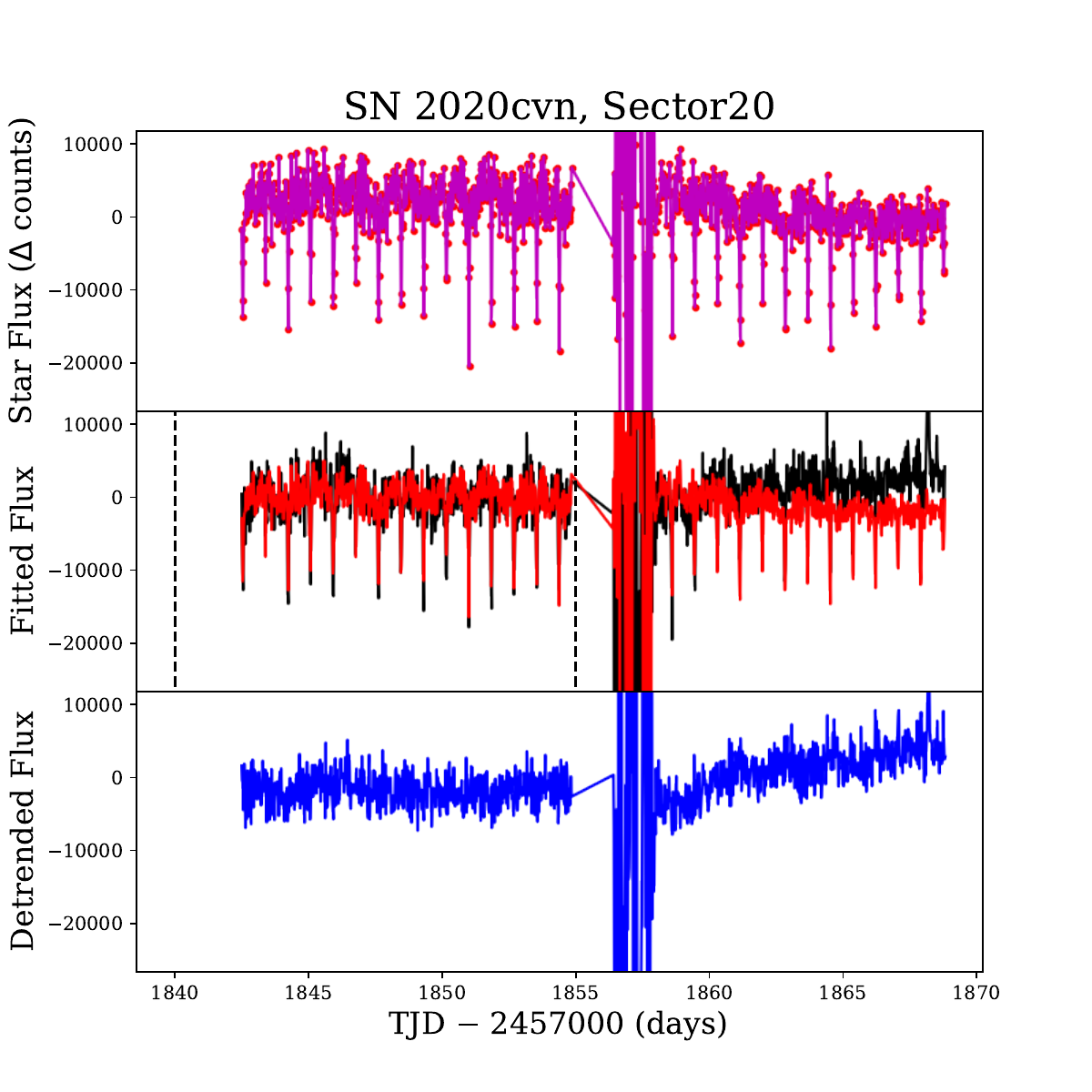}
    \caption{Examples of our procedure for removing variable star signals from the photometric aperture, as described in \S\ref{sec:systematics}.  The top panels show the \texttt{ISIS} light curves of the variable star in red, and the 1-day smoothed light curves in magenta.  The middle panels show the raw SN light curves in black and the smoothed variable star light curves that are fit to the data in red.  The bottom panels show the results of subtracting the fitted variable star light curves from the raw SN light curves.  The complete figure set is available in the online journal article.}
   \label{fig:deblend}
\end{figure*}

\subsection{Systematic Errors\label{sec:systematics}}

Although the remaining light curves show clear SN signals, residual systematic errors are still present.  The primary issue is time-variable scattered light from the Earth and Moon when they are above the \tess\ sunshade.  In the worst cases, the sky background is brighter than 14th \tess\ magnitude per pixel, while the early time SN light curves are 19--21st magnitude.  When the background is this high, the background corrections must have a relative uncertainty of $\leq$\,0.1\% to  isolate the SN signal.  In practice, we have found that strong glints with high-frequency spatial features are the most difficult systematic error to remove.  An example of a strong scattered light signal is shown in Figure~\ref{fig:bkg_clip} for SN2020tld.  

We addressed these issues with two methods. First, we removed data contaminated by the strongest scattered light signals.  We identified these epochs by performing three rounds of iterative 5$\sigma$ clipping  (an example is shown in Figure~\ref{fig:bkg_clip}).  On average, $\sigma$-clipping removes about 9\% of the data in the light curve, though the exact number ranges between 0.5\% and 52\% depending on the specific \tess\ observing sector.  Second, we estimated residual background errors in the photometric aperture by filtering the difference images to remove point sources such as the SN and variable stars.  We used a median filter with a width of 100 pixels, and we applied the filter to the difference images first column-by-column and then row-by-row.  We then produced a "background model" light curve by rerunning forced photometry on the filtered images,  and we smoothed the result over 1 day to reduce the statistical noise.  An example of this process is shown in Figure~\ref{fig:bkg_mod}. If the smoothed background model light curve was well-correlated with the systematic errors in the SN light curve, we subtracted the background model light curve from the original SN light curve.  We flagged SNe for which we applied the background model correction in Table~\ref{tab:physical_data}. For SN2020lsj, the background varies on a faster timescale and we construct the background model light curve by smoothing over 11 consecutive data points instead of 48.

We also removed data taken during periods of  large pointing jitter.  We used three rounds of 5$\sigma$ clipping, based on the average spacecraft guiding offsets and their standard deviations over the duration of each FFI exposure.   Figure~\ref{fig:SN_lc} shows the epochs removed based on backgrounds and spacecraft jitter for each SNe, as well as the background model light curves and corrections.

The final systematic error that frequently affects \tess\ SNe light curves are blended variable stars and asteroids.  For bright stars up to 5 pixels away, the flux from the wings of the PSF may dominate over the SN signal.  In these cases, we performed PSF photometry on the nearby star, smoothed the resulting light curve, and correlated the smoothed light curve with the SN light curve.  If there is a good match between the smoothed variable star light curve and the SN light curve, we rescaled the variable star light curve to match the signal in the SN light curve and subtracted the variable star signal.  Table~\ref{tab:physical_data} marks the 23 sources for which we removed variable star signals, and we show all blended variable star corrections in  Figure~\ref{fig:deblend}.  There are 3 SN (SN2020abdk, SN2021bzm, and SN2021abzn) that show a high-frequency variation in the light curve superimposed on the slowly evolving SN signal.  We could not isolate the variable star signal in the \tess\ FFIs for these light curves, suggesting that the high frequency signal is caused by a blended background star in the photometric aperture.  Because of the differences in time scale, it is still possible to extract meaningful parameters for the SNe and so we include these sources in our analysis in \S\ref{sec:analysis}.  These SN are marked in Table~\ref{tab:physical_data} with the same flag as the other stars with variable star signals that we detrended.  For asteroids, we identified 30 affected light curves based on visual inspection of the light curves and difference images (which clearly show a moving source associated with the asteroid).  Light curves affected by asteroids are flagged in Table~\ref{tab:physical_data}.  Similar to the sources with high frequency background variable star signals, the asteroid signals take place over a short time and are easy to separate from the SN signal.  We therefore include these sources in our analysis in \S\ref{sec:analysis}.

 \begin{figure}
    \centering
    \includegraphics[width=0.75\textwidth]{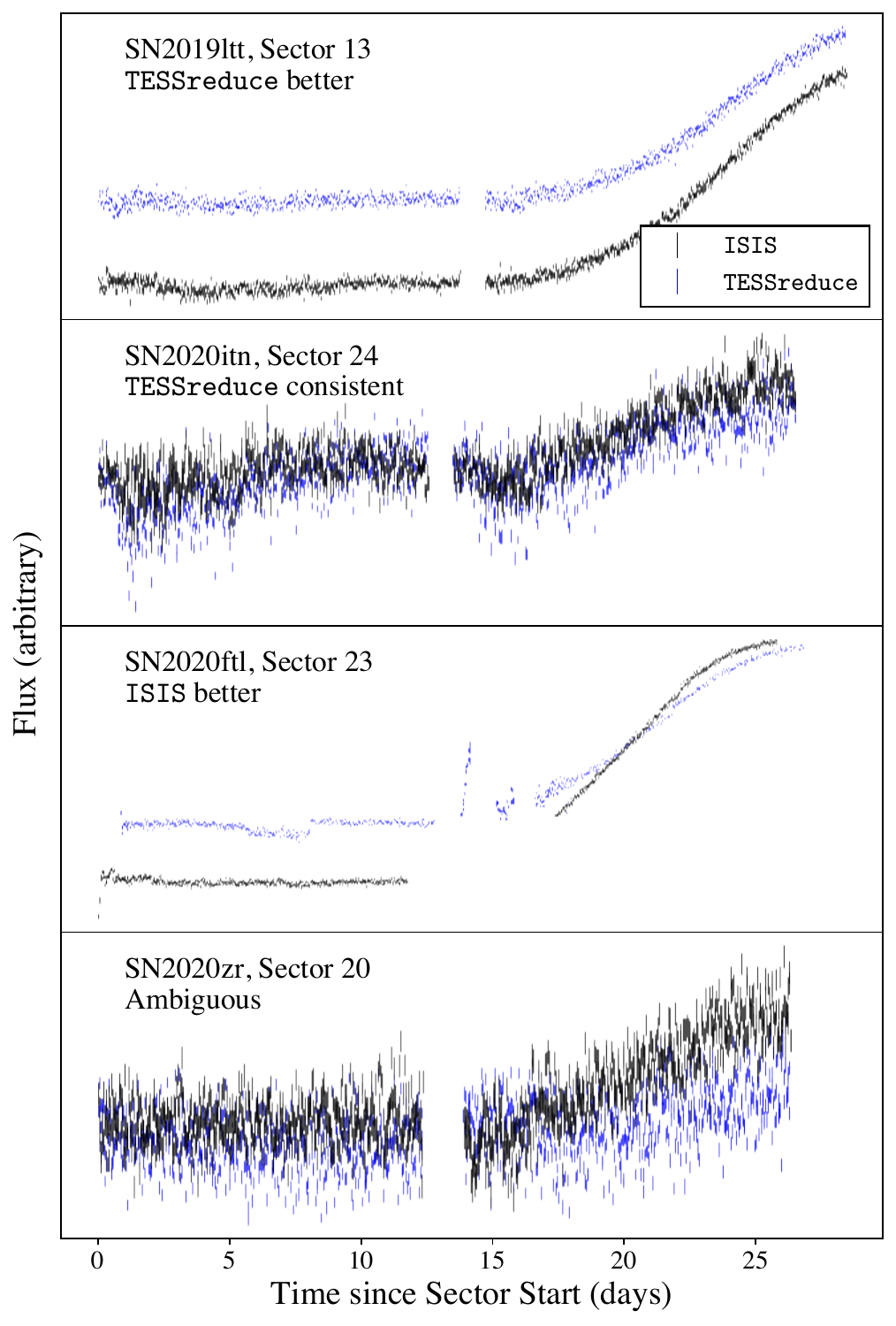}
    \caption{Comparison of \texttt{TESSreduce} light curves (blue)  with our difference imaging \texttt{ISIS} light curves (black).  Both light curves are in consistent units but offset for clarity.  The top panel shows an example of a SN with a better light curve from \texttt{TESSreduce} than from \texttt{ISIS}; there are 20 such sources, and we use the \texttt{TESSreduce} light curves in these cases.  The other panels shows cases where the \texttt{TESSreduce} and \texttt{ISIS} light curves are consistent, the \texttt{ISIS} light curve is better, or both light curves are of comparable quality but have different shapes and amplitudes.  See \S\ref{sec:tessred} for details.}
   \label{fig:TESSred}
\end{figure}

\subsection{Comparison with TESSreduce\label{sec:tessred}}

Our sample of SN light curves have a range of quality.  Frequently, there are systematic errors in the light curves caused by imperfect removal of scattered light, blended stars, or ambiguities caused by quickly varying backgrounds that slip through the $\sigma$-clipping procedure.  To investigate if we could improve the light curves, we flagged sources with a clear SN signal that were affected by strong systematic errors, and extracted light curves using \texttt{TESSreduce} \citep{Ridden-Harper2021}.  \texttt{TESSreduce} performs difference imaging in a different way than \texttt{ISIS}, first by iteratively modeling the backgrounds and interpolating the images to take out pointing shifts, and then by subtracting the shifted images from a reference image and performing aperture photometry at the location of the SN.  We ran \texttt{TESSreduce} on 103 sources that were subject to the very strong systematic errors, such that analysis of the SNe signal was especially difficult.  \texttt{TESSreduce} failed to produce light curves for 15 of these sources.  For 20 sources, \texttt{TESSreduce} produced a better light curve than \texttt{ISIS}.  In these cases, we adopted the \texttt{TESSreduce} light curves for our analysis, and these sources are marked in Table~\ref{tab:physical_data}.  For 32 sources, the \texttt{TESSreduce} light curves were consistent with the \texttt{ISIS} light curves.  For 29 sources, the \texttt{ISIS} light curves had fewer systematic errors.  The remaining 7 light curves were ambiguous, in that both the \texttt{TESSreduce} and \texttt{ISIS} light curves were of comparable quality, but showed differing shapes and amplitudes for the SN signal.  In these cases, we use the difference imagining light curves for consistency, although we also mark these sources in Table~\ref{tab:physical_data}.  Figure~\ref{fig:TESSred} shows examples of light curves from \texttt{TESSreduce} and \texttt{ISIS}.  However, a full comparison between the two pipelines is beyond the scope of this contribution.
 \clearpage
 
\section{Fitting Procedure and Model Comparison\label{sec:fits}}

\begin{figure}
    \centering
	    \includegraphics[width=0.45\textwidth]{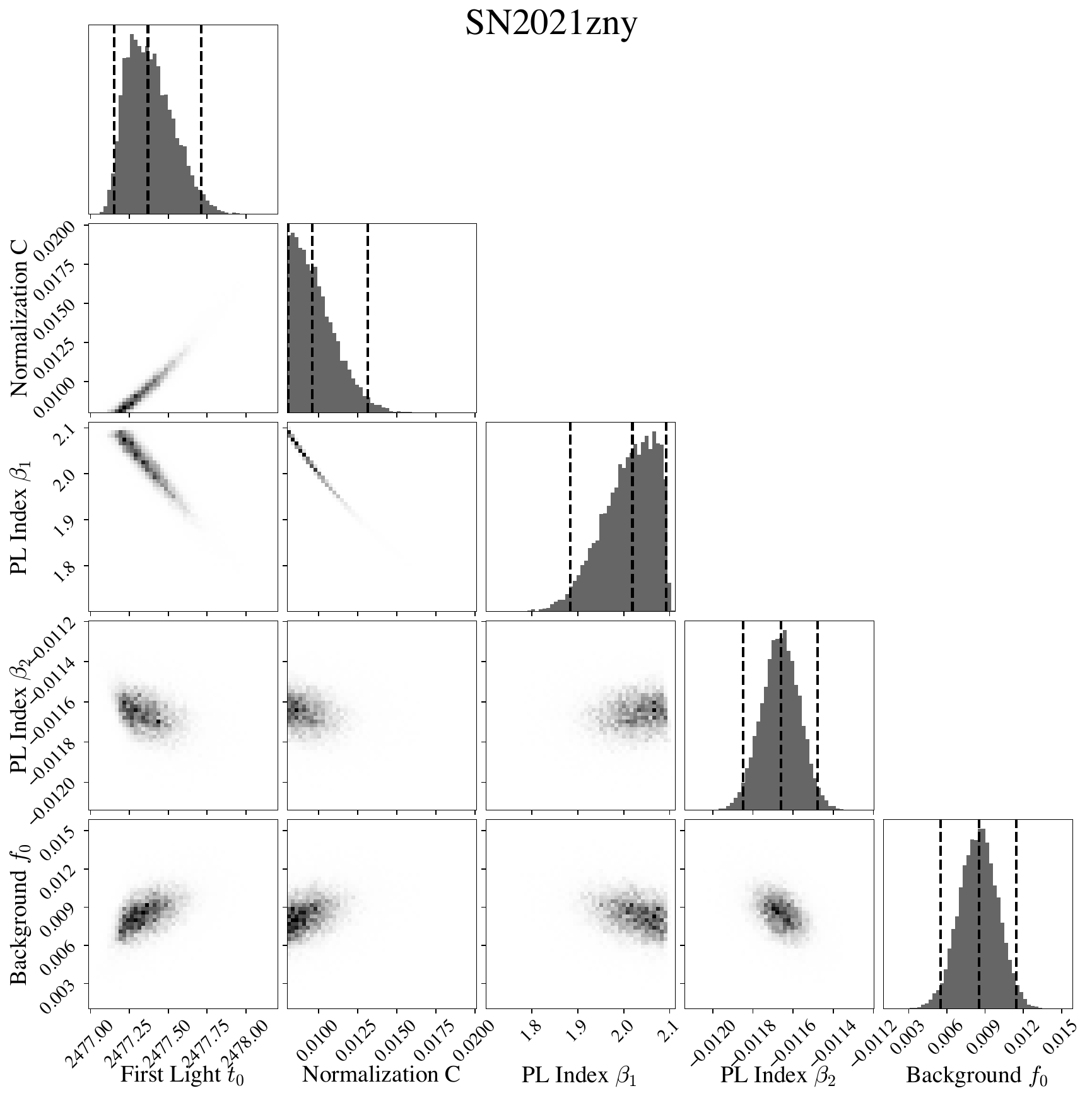} 
	 \includegraphics[width=0.45\textwidth]{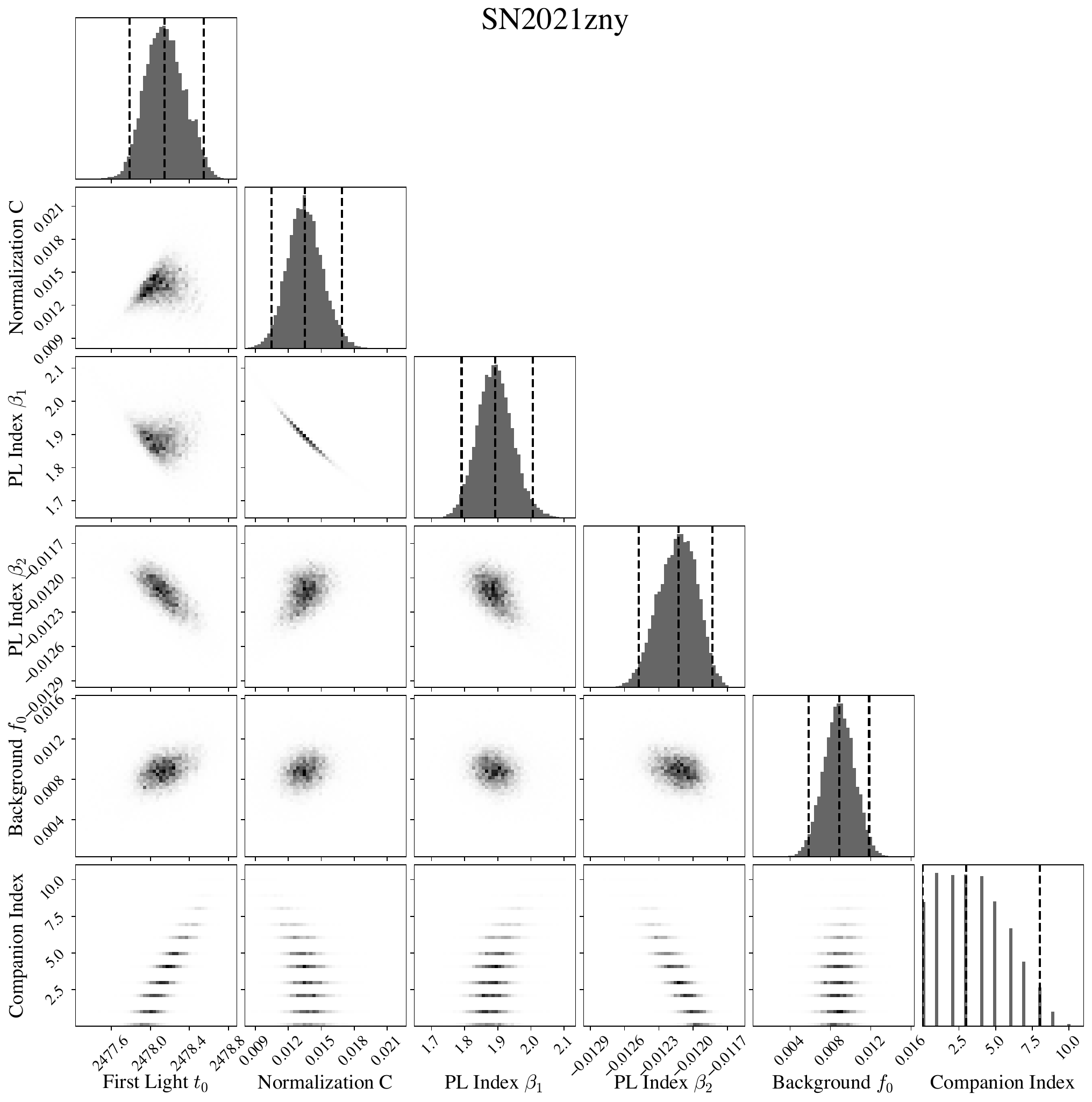} 
    \caption{Example of the posterior parameter distributions for SN2021zny.  The vertical lines in the 1D histograms mark the medians and 95\% credible regions of the distributions.  There are correlations between the time of first light $t_0$, power law index $\beta_1$, and normalization C, which we discuss in \S\ref{sec:results}.  The complete figure set is available in the online journal article.}
   \label{fig:cornerplot}
\end{figure}

We sampled the posterior distributions of the model parameters and estimated the Bayesian evidence of each model using \texttt{dynesty}.  \texttt{dynesty} uses Dynamic Nested Sampling to explore the posterior parameter distributions, identifying contours of constant likelihood by sampling from nested shells of the posterior distributions with monotonically increasing likelihood.  Along the way, Dynamic Nested Sampling estimates the Bayesian evidence by numerically integrating the posterior distribution.  Regions of very high likelihood may correspond to a very narrow region of parameter space and so contribute little to the Bayesian evidence, while regions of lower likelihood that encompass a large portion of the prior parameter space can dominate the Bayesian evidence.  Model comparisons based on the Bayesian evidence use these volumes to penalize additional parameters, which is the main difference from model comparison tests based on maximum likelihood ratios or varieties of Bayesian information criteria.  Besides estimating the Bayesian evidence, Dynamic Nested Sampling carefully explores the parameter space defined by the prior parameter distributions, identifying the global maximum of the likelihood function and any important modes of the posterior distributions.  Figure~\ref{fig:cornerplot} shows the posterior parameter distributions for each SN.

We used uniform priors for the model parameters except for the normalization $C$  and the companion separation $a$. For $C$, we used a uniform prior in $\log C$ because this is a scaling parameter, while for $a$ our logarithmically spaced grid corresponds to a uniform prior on $\log a$.  Table~\ref{tab:priordists} gives the ranges of these prior parameter distributions.  We used a Gaussian likelihood function for the Dynamic Nested Sampling.  We fit the light curves at the original cadence (either 30 minutes or 10 minutes), and we fit from the start of the \tess\ light curve up to $t_{\rm peak} + 0.5$ days, where $t_{\rm peak}$ is given in Table~\ref{tab:physical_data} (see \S\ref{sec:sample}).

For our model comparison tests, we define the Bayes factor $K$ as the ratio of the Bayesian evidence $Z$ for the model with companion signatures to that of the model with no companion signatures
\begin{align}
K = \frac{Z_1}{Z_2} = \frac{ \int \mathcal{L}(\theta_1)\pi(\theta_1)\,d\theta_1}{\int \mathcal{L}(\theta_2)\pi(\theta_2)\,d\theta_2}
\end{align}
where $\mathcal{L}$ is the likelihood,  $\pi$ is the prior distribution, $\theta_1$ is the vector of model parameters with a companion signature and $\theta_2$ is the vector of model parameters without a companion signature.  The $\pi(\theta)$ are normalized probability distributions, so large prior volumes tend to decrease the product of $\mathcal{L}(\theta)\pi(\theta)$.  This definition of the Bayes factor $K$ therefore penalizes models with larger prior volumes, due to either a large number of parameters or broad priors.  

We also compared the Bayesian Information Criteria (BIC) of the two models
\begin{align}
{\rm BIC} = k\ln N  - \ln \mathcal{L}
\end{align}
where $k$ is the number of parameters, $N$ is the number of data points in the fit, and $\mathcal{L}$ is the likelihood.  The BIC explicitly penalizes models with additional parameters, which may be an advantage over the Bayes factor if the change in prior volume introduced by extra parameters is relatively small.  If the penalty for additional parameters is relatively small, then modest improvements in the likelihood will appear significant, which might be the case for companion  interaction models with small amplitudes that cannot, by definition, substantially change the fit.  The BIC test is therefore a useful check on how strongly the data prefer a companion interaction model, especially for very low amplitude signals.

\begin{deluxetable}{lrr} 
\tablewidth{0pt}
\tablecaption{Ranges of Parameter Priors.\label{tab:priordists}}
\tablehead{\colhead{Parameter}&\colhead{Symbol} & \colhead{Range}}
\startdata
Time of first light & $t_0$   & Time range of first sector\\
Normalization at 1 day &$C$      & 0.008 -- 1.0\\
Early-time rising index & $\beta_1$   &0 -- 4.0 \\
Late-time curving index & $\beta_2$  &$-$0.1 -- 0  \\
Residual baseline flux & $f_0$ & $-$0.2 -- 0.2\\
Companion Separation & $a$ & (0.02 -- 4)$\times 10^{13}$ cm\\
Calibration offset & $c$ & $-$1.0 -- 1.0\\
\enddata
\tablecomments{Uniform priors were assumed for all prior distributions except for the normalization $C$ and companion separation $a$. The normalization prior was uniform in $\log C$, and restricted at the low end to capture our highest dynamic range light curves (just over 5 magnitudes). Companion separations were drawn from a grid of 20 models evenly spaced in the logarithm of the separation, which corresponds to a uniform prior on $\log a$.  Fluxes are in units relative to peak.}
\end{deluxetable}

\begin{figure}
    \centering
    \includegraphics[width=0.46\textwidth]{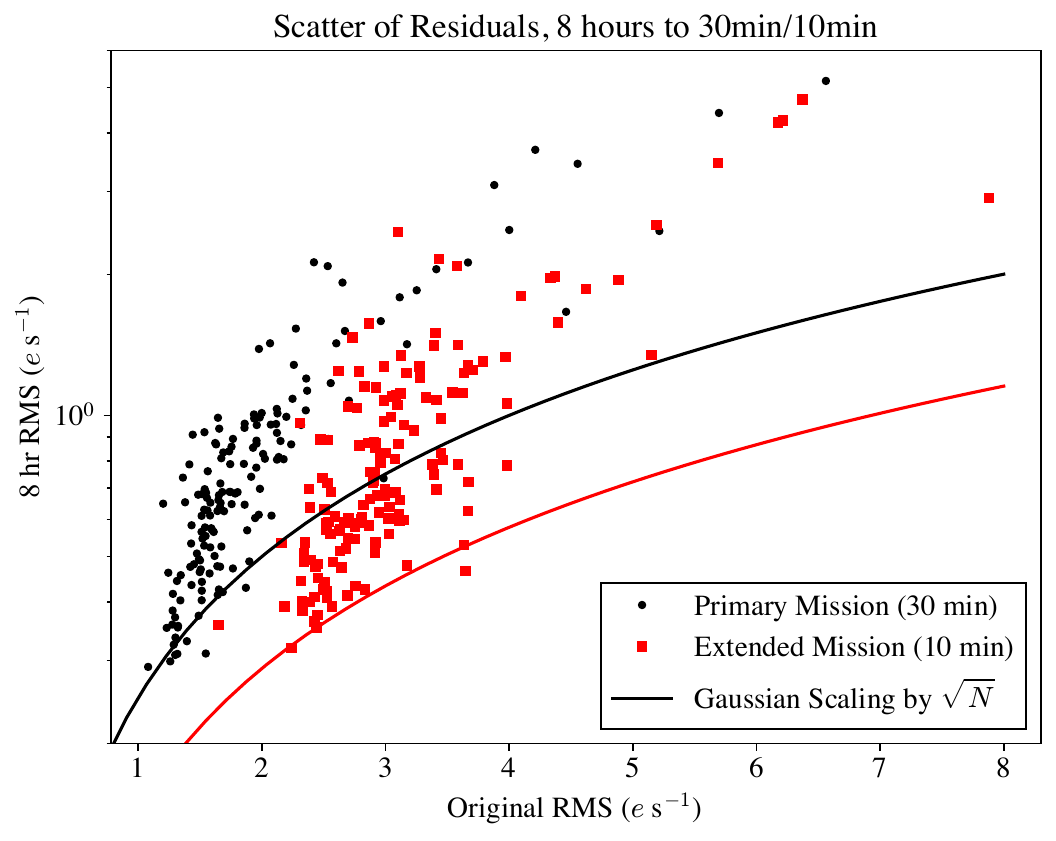}
    \includegraphics[width=0.53\textwidth]{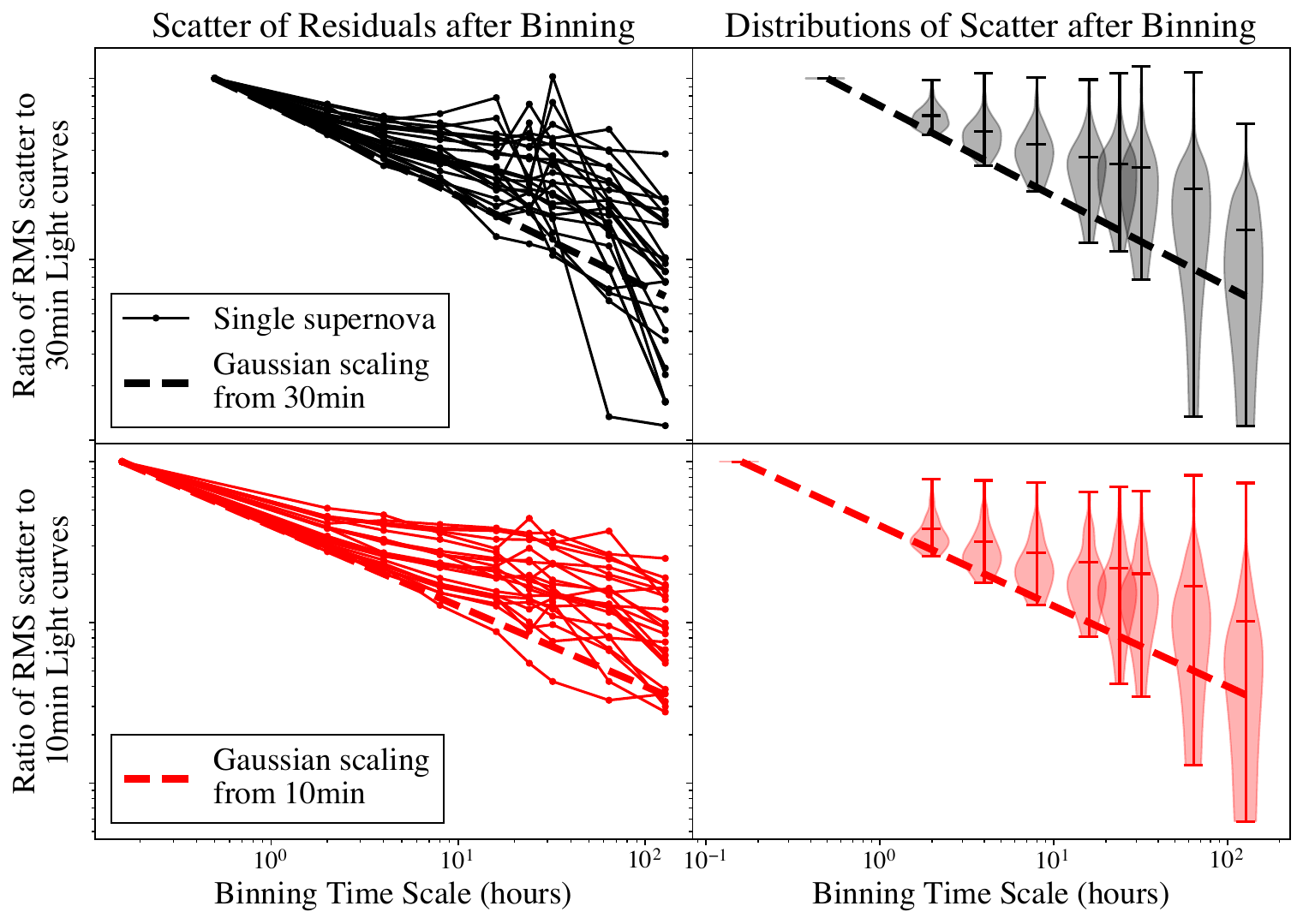}
    \caption{Analysis of residual systematic errors after fitting a curved power law model.  As an example, the left panel shows the root-mean-square (RMS) scatter of the residuals on 30 or 10 minute time scales and on 8 hour time scales.  For Gaussian noise, the ratio of the RMS scatter on 30 or 10 minute time scales and 8 hour time scales should be $\sqrt{16}$ or $\sqrt{48}$, respectively.  These ratios are shown with the solid lines.  The middle panel shows the  RMS scatter scaled to the observed 30 or 10 minutes scatter for supernova light curve residuals after binning to 2, 4, 8, 16, 24, 32, 64 and 128 hours (the solid lines connect points for individual supernovae---only 50 random supernovae are shown for clarity). The thick line shows the expected trend for Gaussian scaling.  We calculate the ratio of the Gaussian expectation to the  observed RMS scatter at each time scale, and take the maximum across timescales for an individual supernova as a metric for the quality of the light curve. The right panel shows the distributions of the scaled RMS scatter at each timescale (the distributions do not show how individual supernova light curve residuals vary across time scales).  The tails of the distribution below the expectation for Gaussian noise are due to sampling error, as there are less than 20, 10, and 5 points after binning to 32, 64, and 128 hours.}
   \label{fig:noise}
\end{figure}

\section{Subsamples\label{sec:subsamples}}
There are four subsamples of TESS SNe light curves.  The first  subsample consists of cases where \tess\  did not observe the peak flux, and so we cannot estimate the value of $t_{\rm rise}$.    The second subsample consists of cases where fits for the time of first light  $t_0$ place it in data gaps, introducing uncertainty into the early time light curve shapes.  We refer to this subset as the "No First Light" subsample.  The last two subsets are the "High Quality" subsample and the "High Dynamic Range" subsample, which we discuss in detail here.  SNe from each subsamples are marked in Tables~\ref{tab:physical_data}, \ref{tab:fit_results_no_comp}, and \ref{tab:fit_results}.

\subsection{Residual Noise Properties and High Quality Subsample\label{sec:noise}}

We inspected the residuals of each SN fit to investigate the quality of the fits.  We found that the residuals are often correlated in time, typically on time scales of 8 hours or shorter.  Correlated residuals suggest the presence of residual systematic errors in some of our light curves.  The correlated noise is not stationary in time and appears at a variety of amplitudes and on different time scales---the  shapes and amplitudes depend primarily on the variable scattered light signals from the Earth and Moon that are specific to each \tess\ observing sector.

Correlated noise can introduce ambiguity into the shape of the early time light curves derived from our fits.  To quantify the impact of correlated noise on our results, we calculated the RMS scatter of the residuals for the original cadence at 30 minutes or 10 minutes and after binning to 2, 4, 8, 16, 32, 64, and 128 hours  (we include 24 hours because we are specifically worried about scattered light signals from the Earth).  In the limit of random Gaussian noise, the RMS scatter of the residuals should decrease by a factor of $\sqrt{N}$, where $N$ is the number of FFIs taken within a given time interval.   If correlated noise is present, the RMS scatter of the residuals binned to 8 hours will decrease more slowly than for the Gaussian limit, indicating the presence of residual systematic errors.

As an example, the left panel of Figure~\ref{fig:noise} shows the 8 hour residual RMS scatter as a function of the original 30 or 10 minute residual RMS scatter, with the theoretical ratios for Gaussian noise are shown with the solid lines.  The RMS scatter spans a wide range of values, but a subset of SN cluster near the theoretical Gaussian limits.  The middle panel shows a generalization of this investigation to multiple timescales, by scaling the observed RMS scatter at a given timescale to that observed at 30 or 10 minutes.  Averaging over longer and longer timescales reduces the random component of the noise, until systematic errors start to limit the observed scatter. The right panel shows the distributions of the scaled RMS scatter, to give an sense of the population as a whole relative to the Gaussian limit. Note that on timescales of 32  hours (1.3 days), there are less than 20 points in the light curves and the estimate of the RMS scatter becomes noisy due to sampling error.  Sampling error likely explains why the RMS scatter of some light curves increase on longer timescales, or decreases to be well below the Gaussian limit.

We define a "noise metric" for each light curve as follows: First, we calculate the RMS scatter of the light curve residuals at each timescale, and then divide the observed RMS scatter by the RMS scatter at the native 30 or 10 minute samples.  This ratio RMS$_t$/RMS$_{\rm native} = {\rm rms}_t$ gives the rate at which random noise is averaging out of the light curves on longer and longer timescales.  We compare this observed scatter rms$_{t,{\rm observed}}$ to the Gaussian expectation rms$_{t,{\rm Gauss}} = N^{-1/2}$, where $N$ is the number of FFIs in a given time bin. We then divide the Gaussian scaled scatter rms$_{t,{\rm Gauss}}$ by the observed scaled scatter rms$_{t,{\rm observed}}$, and the noise metric is the minimum across timescales min(rms$_{t,{\rm Gauss}}/$rms$_{t,{\rm observed}}$) = rms$_{\rm Gauss}/$rms$_{\rm observed}$. The metric rms$_{\rm Gauss}/$rms$_{\rm observed}$ captures the largest departure of the SN light curve from Gaussian noise, and should be sensitive to systematic errors appearing on different timescales.  This metric also has an intuitive scaling from 0 to 1, where values near zero indicate very little improvement from binning (systematic errors dominate over random noise), while 1 indicates perfect Gaussian noise scaling (random noise only).

We then define a "High Quality" subsample as the SNe with noise metrics in the top quartile.  This threshold is at  rms$_{\rm Gauss}/$rms$_{\rm observed} \ge 0.49$, i.e., within a factor of about 2 of the Gaussian expectation.  There are 77 SNe in this subsample, which are the least affected by correlated noise. The "High Quality" SN sample is marked by a flag in Table~\ref{tab:physical_data} and Table~\ref{tab:fit_results}, and rms$_{\rm Gauss}/$rms$_{\rm observed}$ for each light curve is given in Table~\ref{tab:physical_data}. We give more details and show the distributions of the noise properties in Appendix~\ref{sec:noise}.

The correlated noise in the rest of the supernova sample is caused by residual systematic errors from complicated backgrounds that were not removed by  the $\sigma$-clipping or corrected by the background model procedure, or by blends with background  stars for which we could not isolate the variable star signal. However, the SN signal is clearly present in these light curves and so some information about the early rise of these SNe is still available. We show results based on both the full sample of SN light curves and results from the "High Quality" sample alone. The results based on the full sample are consistent with results from the "High Quality" sample.

\begin{figure*}
    \centering
    \includegraphics[width=\textwidth]{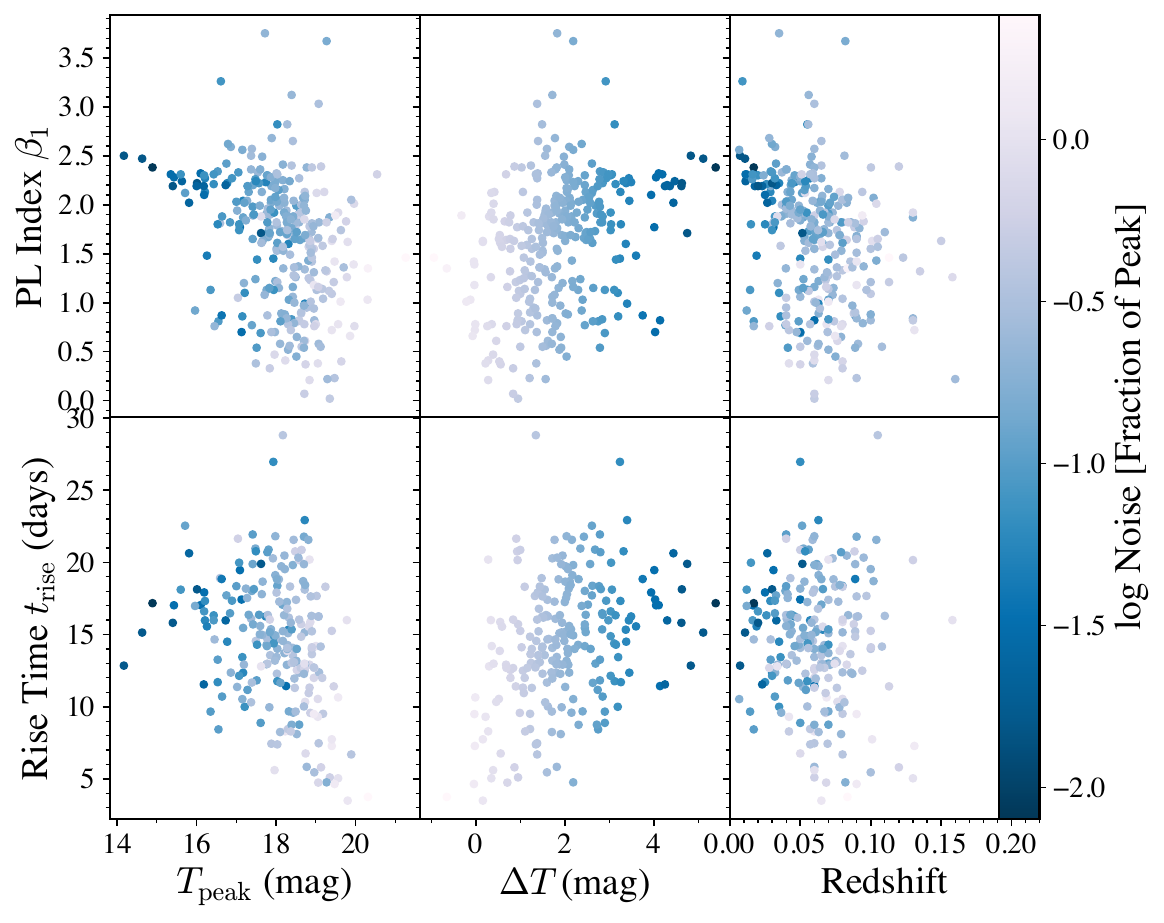}
    \caption{Correlations of power law index $\beta_1$ and rise time $t_{\rm rise}$ with peak magnitude $T_{\rm peak}$, dynamic range $\Delta T$, and redshift.  We only show the \NsupernovaHighDR\ SNe in the "High Dynamic Range" sample, and we only show values of $t_{\rm rise}$ for the \NsupernovaPeak\ SNe that  \tess\ observed the peak (see \S\ref{sec:sample} and Appendix~\ref{sec:subsamples}). The color bar gives the noise amplitude relative to peak, showing that light curves with large noise amplitudes tend to have lower values of $\beta_1$  and $t_{\rm rise}$.  This correlation is a measurement bias found by \citet{Miller2020b}, which we discuss in  Appendix~\ref{sec:obs_bias}.  }
   \label{fig:obscorr}
\end{figure*}

\begin{figure*}
    \centering	
        \includegraphics[width=\textwidth]{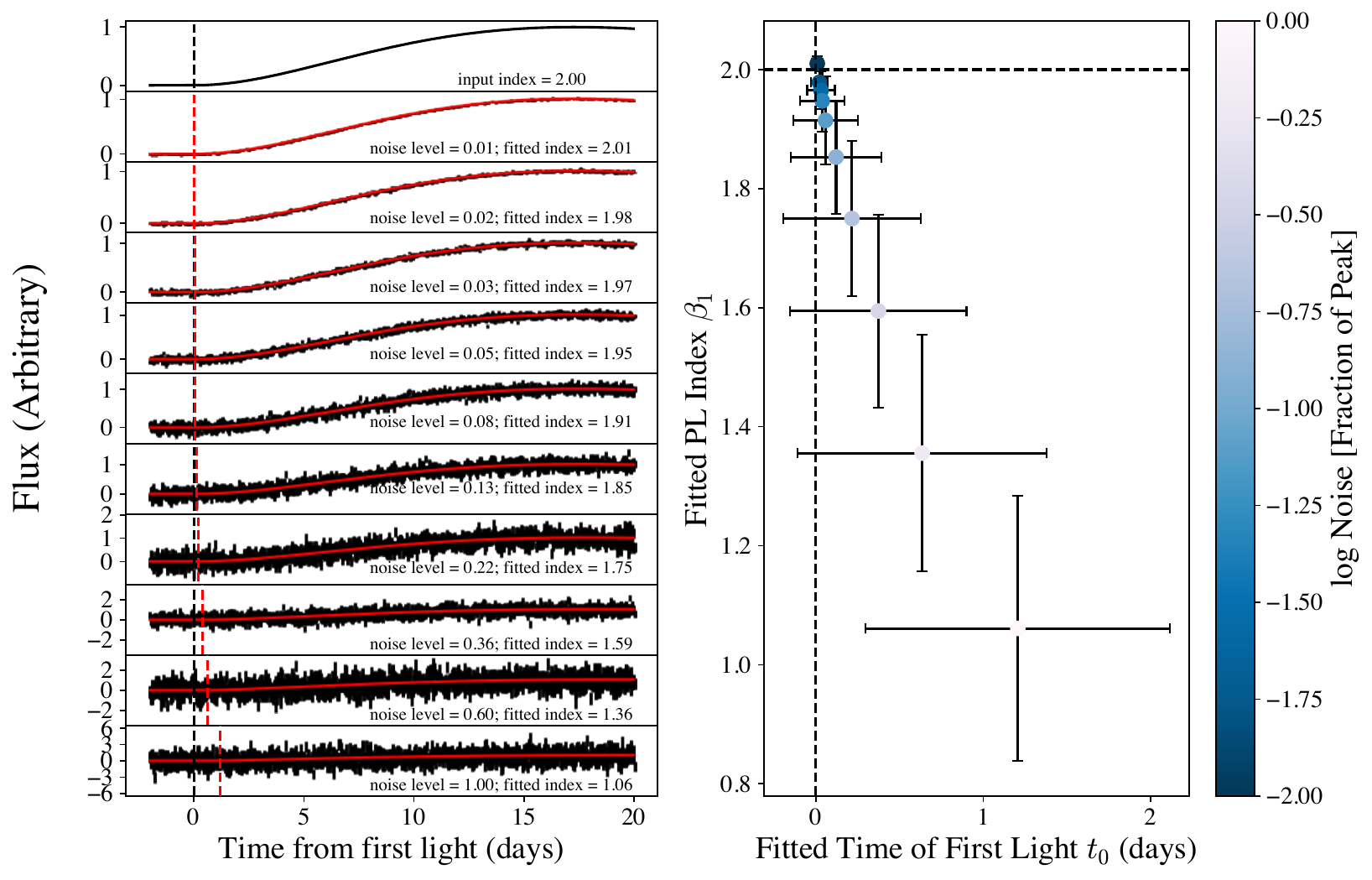}
    \caption{Simulations used to quantify the observational bias discussed in Appendix~\ref{sec:obs_bias}.  The left panels show simulated light curves with different levels of noise.  The input model is shown in the top panel, and has a time of first light $t_0 = 0.0$, power law index $\beta_1 = 2.0$, power law index $\beta_2 = -0.015$, and rise time $t_{\rm rise} = 17.25$ days.  The light curve is sampled at 30 minute cadence and scaled so that the peak flux is unity.  Different levels of noise are added in the lower panels, logarithmically spaced between 1\% and 100\% of the peak flux.  The maximum likelihood models from \texttt{dynesty} are shown with red lines, and the recovered value of $t_0$ is shown with the vertical dashed red lines.  The right panel shows the recovered values of power law index $\beta_1$ as a function of the recovered time of first light $t_0$, with the true values of the input model shown by dashed black lines.  The color coding gives the logarithm of the noise amplitudes from the left panel.  There is a  bias in the recovered parameters for light curves with the noise amplitude $\gtrsim$\,10\% of peak. }
   \label{fig:simdetectthresh}
\end{figure*}

\subsection{Observational Biases and High Dynamic Range Subsample\label{sec:obs_bias}}

\citet{Miller2020b} found an observational bias such that the rising power law index and rise time are correlated with redshift.  This bias is easy to understand as a result of signal-to-noise ratio:  the lower the signal-to-noise ratio, the later the SN will be detected during its initial rise.   A later detection results in a later value of $t_0$ and a bias in $t_{\rm rise}$  to smaller values. Furthermore, $t_0$ is correlated with the fitted power law index $\beta_1$, and so a later value of $t_0$ biases fits for $\beta_1$ to smaller values.

Figure~\ref{fig:obscorr} shows the same biases in power law index $\beta_1$ and rise time $t_{\rm rise}$  for our fits to the \tess\ SN light curves.  Brighter SNe at peak tend to have higher values of $\beta_1$ and smaller values of $t_{\rm rise}$.  Brighter SNe also tend to have larger values of $\Delta T$ and smaller redshifts, so $\beta_1$ is correlated with these observables as well.  We also checked for relationships between $\beta_2$ and the observed properties (peak magnitude $T_{\rm peak}$,  dynamic range $\Delta T_{\rm peak} $, and  redshift), but did not find any significant correlations.

The color bar in Figure~\ref{fig:obscorr} shows the 30 minute RMS scatter (calculated from the first 2 days of each light curve) in units relative to peak flux, which quantifies the signal-to-noise ratio of the light curve.  This noise amplitude relative to peak  is independent of the SN limiting magnitude $T_{\rm lim}$, peak magnitude $T_{\rm peak}$, and redshift, but is closely related to the dynamic range $\Delta T$.    SN with large noise amplitudes relative to peak have smaller power law index fits in each panel of Figure~\ref{fig:obscorr}.

Figure~\ref{fig:simdetectthresh} quantifies the biases in the time of first light $t_0$ and power law index $\beta_1$ by showing simulated light curves with different levels of noise relative to peak and the results of fits to these simulations.  The input model is a curved  power law, with $t_0 = 0$, $\beta_1 = 2.0$, and $\beta_2 = - 0.015$.  The resulting rise time for this model is 17.25 days.  The light curve is sampled at 30 minute cadence and renormalized so that the peak flux equals unity.  We then add Gaussian noise of different amplitudes relative to peak, logarithmically spaced between 1\% and 100\%, as indicated in the left hand panels of Figure~\ref{fig:simdetectthresh}.  Finally, we run \texttt{dynesty} on the simulated light curves and report the median values of the posterior distributions for $t_0$ and $\beta_1$.  The right hand panel shows the recovered value of $\beta_1$ as a function of the recovered value of $t_0$, where the input model had $\beta_1 = 2.0$ and $t_0 = 0$.  Sources with noise amplitudes $\gtrsim$\,10\% of peak have inferred values of $t_0$ and $\beta_1$ significantly biased from the true value.

We therefore restrict our sample to SNe with noise amplitudes $\leq$\,10\% of peak flux on 30 minute time scales, which we refer to as the "High Dynamic Range" subsample.  If \tess\ did not observe the peak of the light curve, we estimate the relative noise amplitude by assuming an absolute magnitude peak of $T_{\rm peak } = -19.5$, calculating the apparent magnitude based on the luminosity distance, and re-normalizing the observed noise floor to the new peak flux.  In total, \NsupernovaHighDR\ SNe have a noise floor $<$\,10\% of peak,  \NsupernovaFL\ of which have observations at the time of first light, \NsupernovaPeak\ of which have observations of the peak (and so valid estimates of $t_{\rm rise}$), and \NsupernovaHighDRHQ\ of which are in the "High Quality" sample (\S\ref{sec:noise}).  These SNe are flagged in Table~\ref{tab:fit_results}. 

The noise amplitude relative to peak is inversely related to the dynamic range $\Delta T$, because $\Delta T$ gives the ratio of the peak flux to the detection limit and the detection limit is based on the observed noise in the early light curve.  The 30 minute RMS scatter relative to peak is roughly related to our definition of $\Delta T$ by ${\rm RMS}_{\rm 30min} = 10^{-0.4\Delta T}\sqrt{16}/3$, where the factor of $\sqrt{16}$ converts from an 8 hour time scale to 30 minutes and the factor of 1/3 converts from a 3$\sigma$ limit to a 1$\sigma$ characteristic amplitude.  For the requirement of 10\% noise relative to peak, unbiased SN parameters can be obtained from \tess\ light curves with $\Delta T> 2.8$\, mag. 

\section{Candidates\label{sec:candidates}}
\begin{figure*}
    \centering
	\begin{tabular}{c}	
	    \includegraphics[width=0.70\textwidth]{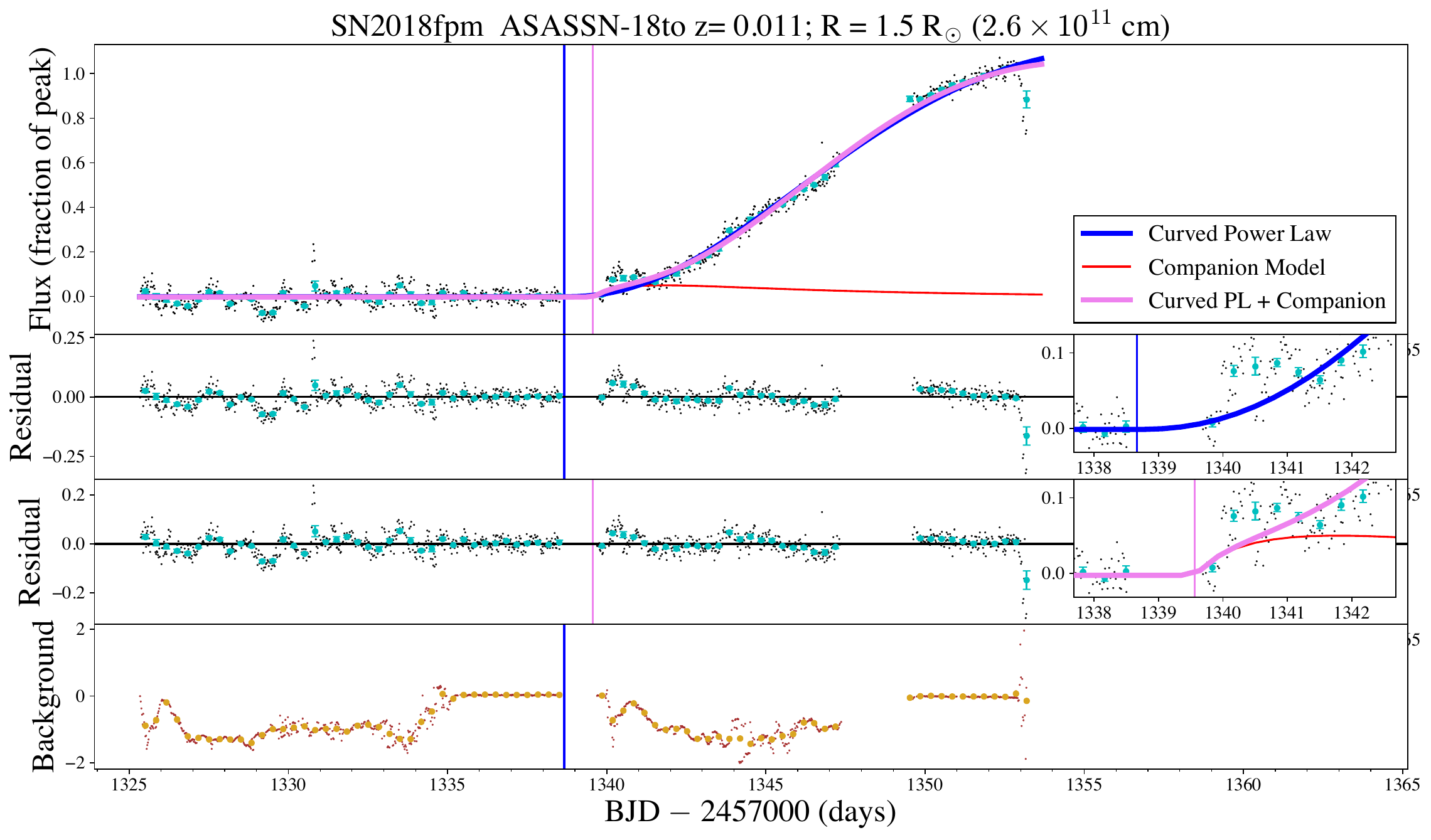}  \\
	    \includegraphics[width=0.70\textwidth]{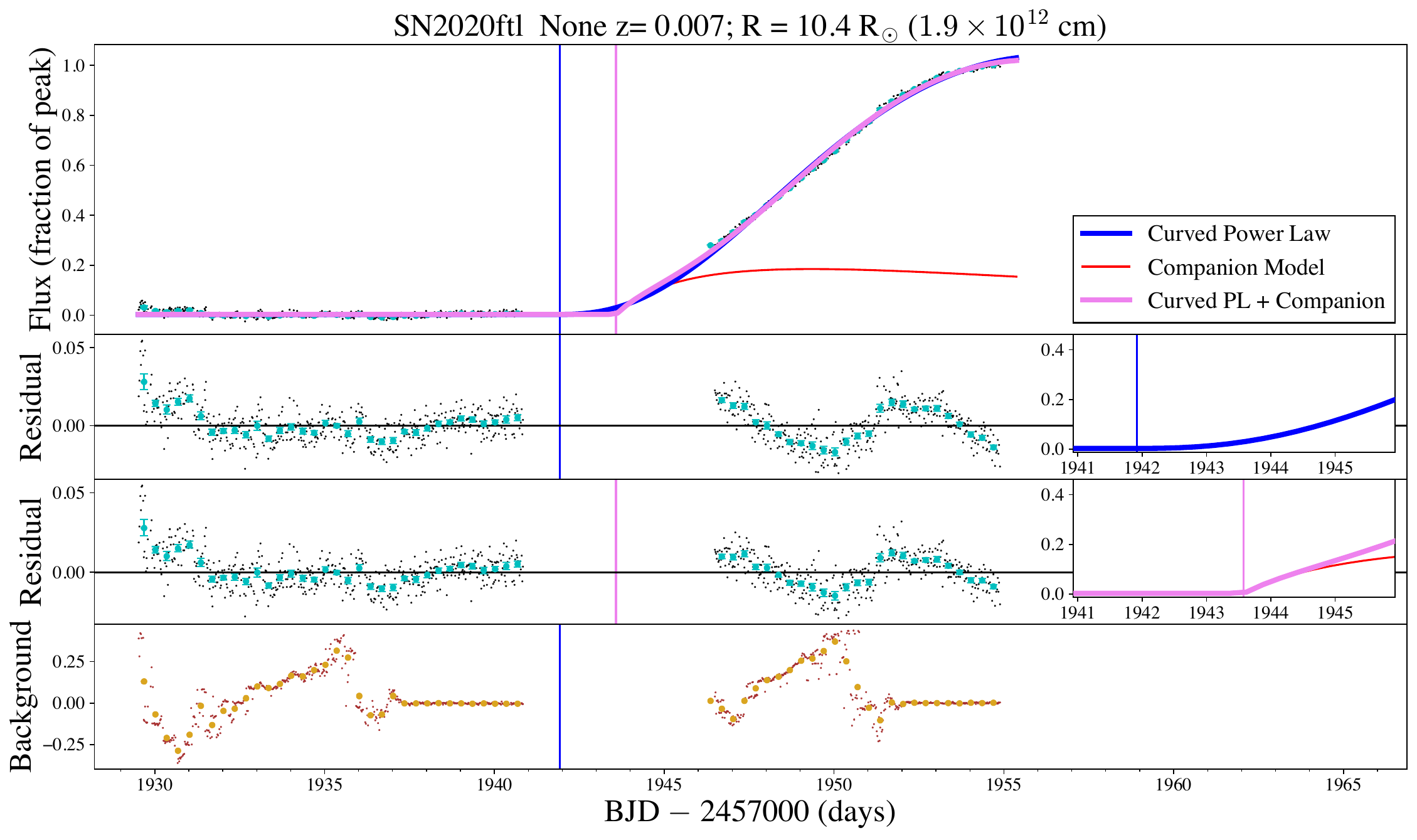}  \\
	    \includegraphics[width=0.70\textwidth]{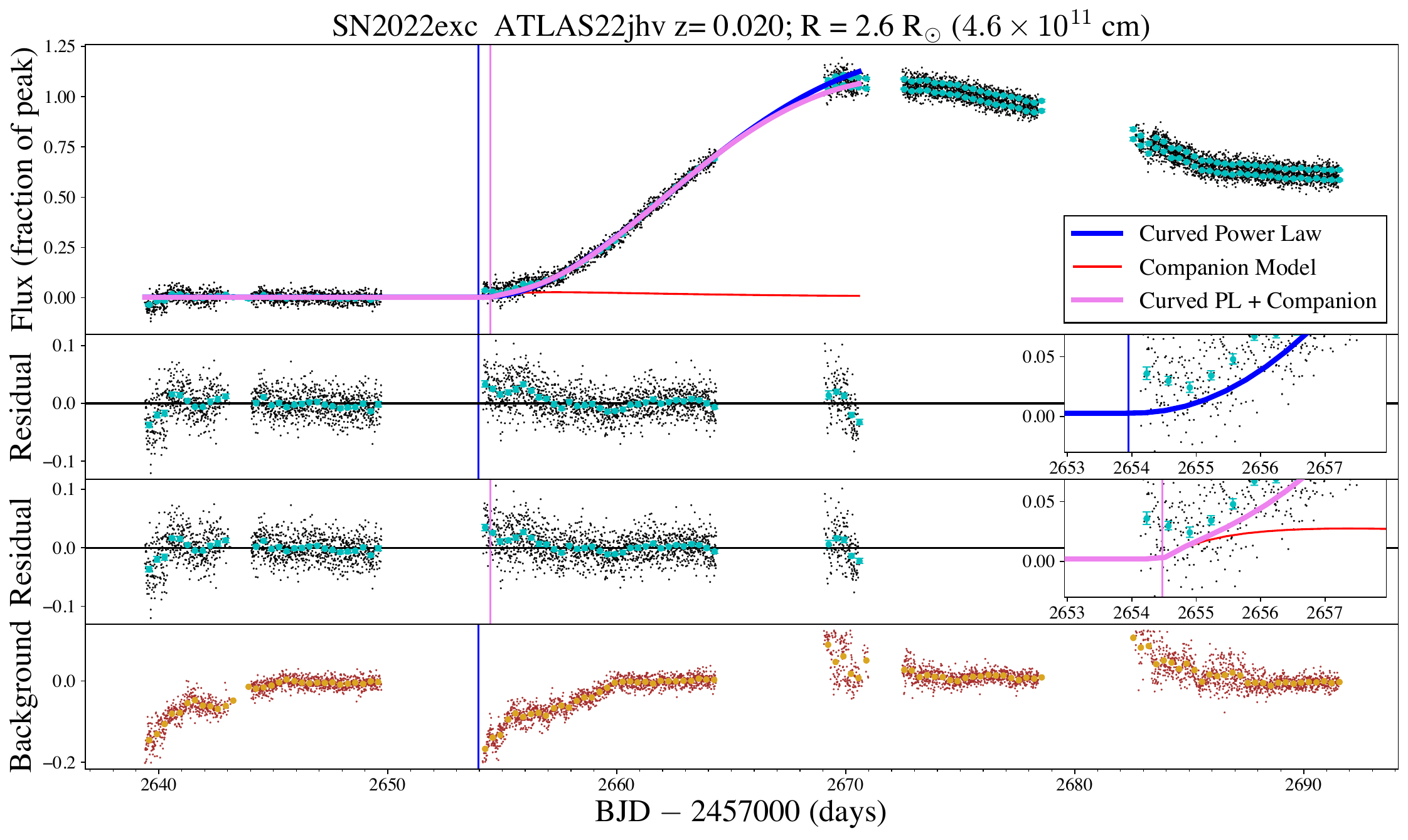}\\
	    \end{tabular}
    \caption{SN light curves with very large Bayes factors or BIC values that strongly prefer companion interactions, but are false positives due to red noise and/or data gaps.  The panels are formatted and marked in the same way as Figure~\ref{fig:SN_model_fits}.}		
   \label{fig:falsepositives}
\end{figure*}

\begin{figure*}
    \centering
	\begin{tabular}{c}	
	        \includegraphics[width=0.70\textwidth]{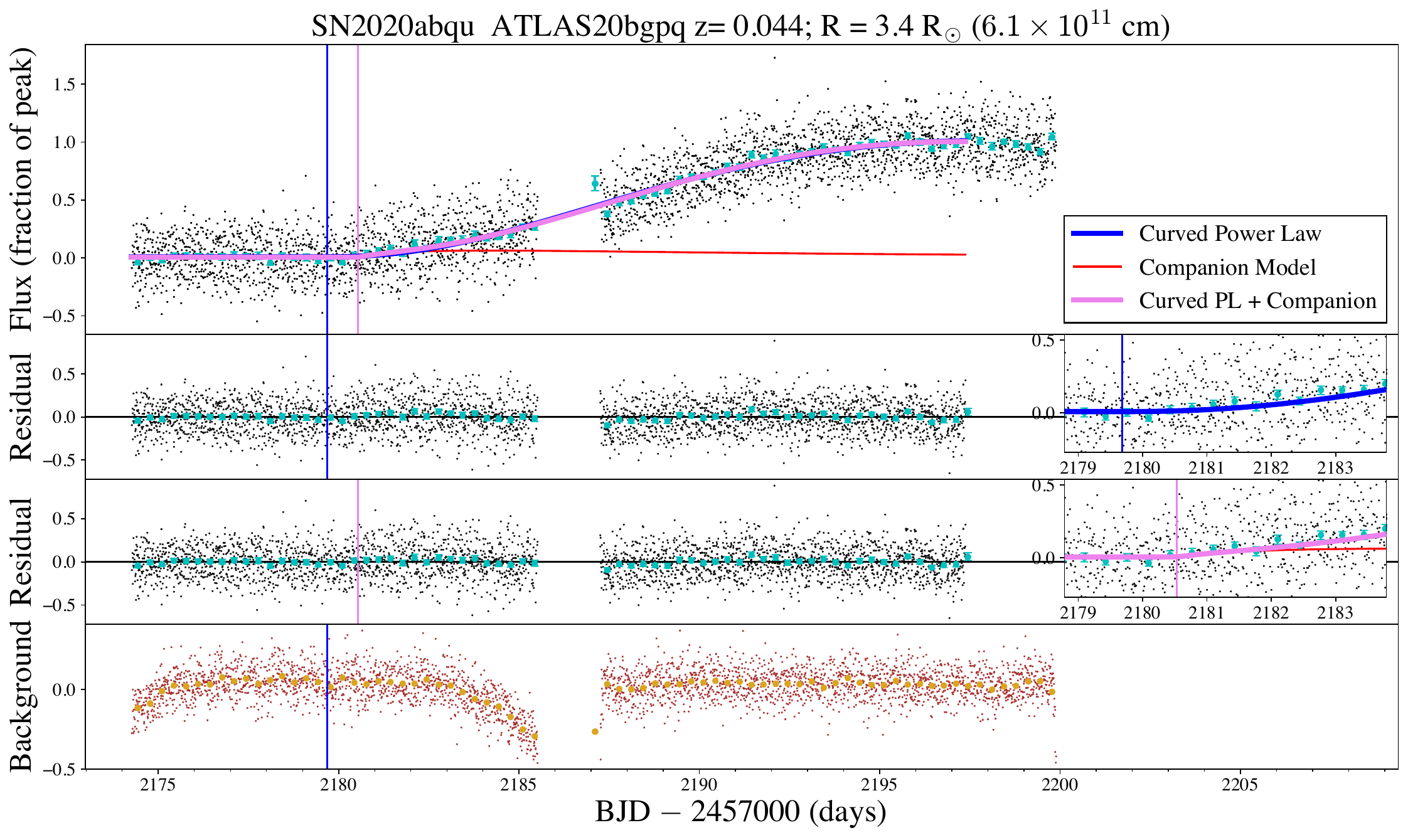}\\
	\includegraphics[width=0.70\textwidth]{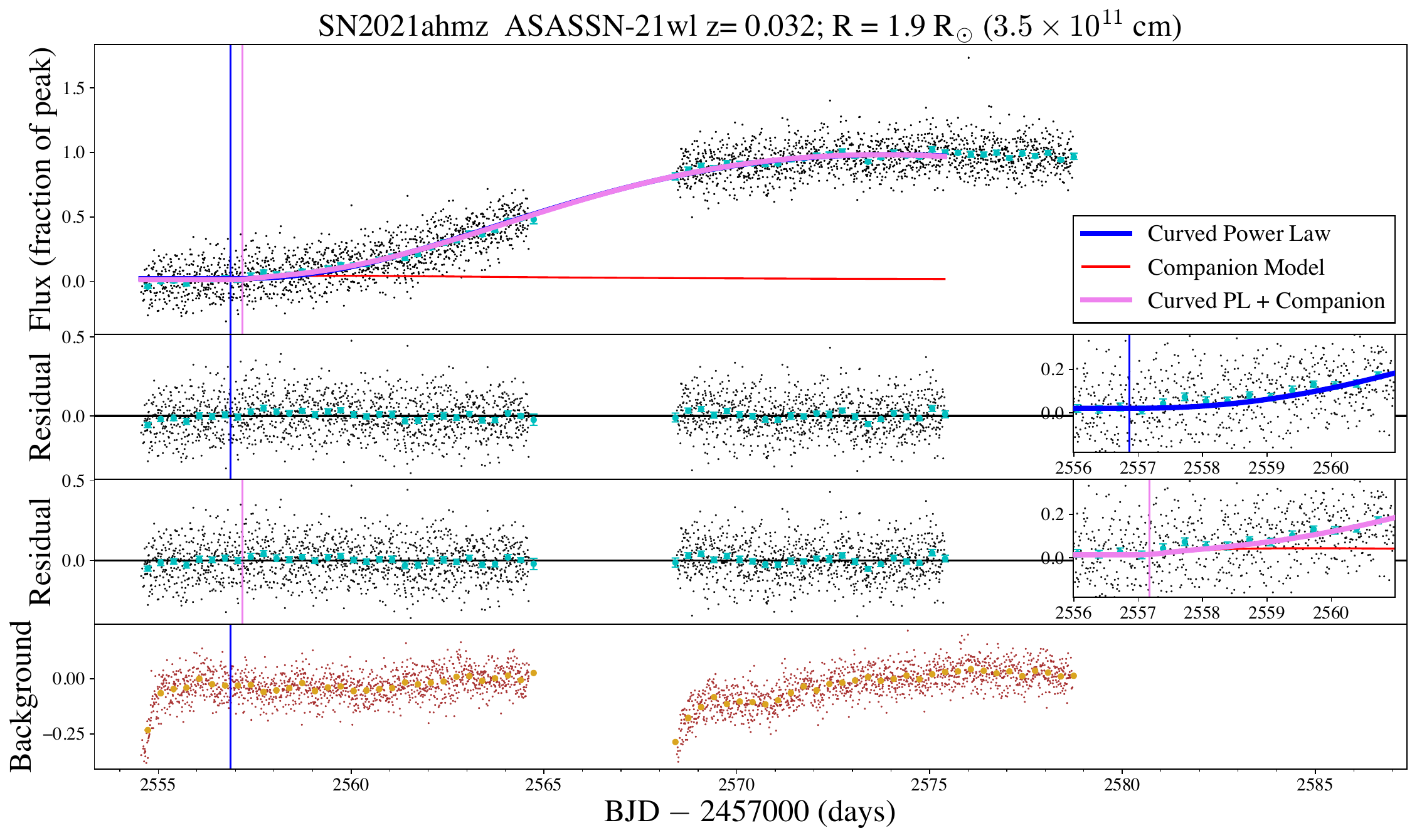} \\
		\includegraphics[width=0.70\textwidth]{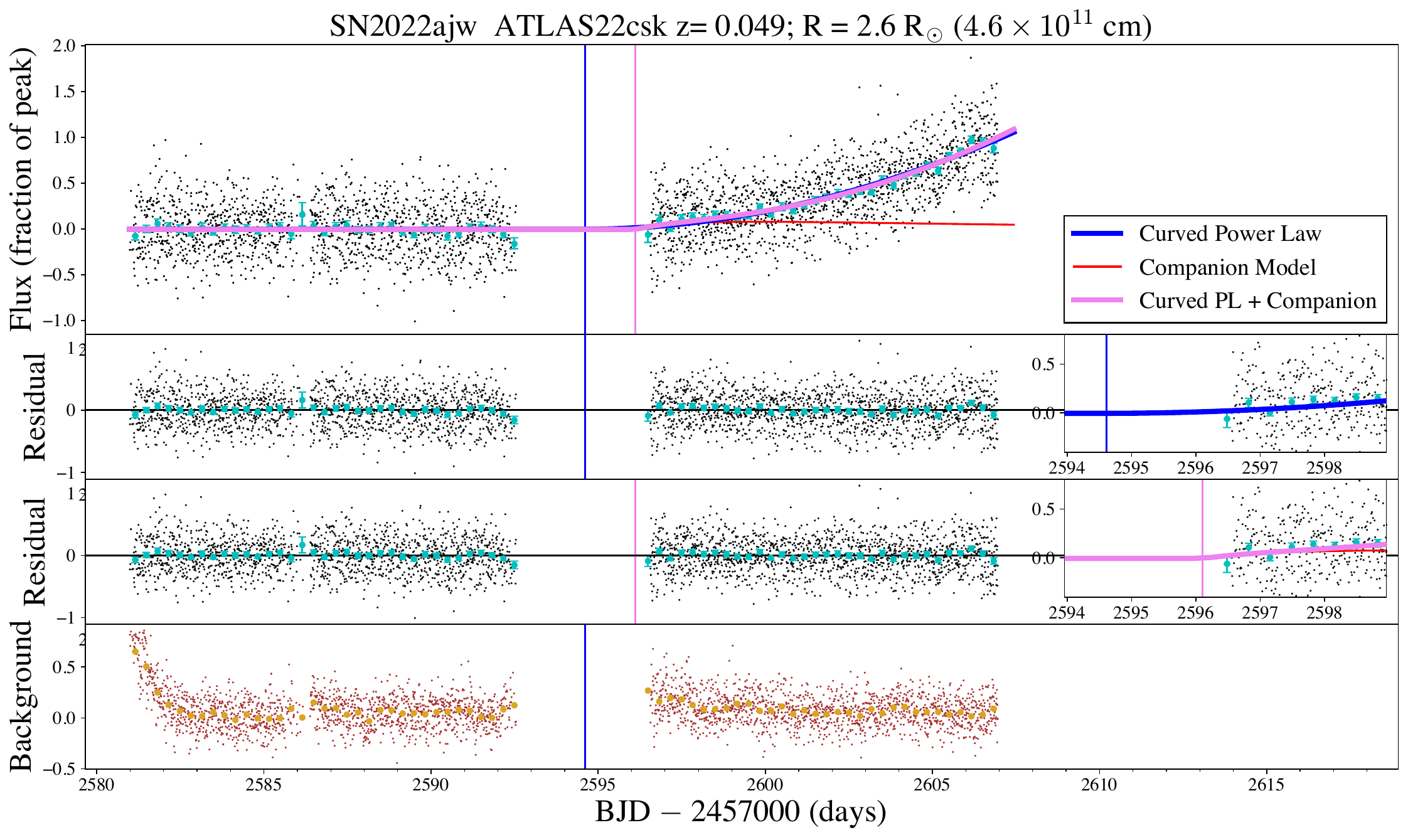} \\
	    \end{tabular}
    \caption{SN light curves approaching Gaussian statistics that prefer companion signatures in addition to a curved power law model.  Note that the evidence for the companion interaction vanishes when using the BIC for all three SN. For SN2022ajw, the Bayes factor is just below our fiducial threshold. The panels are formatted and marked in the same way as Figure~\ref{fig:SN_model_fits}.  }
   \label{fig:prefercomp}
\end{figure*}

\begin{figure*}
    \centering
	\begin{tabular}{c}	
	\includegraphics[width=0.70\textwidth]{models_2020tld.pdf}  \\
	\includegraphics[width=0.70\textwidth]{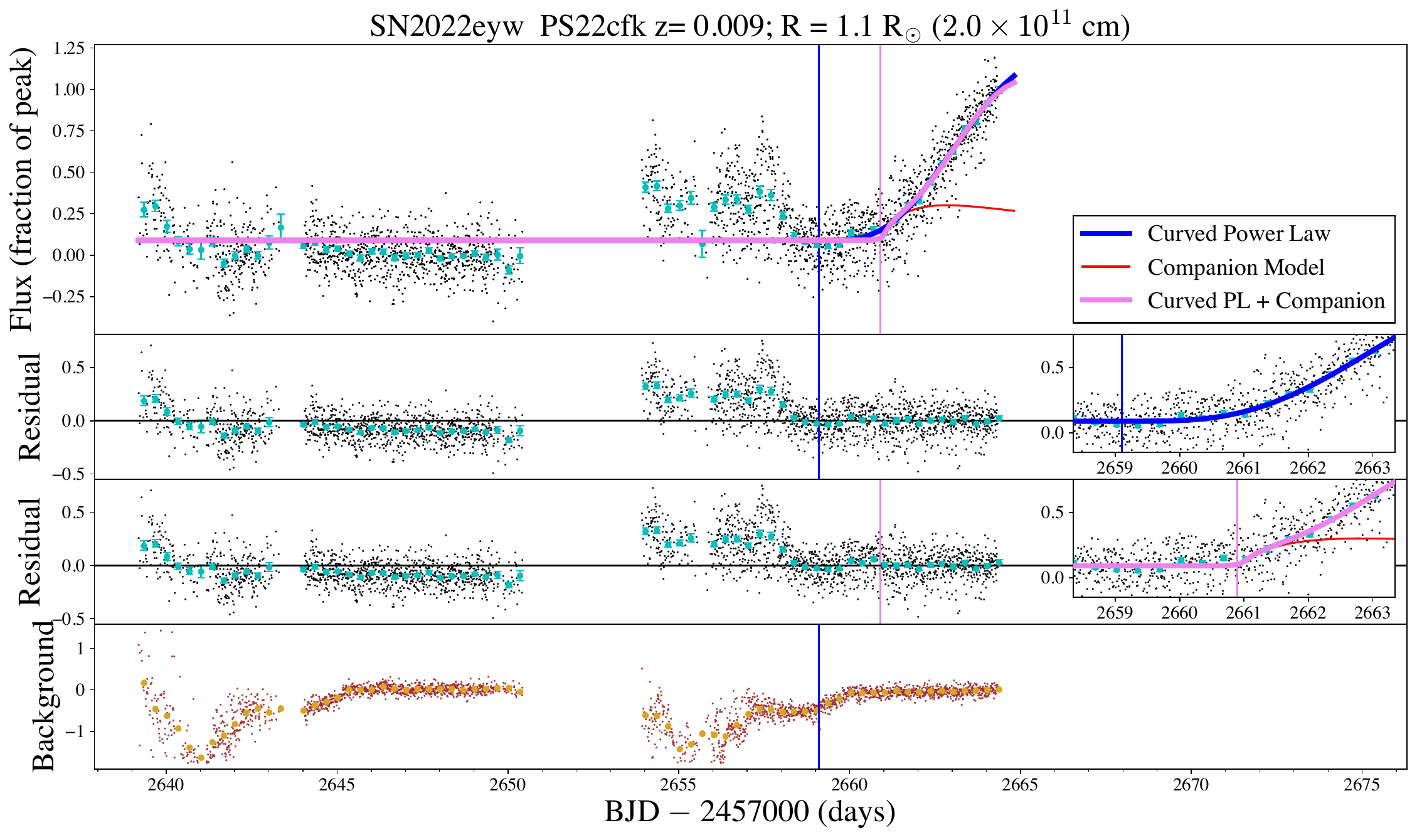}\\
	    \end{tabular}
    \caption{SN light curves approaching Gaussian statistics that disfavor companion signatures.  The panels are formatted and marked in the same way as Figure~\ref{fig:SN_model_fits}.   SN2022eyw has excess noise  from scattered light before the time of first light, but the early rise of the SN itself is reliable.}
   \label{fig:disfavorcomp}
\end{figure*}

\begin{figure*}
    \centering
	\begin{tabular}{c}	
	\includegraphics[width=0.70\textwidth]{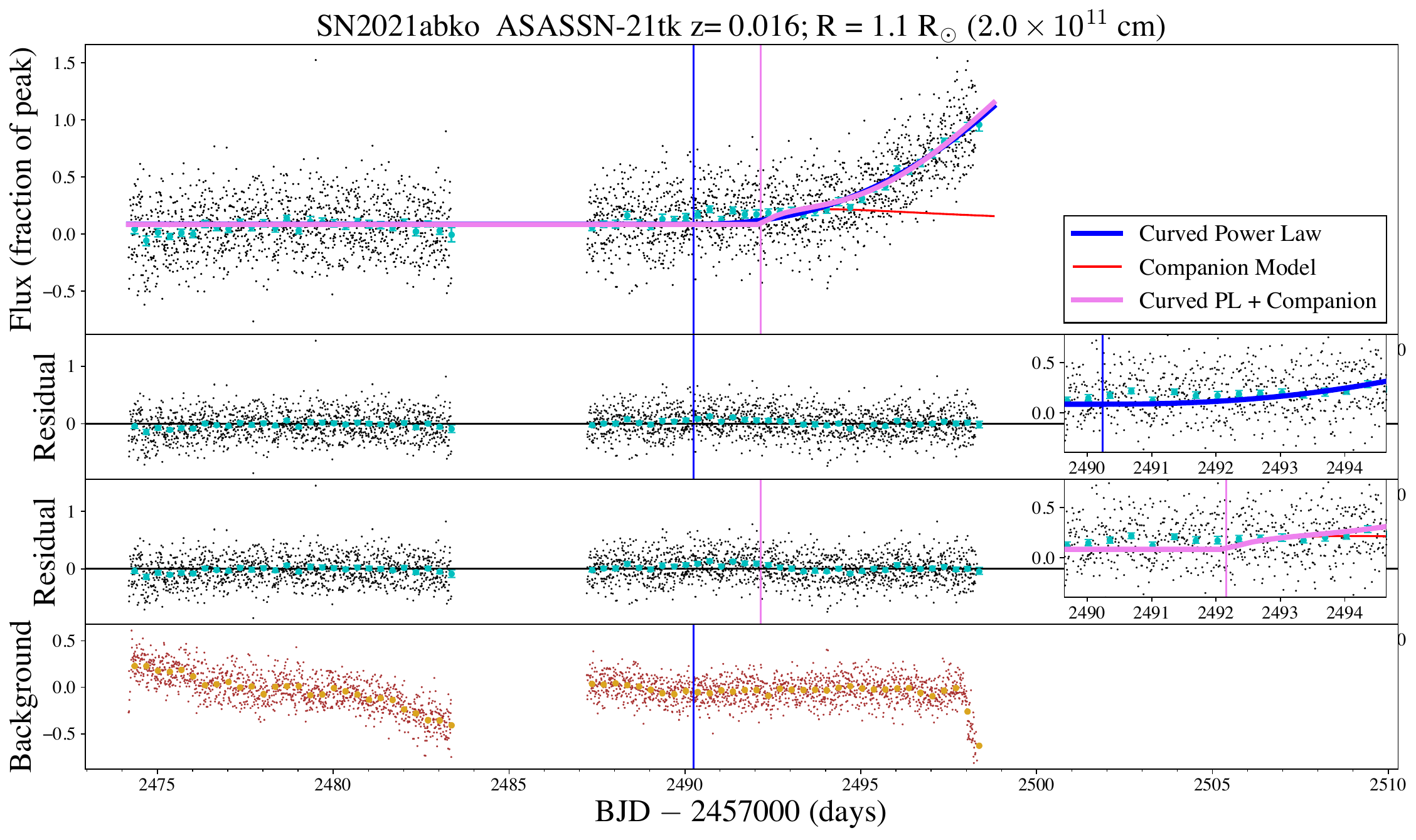}  \\
	    \end{tabular}
    \caption{Light curve for SN2021abko that disfavor companion signatures but is affected by residual systematic errors.  The panels are formatted and marked in the same way as Figure~\ref{fig:SN_model_fits}.   }
   \label{fig:baddisfavorcomp}
\end{figure*}

In this section, we show the light curves discussed in \S\ref{sec:model_comp}.  This includes the light curves that strongly prefer \citet{Kasen2010} models but are false positives due to red noise and/or data gaps (SN2018fpm, SN2020ftl, and SN2022exc, Figure~\ref{fig:falsepositives}), the light curves within a factor of 3 of Gaussian statistics that show a slight preference for adding a \citet{Kasen2010} model in addition to a curved power law model (SN2020abqu, SN2021ahmz, and SN2022ajw,  Figure~\ref{fig:prefercomp}),  the light curves approaching Gaussian statistics that disfavor \citet{Kasen2010} models (SN2020tld and SN2022eyw, Figure~\ref{fig:disfavorcomp}), and  the light curve that disfavors the \citet{Kasen2010} models but is affected by correlated red noise (SN2021abko, Figure~\ref{fig:baddisfavorcomp}).

\end{document}